\documentclass[aps,
reprint,nofootinbib,nobibnotes,notitlepage,superscriptaddress,twocolumn,prl,
 amsmath,amssymb
]{revtex4-2}

\usepackage{graphicx}
\usepackage{dcolumn}
\usepackage{bm}

\usepackage{fix-cm}
\usepackage{amsmath}
\usepackage{soul}
\usepackage{hyperref}
\usepackage{CJKutf8}
\usepackage{appendix}

\usepackage{graphicx}
\usepackage{duckuments}
\usepackage{tikzducks}
\usepackage{orcidlink}
\usepackage{booktabs}
\usepackage{siunitx}

\newcommand{\hyw}[1]{{\color{black}{#1}}}

\graphicspath{{./}{figures/}}

\begin{document}


\preprint{APS/123-QED}


\title{Galactic-scale Feeding Reveals Warped Hypermagnetized Multiphase Circumbinary Accretion Around Supermassive Black Hole Binaries }

\author{Hai-Yang Wang \orcidlink{0000-0001-7167-6110}}
\email{hw609@cantab.ac.uk}
\affiliation{TAPIR, Mailcode 350-17, California Institute of Technology, Pasadena, CA 91125, USA}
\affiliation{Walter Burke Institute for Theoretical Physics, California Institute of Technology, Pasadena, CA 91125, USA}

\author{Minghao Guo \orcidlink{0000-0002-3680-5420}}
\affiliation{Department of Astrophysical Sciences, Princeton University, Princeton, NJ 08544, USA}

\author{Elias R. Most \orcidlink{0000-0002-0491-1210}}
\affiliation{TAPIR, Mailcode 350-17, California Institute of Technology, Pasadena, CA 91125, USA}
\affiliation{Walter Burke Institute for Theoretical Physics, California Institute of Technology, Pasadena, CA 91125, USA}

\author{Philip F. Hopkins \orcidlink{0000-0003-3729-1684}}
\affiliation{TAPIR, Mailcode 350-17, California Institute of Technology, Pasadena, CA 91125, USA}
\affiliation{Walter Burke Institute for Theoretical Physics, California Institute of Technology, Pasadena, CA 91125, USA}

\author{Aretaios Lalakos \orcidlink{0000-0002-6883-6520}}
\affiliation{TAPIR, Mailcode 350-17, California Institute of Technology, Pasadena, CA 91125, USA}
\affiliation{Walter Burke Institute for Theoretical Physics, California Institute of Technology, Pasadena, CA 91125, USA}

\date{\today}

\begin{abstract}

Supermassive black hole (SMBH) binaries 
in gaseous and stellar environments
are prime targets for next-generation space-based gravitational wave detectors.
Yet, realistic accretion conditions under which these binary systems evolve are not fully understood. In this work, we demonstrate the hypermagnetized multi-phase nature of the surrounding accretion flow formed by large-scale feeding from a galaxy background. 
Our simulations indicate that the hypermagnetized circumbinary disk is eccentric and warped, 
 hosting a hot gas core for a parsec-scale separated binary. We also observe collimated bipolar magnetic tower-like outflows launched from each SMBH.

\end{abstract}

\maketitle


\begin{figure*}
    \centering
    \includegraphics[width=0.95\linewidth]{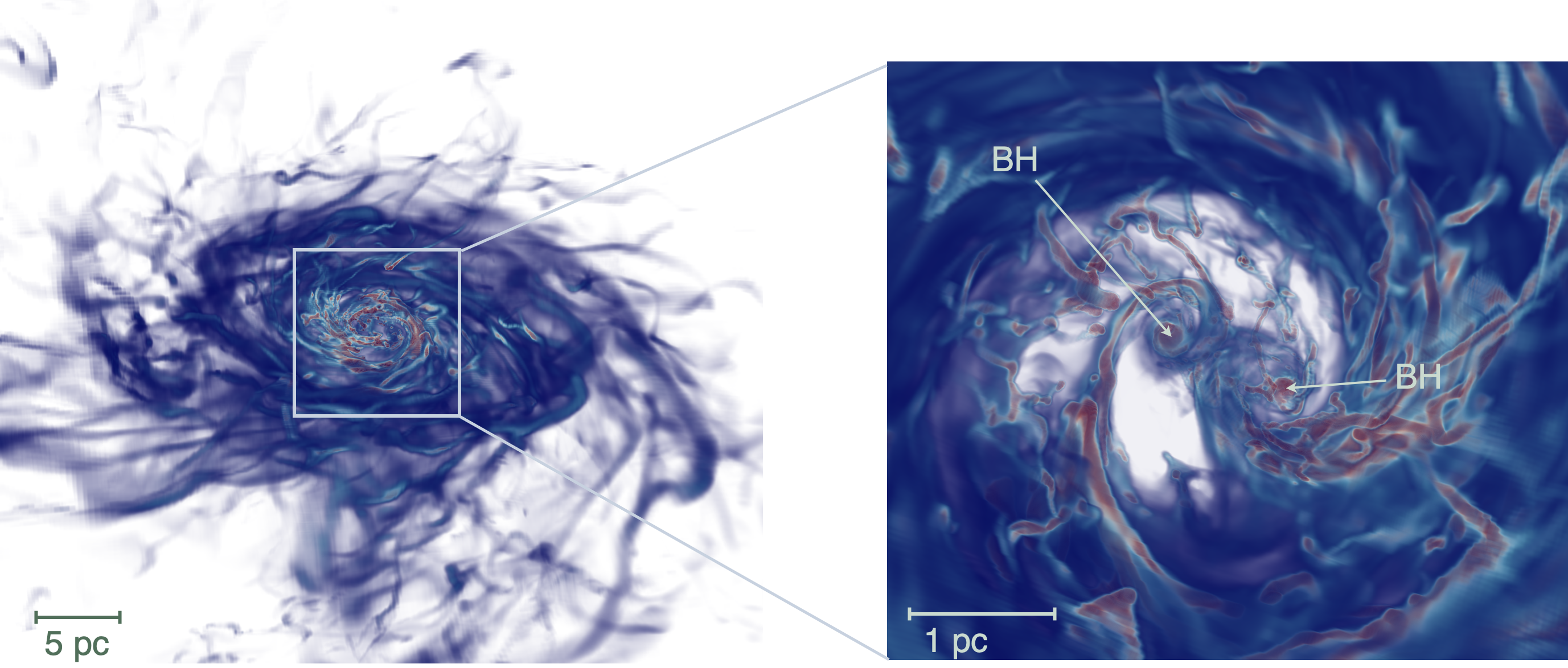}
    \caption{Circumbinary disk formation on parsec scales. (\textit{Left}) The outer boundary of the circumbinary disk is shown, which forms as cold magnetized streams gradually circularize. The relative inclination, or warp, of the circularizing stream and the inner part of the disk is evident. (\textit{Right}) Transition between the cold (ordered) disk and a hot low-density turbulent accretion flow, which occupies central cavity region (white/transparent). Each black hole is surrounded by a mini-disk (indicated by white arrows). The binary's orbital axis is in polar alignment (perpendicular to) the disk. }
    \label{fig:1}
\end{figure*}



\textit{Introduction}. 
Significant theoretical and observational progress has been made in studying the coevolution of single supermassive black holes (SMBHs) and their host galaxies \cite{Kormendy:2013dxa}. It is now understood that almost all galaxies harbor SMBHs at their centers \cite{Kormendy:2013dxa}, which co-evolve with their host galaxies through accretion and feedback processes \cite{Gaspari:2020pgo}. Advances in accurately modeling SMBH and intermediate-mass BHs evolution have been made possible thanks to various multi-scale zoom-in and zoom-out techniques \cite{2013MNRAS.432.3401G,Ressler:2018yhi,Hopkins:2023ipv,Hopkins2023,Guo2023,Guo2024,Cho:2023wqr,Cho:2024wsp,Shi:2024wgh,Shi2024}, or ab-initio conditions \cite{Kaaz:2024jxl} assuming small scale seperation.
Recently, gas feeding from galactic scales down to the event horizon — spanning nine orders of magnitude, from megaparsecs to milliparsecs — has unveiled the multiphase nature of the accretion processes onto the SMBH, and found the hypermagnetized cold accretion disk that not been studied in the traditional framework \cite{Gaburov:2012jd,Hopkins2023,Guo2024}.

On the other hand, SMBH binaries are a key focus for next-generation gravitational wave detectors such as LISA \cite{lisa2023}, TianQin \cite{tianqin}, and Pulsar Timing Arrays such as NANOGrav\cite{Nanograv2023a,Nanograv2023b,InternationalPulsarTimingArray:2023mzf,NANOGrav:2023pdq,NANOGrav:2023vfo,NANOGrav:2023hde,NANOGrav:2023ctt,NANOGrav:2023tcn}. The coevolution of SMBH binaries and their host galaxies, driven by galaxy mergers, is mostly studied on galactic scales \cite{Toomre:1972vt,Barnes:1992rm,Boylan-Kolchin:2005cvz,Cox:2006hd,Naab:2005wz,Conselice:2014joa,Fiacconi:2014lba,Talbot:2023kfn}. Before entering the stage when the orbital evolution of the binary is dominantly driven by gravitational wave radiation \cite{Peters:1964zz}, the binary's surrounding astrophysical environment including dark matter \cite{Alonso-Alvarez:2024gdz,NANOGrav:2024nmo}, stars \cite{Quinlan:1996vp,Milosavljevic:2001vi,2003AIPC..686..201M,Berczik:2006tz,Sesana:2006xw,Rantala:2016rng}, tertial black hole \cite{Hoffman:2006iq,Bonetti:2017dan,Mannerkoski:2021lal}, and gas \cite{Lai:2022ylu} could significantly alter their evolution path. 
Despite the complexity of the environment, it is generally believed that a circumbinary accretion disk (CBD) forms around the central SMBH binary \cite{Mayer:2007vk,Dunhill:2014oka,Goicovic:2015kda,Goicovic:2016dul,Goicovic:2018xxi}, as the product of efficient gas funnelling into the center of the merger remnant of the gas-rich galaxies \cite{Barnes:1991zz,Barnes:1992rm,Mihos:1995ri,Barnes:2002sh,DiMatteo:2005ttp,Hopkins:2005fb,Hopkins:2012fd,Mayer:2007vk,Cox:2007mn,Johansson:2008ib,Capelo:2014gqa,Capelo:2016vif}. 
However, the detailed evolution of the SMBH pair and the surrounding circumbinary disk (CBD) remains unclear, mainly due to limitations in computational resources and a lack of understanding of the physical processes occurring at different scales.

A central topic in the study of circumbinary disks and SMBH binaries in recent years is the `final-parsec problem' \cite{1980Natur.287..307B}. This problem focuses on how to efficiently reduce the separation of SMBH binaries from kiloparsecs to milliparsecs, particularly how to overcome the orbital migration barrier around one parsec\footnote{Depending on the SMBH binary mass}. 
Proposed solutions include stellar dynamical friction \cite{2005ApJ...630..152E}, angular momentum extraction from circumbinary gaseous accretion disks 
\cite{Cuadra:2008xn,Nixon:2010by,Roedig:2012nc,DOrazio:2012vqt,Nixon:2013qfa,Farris:2013uqa,Miranda2017,Moody:2019nes,Munoz:2018tnj}, and interactions with dark matter \cite{Alonso-Alvarez:2024gdz,NANOGrav:2024nmo}. 
One key and so far lacking ingredient, therefore, are realistic accretion conditions under which the binary's orbit evolves.

In most numerical studies of accretion disks, investigations of binary-disk interactions typically begin with an idealized disk setup (e.g., \cite{Munoz:2018tnj}), where the disk's tidal truncation radius is several times the binary separation \cite{Artymowicz:1994bw}. Furthermore, such problems are usually explored using two-dimensional hydrodynamical simulations, facilitating the exploration of a multi-dimensional parameter space (such as 
binary mass ratio \citep{Duffell2020,Derdzinski:2020wlw,Siwek:2022xhf,Siwek:2023rlk}, 
eccentricity \citep{DOrazio2021,Siwek2023b,Siwek2023a}, 
disk aspect ratio  \citep{Tiede2020,Dittmann2023b}, 
viscosity \cite{Dittmann2022,Dittmann2023b}, 
and equation of state \cite{Sudarshan2022,Wang2022,Wang2023,Tiwari:2025imm}). 
However, recent work has shown that the inclusion of magnetic fields in full three dimensions alters the accretion properties \cite{Shi:2011us,Noble:2012xz,Noble:2021vfg}, even in ways that cannot be captured in two-dimensional simulations \cite{Most:2024qus,Most:2024onq,Tiwari:2025imm,Ennoggi:2025nht}.
This includes a magnetically arrested state regulated by flux eruptions \cite{Most:2024qus,Most:2024onq} (see Refs. \cite{Narayan2003,Tchekhovskoy:2011zx,Ripperda:2021zpn} for single SMBHs, see also Ref. \cite{Liska:2022jdy}).\\
In this study, we utilize novel multi-scale zoom-in three-dimensional magnetohydrodynamical (MHD) simulations \cite{Guo2024} to model the formation of the CBD and the long-duration binary-disk interaction driven by galaxy-scale feeding. 
We show that a CBD forms spontaneously through the circularization of cold accretion streams \hyw{from} kiloparsec scales; 
the disk is intrinsically eccentric, warped \cite{Fragner:2009mk} {(see also Ref. \cite{Ressler:2023ptc} similar observations in wind-fed scenarios of the galactic center)}, hypermagnetized and multiphase, depending on the distance to the central binaries.
The accretion environment and dynamics our simulations reveal are unlike those commonly considered in the context of SMBHB.

\begin{figure*}
    \centering
    \includegraphics[width=0.98\linewidth]{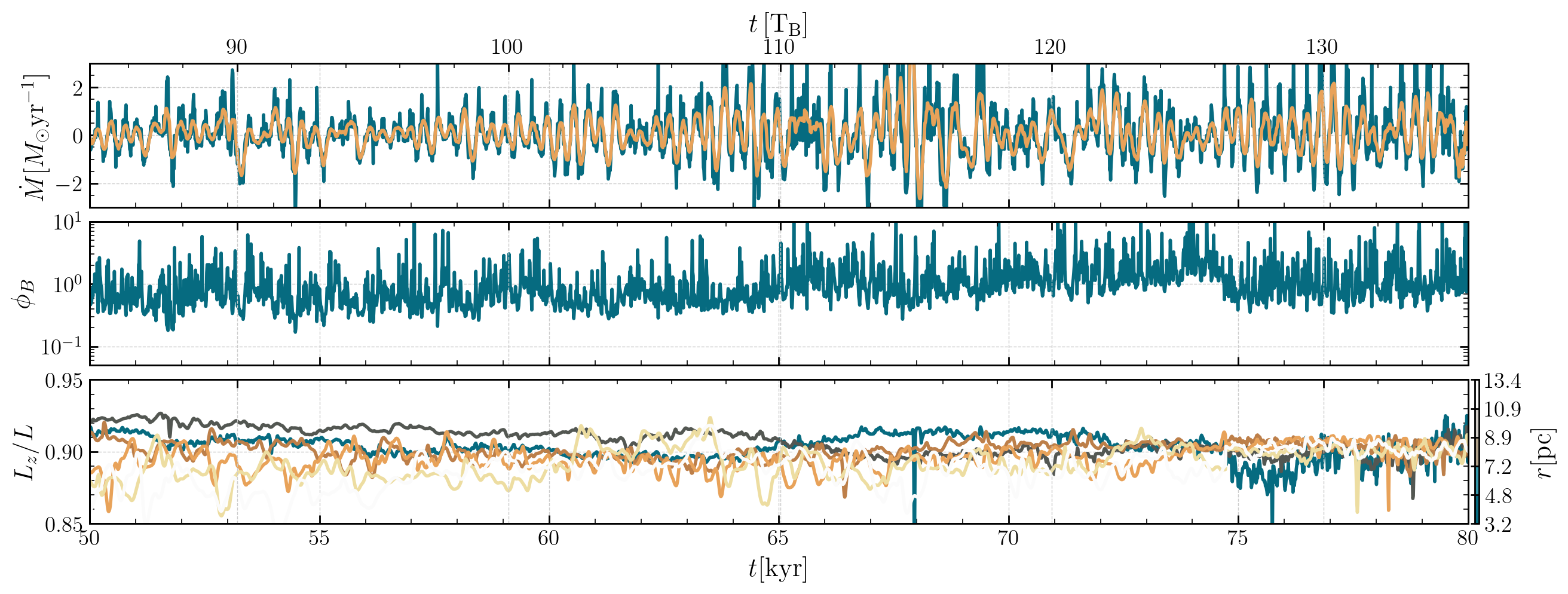}
    \caption{Evolution of accretion properties through the cavity. {\it (Top)} Mass accretion rate $\dot{M}$ (blue (instantaneous), orange (smoothed)). The accretion rate variability is dominated by binary orbital period, $T_B$.
    {\it (Center)} Dimensionless magnetic flux $\phi_B$. ({\it Bottom}) Angular momentum fraction, $L_z/L$, aligned with the average disk angular momentum, $L_z$, at various radii, $r$ (colored).  The hypermagnetized disk is warped (radius-dependent alignment) with realignment happening on a secular evolution timescale. }
    \label{fig:3}
\end{figure*}


\textit{Methods.} 
In this work, we simulate the formation of a hypermagnetized circumbinary accretion disk assembled from the galactic scale spanning over six orders of magnitude. We solve the ideal Newtonian MHD equations, utilizing \texttt{AthenaK} \cite{2024arXiv240916053S}, the performance portable version of the \texttt{Athena++} code \cite{2020ApJS..249....4S}. 
The cooling and heating function, initial and boundary conditions, gravitational field, numerical floors, first-order flux corrections, and sink prescriptions are identical to the high-resolution simulation of single black hole feeding in \cite{Guo2024}, where the single point mass has been replaced with a fixed circular orbit SMBH binary at parsec scale separation. In brief, the gravitational potential
includes a double \hyw{analytic} Navarro-Frenk-White (NFW) profile \cite{1997ApJ...490..493N} for both star and dark matter distributions, and SMBH (binary) mass $M_{\rm{BH}}=6.5\,\times10^9M_\odot$ \cite{EventHorizonTelescope:2019pgp}.
We also adopt a cooling scheme including optically thin bremsstrahlung and line cooling \cite{Guo2023}, and a heating scheme \hyw{shell by shell} maintaining global equilibrium \hyw{but allowing local thermal instability}. To imitate interstellar turbulence, the simulation is initialized with isobaric random density perturbation and entangled magnetic vector potential with plasma $\beta \approx 100$ on large scales.

To bridge the large scale disparity, in the first step, we start from galactic scale with an outer radius of $40\,\rm kpc$ and adopt a stage-wise nested zoom-in approach {with 7 initial levels of mesh refinement}\cite{Guo2023,Guo2024}.
In each stage, we run the simulation till a quasi-steady state at the inner boundary is reached. 
Then we add 4 more levels of mesh refinement at the center. We repeat this process twice. The finest resolution achieved this way is $0.48\,\rm{mpc}$, with the inner radius of the central sink region (central black hole) being $30\,\rm mpc$, with 15 levels of static mesh refinement. The base grid is a cubic box with a resolution of $256^3$ grid points. Compared to Ref. \cite{Guo2024}, we stopped {two stages earlier.}
{The highest resolution domain covers a cube of $1.27\,\rm{pc}^3$ with 1024 grid points per direction.}
Taking this as a quasi-steady state for the inner accretion disk, in the second step, we replace the central black hole with a pair of equal-mass black holes on a circular orbit, with separation $a=1\,\rm{pc}$. 
This value is chosen as it is close to the transition radius of $0.3\, \rm pc$ from the outer cold thin accretion disk and the inner hot turbulent flow (see Results section for details). 
The simulation evolves for another 135 binary orbits (equivalent to 80 $\rm{kyr}$ for the galaxy parameters we adopt). \\
The simulation was carried out on the OLCF Summit system for a total of $350,000$ V100-GPU hours.

\textit{Results.}
In this work, we study the dynamical formation of a circumbinary accretion disk from accretion over six orders of magnitude. The flow on large scales is turbulent and forms cold dense streams falling toward the 
central gravitational source \cite{Guo2023}. Ultimately, these streams collide, circularize and form a circumnuclear disk on a tens of parsec scale. 
This scenario has been extensively investigated by Ref. \cite{Guo2024} in the case of accretion in an M87-like elliptical galaxy background.
As such, we only briefly recap the large-scale accretion dynamics in the following, and mainly focus on the
changes in the accretion environment due to binary-disk interaction.
Following Ref. \cite{Guo2024}, we split the accretion dynamics into three main stages. 
On kpc scales, accretion proceeds quasi-spherically symmetric as a \textit{diffuse hot halo} with infalling diffuse gas ($n\sim0.1\,\mathrm{cm}^{-3}$) with virial temperatures ($T\sim10^7\,\mathrm{K}$). The magnetic field 
on large scales has low initial magnetization ($\beta\sim100$). As the matter falls inwards, strong cooling leads to the formation of 
\textit{chaotic cold magnetic filaments}. These extend down to about $30\, \rm{pc}$ ($10^5\,r_\mathrm{g}$, where $r_\mathrm{g}=GM_\mathrm{BH}/c^2$ is the gravitational radius), are moderately magnetized ($\beta\sim1$), and will ultimately form the circumbinary disk at $30\,\rm{pc}$ truncated at parsec scale (Fig. \ref{fig:1}). 
\hyw{These are similar to the cold accretion streams \cite{2012ApJ...747...26L,2013MNRAS.432.3401G} but appears to be more filamentary \cite{2020MNRAS.493.4065W}.}

\textit{Formation of a warped circumbinary disk.}
%
%
The orientation of the circumbinary disk is (pre-)determined by the angular momentum budget carried by the cold accretion streams at large scales. As a consequence, the angular momentum axis of the disk and the central SMBH binary can be significantly misaligned.
Indeed, as in Fig. \ref{fig:1}, the angular momentum axis is almost perpendicular to that of the inner disk, forming a near-{\it polar alignment} geometry \cite{Aly:2015vqa,Rebecca:2017ApJ...835L..28M} (but see also Ref. \cite{Lepp:2022gni,Childs:2023zsf} for potential limitations on longer timescales and binary eccentricity). 
We can further see, that the disk is warped and globally eccentric, with the outer edge aligning with the incoming cold streams.
We emphasize, however, that the orientation of the disk can drastically vary over the secular timescale at the outer edge, while the orientation of the SMBH binary changes over the viscous timescale at the inner edge. 
Following Ref. \cite{Ressler:2023ptc}, we quantify this explicitly by computing the net angular momentum, $L_z^{\rm binary}$, of the binary orbit aligned with the disk angular momentum, $L$. We find that $L_z^{\rm binary}/L \simeq 0.1$ implying a strong polar alignment 
relative to the binary (see also Fig. \ref{fig:1}, right panel). {Namely, the angular momentum vector of the binary and that of the disk are perpendicular.} 
In addition, we quantify the effective warp of the disk by comparing the angular momentum, $L_z$, averaged at spheres of different radii with the average angular momentum of the disk (Fig.\ref{fig:3}). 
We see that the annuli at different radii have relative inclination, illustrating a globally warped disk (between 3 and 13 pc). Over time, the disk gradually re-aligns to almost co-planar configuration by the end of the simulation. We caution that the warp could be re-excited by new incoming cold streams on longer time scales than we consider.

\begin{figure*}
    \centering
    \includegraphics[width=0.95\linewidth]{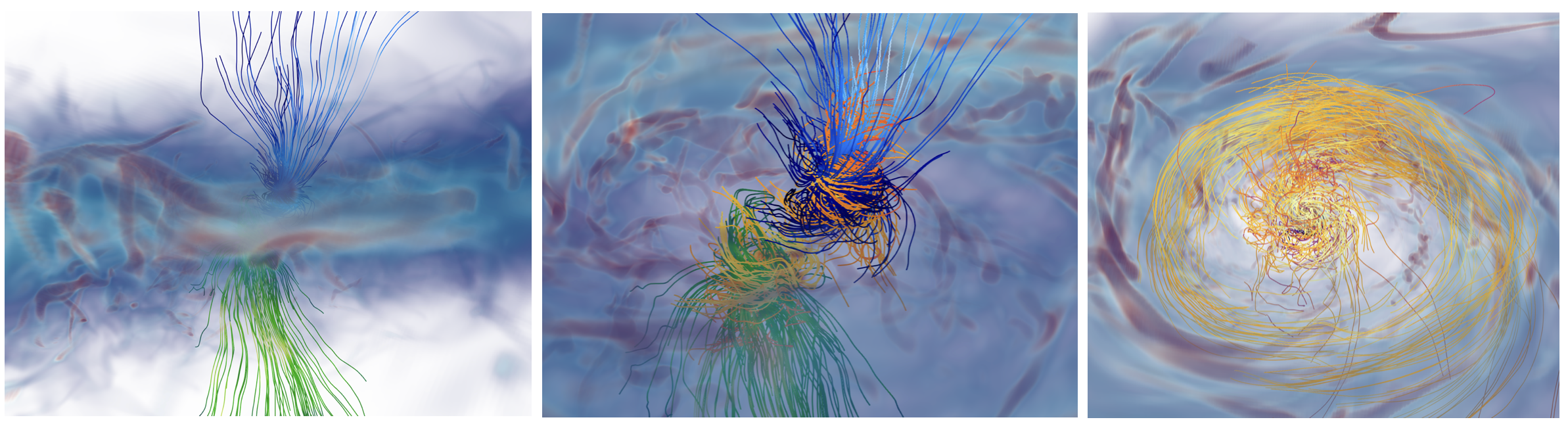}
    \caption{Magnetic field and outflow structure after 135 binary orbits. (\textit{Left}) Large-scale bipolar outflows (dual jets) as indicated by velocity streamlines (green and blue). (\textit{Center}) The outflows are anchored in each mini-disk, and are launched via a magnetic tower mechanism. (\textit{Right}) Strongly ordered toroidal field inside the hypermagnetized accretion disk, whereas the cavity is largely filled with turbulent magnetic field.}
    \label{fig:2}
\end{figure*}


As we show below, the disk is cold and geometrically thick due to magnetic support (as in Refs. \cite{Guo2024} and \cite{Hopkins2023,Gaburov:2012jd}, but see also \cite{Squire:2024yhe} for potential caveats). The cold disk is truncated roughly at $3\,\mathrm{pc}$ in our fiducial run with binary separation $a=1\,\mathrm{pc}$. Unlike the disk surrounding a single SMBH in which the truncation happens at $0.3\,\mathrm{pc}$ \cite{Guo2024}, the binary alters the inner disk morphology significantly. 
In the case of a binary the precise disk truncation is impacted by tidal forces of the binary, and will occur near the inner Lindblad resonance of the disk \cite{Artymowicz:1994bw,2015MNRAS.452.2396M}. In contrast to the widely studied coplanar binary configurations, polar-aligned binaries yield a more irregular accretion process between the circumbinary disk and the mini-disks \cite{Nixon:2013qfa,Moody:2019nes}.
Indeed, we observe a cavity penetrated by two spiral accretion streams connecting the CBD edge and two mini-disks surrounding each SMBH. The mini-disks are inclined \cite{Nixon:2013qfa,2018MNRAS.475.5618B,Moody:2019nes} relative to the binary. The cavity is filled with secondary \hyw{cold} accretion streams \hyw{triggered by thermal instability}.

\begin{figure}
    \centering
    \includegraphics[width=0.48\textwidth]{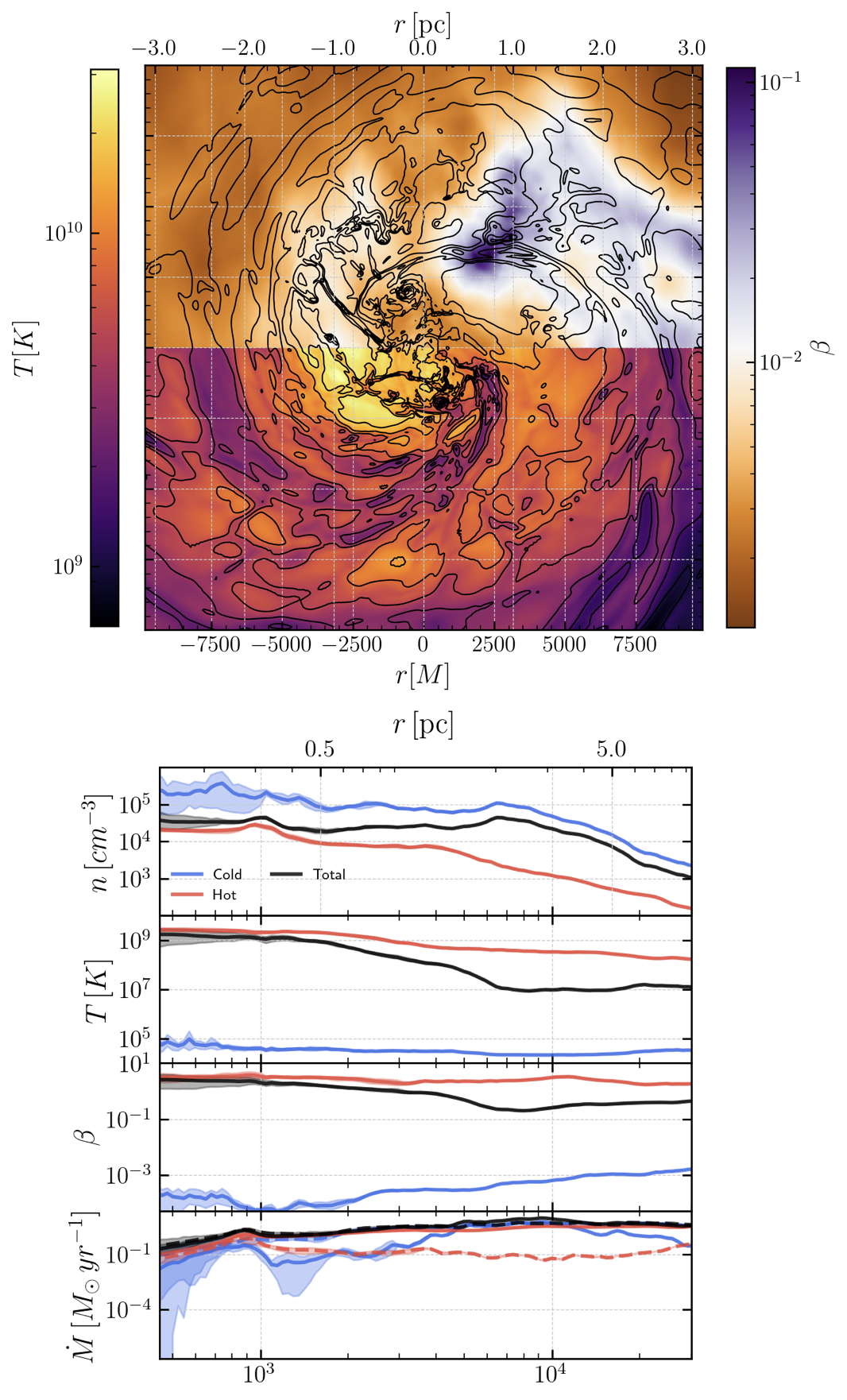}
    \caption{Multi-phase properties of the circumbinary acccretion disk {after $135$ binary orbits.} {\it (Top)} Temperature $T$ and plasma $\beta$ parameter in the disk mid-plane. Solid black lines indicate iso-density contours, highlighting the mini-disks and spiral accretion streams.
    {\it (Bottom)} {Adjusting to the normal coordinate of the disk}, azimuthally averaged radial profiles of disk quantities also showing the density, $T$, $\beta$, and $\dot{M}$ for two representative temperature ranges (cold, below $3\times10^4\,\rm{K}$; and hot, similar to initial condition) \cite{Guo2024} shown in the center panel. 
    For $\dot{M}$, the solid and dashed lines separately represent inflow and outflow $\dot{M}$. Lines show the mean and the shaded ranges show $10\%$ to $90\%$ volume inclusion interval for the final $1\,\rm{kyr}$.
    }
    \label{fig:4}
\end{figure}

\textit{Hypermagnetized multi-phase accretion flow.}
Another noteworthy aspect of the multiphase structure is the magnetic field property (Fig. \ref{fig:2}). Flux freezing facilitates the enhancement of the magnetic field during the formation/condensation of the cold filaments with plasma-$\beta\simeq 1$ (Fig. \ref{fig:4}). This effect is further amplified by the differential rotation within the parsec-scale CBD, which reduces $\beta$ to approximately $10^{-3}$, and leads to a dominance of toroidal magnetic fields — a configuration reminiscent of those observed in disruption scenarios (e.g., \cite{Etienne:2011ea,Most:2021ytn,Izquierdo:2024rbb}).
Similar disk structures have been reported previously around single SMBH formed from disruption of magnetized
interstellar gas clouds \cite{Gaburov:2012jd}.
Inside the cavity, the presence of the binary SMBH drastically changes the magnetic field configuration at parsec scales compared to the multiphase accretion structure formed around single SMBH \cite{Guo2023,Guo2024,Hopkins2023}, since the approximate spherical symmetry of hot turbulent accretion flow is broken. The tidal accretion streams connecting the outer disk and mini-disks impose a preferred orientation on the magnetic flux within the cavity.
In Fig. \ref{fig:3}, we further quantify the evolution of normalized magnetic flux, $\phi_B = \sqrt{\pi} \Phi_B/\sqrt{|\dot{M}| r^2 v_K}$ \footnote{Note that this expressions differs from Ref. \cite{Ressler:2020voz} by the choice of units \cite{Guo2024}.}, using a sphere centered at the center of mass of the binary, with radius $0.94\,\rm{pc}$, where $v_K=\sqrt{GM/r}$ is the local \hyw{analytical} Keplerian velocity, and $\Phi_B(r) = \oint_{\rm d} S_r \left|B^r\right|$.
After an initial period ($t < 40\, \rm kyr)$, the cavity settles into a mini-disk regime consistent with three dimensional simulations of standard and normal (SANE) CBD accretion flows \cite{Combi:2021dks,Ennoggi:2025nht}. As we can see the magnetic flux inside the cavity is continuously {but slowly} growing with $\phi_B \simeq \mathcal{O}(1)$ by the end of the simulation. Whether this growth continues, or whether the system will eventually approach a magnetically arrested (MAD) state \cite{Most:2024qus} (see also Refs. \cite{Narayan2003,Igumenshchev:2007bh,Tchekhovskoy:2011zx}) cannot be conclusively predicted, especially due to {the particular choice of cooling and heating function we adopted. The hypermagnetized nature of the surrounding accretion flow, together with the turbulent but less magnetized cavity, can potentially suppress interchange instabilities \cite{Spruit:1995fr,Kulkarni:2008vk} characteristic of the MAD regime \cite{Begelman:2021ufo}.}

Apart from the magnetization properties of the flow, it is important to also highlight the multi-phase nature of the circumbinary accretion disk. To this end, Fig. \ref{fig:4} shows a {vertically integrated two-dimensional slice through the central accretion region in the normal coordinate of the disk.} We can clearly spot a two-temperature dichotomy: Hot ($T>10^{10}\, \rm K$), turbulent regions of the central accretion flow carry lower magnetizations, compared with cool hypermagnetized, $\beta < 0.01$, regions in the outer CBD. 

We further clarify the disk properties by showing angular averaged radial profiles of the CBD (Fig. \ref{fig:4}, lower panel). This confirms that only the accretion streams and disk are hypermagnetized ($\beta <10^{-3}$), whereas hot regions feature near pressure equipartition ($\beta\simeq 1$) 

\textit{Summary}.
\label{sec:conclusion}
In this work, we have shed light on the nature of a realistic accretion flow around a SMBHB. Using MHD zoom-in simulations over six orders of magnitude, we demonstrate the formation and long-term evolution of a multi-phase circumbinary accretion disk, formed from large-scale feeding at the galaxy scale. 
Gas virializing beyond kpc scale, becomes thermally unstable at $0.03-3\mathrm{kpc}$ and accretes through dense cold magnetized filaments \cite{2012ApJ...747...26L,2013MNRAS.432.3401G}. These filaments circularize at $30\,\mathrm{pc}$ and form a warped circumbinary disk at almost Keplerian rotation (see also Ref. \cite{Fragner:2009mk,Nixon:2013qfa}).
{Long-time high-resolution simulations will be valuable to better understand the dynamics of these warped CBDs (akin to, e.g., \cite{Kaaz:2022fbg,White:2019udt,2007ApJ...668..417F} for single BH accretion).}

This circumbinary disk is cold, dense, magnetically supported by a predominantly toroidal field \cite{Gaburov:2012jd,Hopkins2023,Guo2024}, and drives hot, strong magnetized bipolar outflows. Such outflows from the mini-disks have been previously reported on horizon scales \cite{Palenzuela:2010nf,Combi:2021dks,Ressler:2024tan,Ennoggi:2025nht}, as well as in the context of idealized MAD CBD scenarios \cite{Most:2024onq,Most:2024qus}. In the case of a single SMBH, the use of realistic cooling naturally truncates the circumnuclear disk at $0.3 \rm{pc}$ \cite{Guo2024}, whereas the binary tidally truncates it at $2-3 \rm{pc}$. While embedded in the hot, low-density turbulent environment, the accretion flow onto the binary is always cold and filamentary, before circularizing and forming mini-disks \cite{lin1979,lin1993,Artymowicz:1994bw,Artymowicz:1996zz}.\\
{The nature of turbulent convection at magnetized filamentary feeding scales} leads to a strongly tilted, nearly perpendicular, alignment of the inner circumbinary disk relative to the SMBHB. {For different and more extreme (retrograde) alignments, this may have direct implications for the binaries evolution \cite{Nixon:2011tn}, see also Ref. \cite{Martin:2023dfp}).}\\
Overall, our simulations indicate that commonly considered initial conditions used to inform most studies of circumbinary accretion onto SMBHBs (see, e.g., Ref. \cite{Lai:2022ylu} for a recent review) might not be able to correctly explore realistic accretion geometries expected in these systems. 
\hyw{We will dive into the choice of equation of state, mini-disk dynamics, the interpretation of the different radial scaling, and impact of binary separation in a follow-up work.}

In future work, it is important to integrate these calculations in a cosmological context, akin to recent efforts like \cite{Hopkins2024}, since merging binary SMBHs like those modeled here come from massive merging galaxies \cite{2017MNRAS.464.3131K}. The initial conditions, particularly the nuclear stars (and potentially dark matter), may be important on scales as large as those we model here in such circumstances, and could interact with or shape disk formation. Examining a broader suite of such initial conditions also allows for more a priori prediction of merger orbital parameters and magnetization, which we show play a major role in the disk structure. This is also critical for predicting rates of potential LISA sources \citep{2023LRR....26....2A}. Expanding to more detailed treatments of the gas heating/cooling physics would allow for more complex phase structure, but also more robust predictions of observables from these binary disks. In these stages, there could be unique signatures in emission line structure, jet emission, or variability that indicate a pc-scale SMBH binary even in high-redshift (spatially-unresolved) observations \citep{2019NewAR..8601525D}. There has been tremendous effort to investigate candidate systems observationally at these separations \cite{2014Natur.511...57D,2015ApJ...806..219C,2015ApJ...813..103M,2016MNRAS.463.2145C}, but degeneracies remain between actual binarity and other types of jet/disk structure, and more detailed forward-modeling of observations (of the sort enabled by simulations like these) is critical to break these degeneracies.


%

\textit{Acknowledgement}
The authors are grateful for discussions with Xue-Ning Bai, Kejian Chen, Hongping Deng, Kohei Inoyashi, Wu Jiang, Hui Li, Douglas N. C. Lin, E. Sterl Phinney, Bart Ripperda, A. Tchekhovskoy, Daichi Tsuna, Xing Wei, Huan Yang, Cong Yu, and Feng Yuan.
The simulations were performed on DOE OLCF Summit under allocation AST198.
This research used resources from the Oak Ridge Leadership Computing Facility at the Oak Ridge National Laboratory, which is supported by the Office of Science of the U.S. Department of Energy under Contract No. DE-AC05-00OR22725. PFH was supported by a Simons Investigator Grant.




\bibliography{cbd}

\begin{thebibliography}{138}%
\makeatletter
\providecommand \@ifxundefined [1]{%
 \@ifx{#1\undefined}
}%
\providecommand \@ifnum [1]{%
 \ifnum #1\expandafter \@firstoftwo
 \else \expandafter \@secondoftwo
 \fi
}%
\providecommand \@ifx [1]{%
 \ifx #1\expandafter \@firstoftwo
 \else \expandafter \@secondoftwo
 \fi
}%
\providecommand \natexlab [1]{#1}%
\providecommand \enquote  [1]{``#1''}%
\providecommand \bibnamefont  [1]{#1}%
\providecommand \bibfnamefont [1]{#1}%
\providecommand \citenamefont [1]{#1}%
\providecommand \href@noop [0]{\@secondoftwo}%
\providecommand \href [0]{\begingroup \@sanitize@url \@href}%
\providecommand \@href[1]{\@@startlink{#1}\@@href}%
\providecommand \@@href[1]{\endgroup#1\@@endlink}%
\providecommand \@sanitize@url [0]{\catcode `\\12\catcode `\$12\catcode `\&12\catcode `\#12\catcode `\^12\catcode `\_12\catcode `\%12\relax}%
\providecommand \@@startlink[1]{}%
\providecommand \@@endlink[0]{}%
\providecommand \url  [0]{\begingroup\@sanitize@url \@url }%
\providecommand \@url [1]{\endgroup\@href {#1}{\urlprefix }}%
\providecommand \urlprefix  [0]{URL }%
\providecommand \Eprint [0]{\href }%
\providecommand \doibase [0]{https://doi.org/}%
\providecommand \selectlanguage [0]{\@gobble}%
\providecommand \bibinfo  [0]{\@secondoftwo}%
\providecommand \bibfield  [0]{\@secondoftwo}%
\providecommand \translation [1]{[#1]}%
\providecommand \BibitemOpen [0]{}%
\providecommand \bibitemStop [0]{}%
\providecommand \bibitemNoStop [0]{.\EOS\space}%
\providecommand \EOS [0]{\spacefactor3000\relax}%
\providecommand \BibitemShut  [1]{\csname bibitem#1\endcsname}%
\let\auto@bib@innerbib\@empty
\bibitem [{\citenamefont {Kormendy}\ and\ \citenamefont {Ho}(2013)}]{Kormendy:2013dxa}%
  \BibitemOpen
  \bibfield  {author} {\bibinfo {author} {\bibfnamefont {J.}~\bibnamefont {Kormendy}}\ and\ \bibinfo {author} {\bibfnamefont {L.~C.}\ \bibnamefont {Ho}},\ }\bibfield  {title} {\bibinfo {title} {{Coevolution (Or Not) of Supermassive Black Holes and Host Galaxies}},\ }\href {https://doi.org/10.1146/annurev-astro-082708-101811} {\bibfield  {journal} {\bibinfo  {journal} {Ann. Rev. Astron. Astrophys.}\ }\textbf {\bibinfo {volume} {51}},\ \bibinfo {pages} {511} (\bibinfo {year} {2013})},\ \Eprint {https://arxiv.org/abs/1304.7762} {arXiv:1304.7762 [astro-ph.CO]} \BibitemShut {NoStop}%
\bibitem [{\citenamefont {Gaspari}\ \emph {et~al.}(2020)\citenamefont {Gaspari}, \citenamefont {Tombesi},\ and\ \citenamefont {Cappi}}]{Gaspari:2020pgo}%
  \BibitemOpen
  \bibfield  {author} {\bibinfo {author} {\bibfnamefont {M.}~\bibnamefont {Gaspari}}, \bibinfo {author} {\bibfnamefont {F.}~\bibnamefont {Tombesi}},\ and\ \bibinfo {author} {\bibfnamefont {M.}~\bibnamefont {Cappi}},\ }\bibfield  {title} {\bibinfo {title} {{Linking macro-, meso- and microscales in multiphase AGN feeding and feedback}},\ }\href {https://doi.org/10.1038/s41550-019-0970-1} {\bibfield  {journal} {\bibinfo  {journal} {Nature Astron.}\ }\textbf {\bibinfo {volume} {4}},\ \bibinfo {pages} {10} (\bibinfo {year} {2020})},\ \Eprint {https://arxiv.org/abs/2001.04985} {arXiv:2001.04985 [astro-ph.GA]} \BibitemShut {NoStop}%
\bibitem [{\citenamefont {{Gaspari}}\ \emph {et~al.}(2013)\citenamefont {{Gaspari}}, \citenamefont {{Ruszkowski}},\ and\ \citenamefont {{Oh}}}]{2013MNRAS.432.3401G}%
  \BibitemOpen
  \bibfield  {author} {\bibinfo {author} {\bibfnamefont {M.}~\bibnamefont {{Gaspari}}}, \bibinfo {author} {\bibfnamefont {M.}~\bibnamefont {{Ruszkowski}}},\ and\ \bibinfo {author} {\bibfnamefont {S.~P.}\ \bibnamefont {{Oh}}},\ }\bibfield  {title} {\bibinfo {title} {{Chaotic cold accretion on to black holes}},\ }\href {https://doi.org/10.1093/mnras/stt692} {\bibfield  {journal} {\bibinfo  {journal} {Mon. Not. Roy. Astron. Soc.}\ }\textbf {\bibinfo {volume} {432}},\ \bibinfo {pages} {3401} (\bibinfo {year} {2013})},\ \Eprint {https://arxiv.org/abs/1301.3130} {arXiv:1301.3130 [astro-ph.CO]} \BibitemShut {NoStop}%
\bibitem [{\citenamefont {Ressler}\ \emph {et~al.}(2018)\citenamefont {Ressler}, \citenamefont {Quataert},\ and\ \citenamefont {Stone}}]{Ressler:2018yhi}%
  \BibitemOpen
  \bibfield  {author} {\bibinfo {author} {\bibfnamefont {S.~M.}\ \bibnamefont {Ressler}}, \bibinfo {author} {\bibfnamefont {E.}~\bibnamefont {Quataert}},\ and\ \bibinfo {author} {\bibfnamefont {J.~M.}\ \bibnamefont {Stone}},\ }\bibfield  {title} {\bibinfo {title} {{Hydrodynamic simulations of the inner accretion flow of Sagittarius A* fuelled by stellar winds}},\ }\href {https://doi.org/10.1093/mnras/sty1146} {\bibfield  {journal} {\bibinfo  {journal} {Mon. Not. Roy. Astron. Soc.}\ }\textbf {\bibinfo {volume} {478}},\ \bibinfo {pages} {3544} (\bibinfo {year} {2018})},\ \Eprint {https://arxiv.org/abs/1805.00474} {arXiv:1805.00474 [astro-ph.HE]} \BibitemShut {NoStop}%
\bibitem [{\citenamefont {{Hopkins}}\ \emph {et~al.}(2024{\natexlab{a}})\citenamefont {{Hopkins}}, \citenamefont {{Grudic}}, \citenamefont {{Su}}, \citenamefont {{Wellons}}, \citenamefont {{Angles-Alcazar}}, \citenamefont {{Steinwandel}}, \citenamefont {{Guszejnov}}, \citenamefont {{Murray}}, \citenamefont {{Faucher-Giguere}}, \citenamefont {{Quataert}},\ and\ \citenamefont {{Keres}}}]{Hopkins:2023ipv}%
  \BibitemOpen
  \bibfield  {author} {\bibinfo {author} {\bibfnamefont {P.~F.}\ \bibnamefont {{Hopkins}}}, \bibinfo {author} {\bibfnamefont {M.~Y.}\ \bibnamefont {{Grudic}}}, \bibinfo {author} {\bibfnamefont {K.-Y.}\ \bibnamefont {{Su}}}, \bibinfo {author} {\bibfnamefont {S.}~\bibnamefont {{Wellons}}}, \bibinfo {author} {\bibfnamefont {D.}~\bibnamefont {{Angles-Alcazar}}}, \bibinfo {author} {\bibfnamefont {U.~P.}\ \bibnamefont {{Steinwandel}}}, \bibinfo {author} {\bibfnamefont {D.}~\bibnamefont {{Guszejnov}}}, \bibinfo {author} {\bibfnamefont {N.}~\bibnamefont {{Murray}}}, \bibinfo {author} {\bibfnamefont {C.-A.}\ \bibnamefont {{Faucher-Giguere}}}, \bibinfo {author} {\bibfnamefont {E.}~\bibnamefont {{Quataert}}},\ and\ \bibinfo {author} {\bibfnamefont {D.}~\bibnamefont {{Keres}}},\ }\bibfield  {title} {\bibinfo {title} {{FORGE'd in FIRE: Resolving the End of Star Formation and Structure of AGN Accretion Disks from Cosmological Initial Conditions}},\ }\href {https://doi.org/10.21105/astro.2309.13115} {\bibfield  {journal} {\bibinfo  {journal} {The Open Journal of Astrophysics}\ }\textbf {\bibinfo {volume} {7}},\ \bibinfo {eid} {18} (\bibinfo {year} {2024}{\natexlab{a}})},\ \Eprint {https://arxiv.org/abs/2309.13115} {arXiv:2309.13115 [astro-ph.GA]} \BibitemShut {NoStop}%
\bibitem [{\citenamefont {{Hopkins}}\ \emph {et~al.}(2024{\natexlab{b}})\citenamefont {{Hopkins}}, \citenamefont {{Squire}}, \citenamefont {{Su}}, \citenamefont {{Steinwandel}}, \citenamefont {{Kremer}}, \citenamefont {{Shi}}, \citenamefont {{Grudic}}, \citenamefont {{Wellons}}, \citenamefont {{Faucher-Giguere}}, \citenamefont {{Angles-Alcazar}}, \citenamefont {{Murray}},\ and\ \citenamefont {{Quataert}}}]{Hopkins2023}%
  \BibitemOpen
  \bibfield  {author} {\bibinfo {author} {\bibfnamefont {P.~F.}\ \bibnamefont {{Hopkins}}}, \bibinfo {author} {\bibfnamefont {J.}~\bibnamefont {{Squire}}}, \bibinfo {author} {\bibfnamefont {K.-Y.}\ \bibnamefont {{Su}}}, \bibinfo {author} {\bibfnamefont {U.~P.}\ \bibnamefont {{Steinwandel}}}, \bibinfo {author} {\bibfnamefont {K.}~\bibnamefont {{Kremer}}}, \bibinfo {author} {\bibfnamefont {Y.}~\bibnamefont {{Shi}}}, \bibinfo {author} {\bibfnamefont {M.~Y.}\ \bibnamefont {{Grudic}}}, \bibinfo {author} {\bibfnamefont {S.}~\bibnamefont {{Wellons}}}, \bibinfo {author} {\bibfnamefont {C.-A.}\ \bibnamefont {{Faucher-Giguere}}}, \bibinfo {author} {\bibfnamefont {D.}~\bibnamefont {{Angles-Alcazar}}}, \bibinfo {author} {\bibfnamefont {N.}~\bibnamefont {{Murray}}},\ and\ \bibinfo {author} {\bibfnamefont {E.}~\bibnamefont {{Quataert}}},\ }\bibfield  {title} {\bibinfo {title} {{FORGE'd in FIRE II: The Formation of Magnetically-Dominated Quasar Accretion Disks from Cosmological Initial Conditions}},\ }\href {https://doi.org/10.21105/astro.2310.04506} {\bibfield  {journal} {\bibinfo  {journal} {The Open Journal of Astrophysics}\ }\textbf {\bibinfo {volume} {7}},\ \bibinfo {eid} {19} (\bibinfo {year} {2024}{\natexlab{b}})},\ \Eprint {https://arxiv.org/abs/2310.04506} {arXiv:2310.04506 [astro-ph.HE]} \BibitemShut {NoStop}%
\bibitem [{\citenamefont {{Guo}}\ \emph {et~al.}(2023)\citenamefont {{Guo}}, \citenamefont {{Stone}}, \citenamefont {{Kim}},\ and\ \citenamefont {{Quataert}}}]{Guo2023}%
  \BibitemOpen
  \bibfield  {author} {\bibinfo {author} {\bibfnamefont {M.}~\bibnamefont {{Guo}}}, \bibinfo {author} {\bibfnamefont {J.~M.}\ \bibnamefont {{Stone}}}, \bibinfo {author} {\bibfnamefont {C.-G.}\ \bibnamefont {{Kim}}},\ and\ \bibinfo {author} {\bibfnamefont {E.}~\bibnamefont {{Quataert}}},\ }\bibfield  {title} {\bibinfo {title} {{Toward Horizon-scale Accretion onto Supermassive Black Holes in Elliptical Galaxies}},\ }\href {https://doi.org/10.3847/1538-4357/acb81e} {\bibfield  {journal} {\bibinfo  {journal} {Astrophys. J.}\ }\textbf {\bibinfo {volume} {946}},\ \bibinfo {eid} {26} (\bibinfo {year} {2023})},\ \Eprint {https://arxiv.org/abs/2211.05131} {arXiv:2211.05131 [astro-ph.HE]} \BibitemShut {NoStop}%
\bibitem [{\citenamefont {{Guo}}\ \emph {et~al.}(2024)\citenamefont {{Guo}}, \citenamefont {{Stone}}, \citenamefont {{Quataert}},\ and\ \citenamefont {{Kim}}}]{Guo2024}%
  \BibitemOpen
  \bibfield  {author} {\bibinfo {author} {\bibfnamefont {M.}~\bibnamefont {{Guo}}}, \bibinfo {author} {\bibfnamefont {J.~M.}\ \bibnamefont {{Stone}}}, \bibinfo {author} {\bibfnamefont {E.}~\bibnamefont {{Quataert}}},\ and\ \bibinfo {author} {\bibfnamefont {C.-G.}\ \bibnamefont {{Kim}}},\ }\bibfield  {title} {\bibinfo {title} {{Magnetized Accretion onto and Feedback from Supermassive Black Holes in Elliptical Galaxies}},\ }\href {https://doi.org/10.48550/arXiv.2405.11711} {\bibfield  {journal} {\bibinfo  {journal} {arXiv e-prints}\ ,\ \bibinfo {eid} {arXiv:2405.11711}} (\bibinfo {year} {2024})},\ \Eprint {https://arxiv.org/abs/2405.11711} {arXiv:2405.11711 [astro-ph.HE]} \BibitemShut {NoStop}%
\bibitem [{\citenamefont {Cho}\ \emph {et~al.}(2023)\citenamefont {Cho}, \citenamefont {Prather}, \citenamefont {Narayan}, \citenamefont {Natarajan}, \citenamefont {Su}, \citenamefont {Ricarte},\ and\ \citenamefont {Chatterjee}}]{Cho:2023wqr}%
  \BibitemOpen
  \bibfield  {author} {\bibinfo {author} {\bibfnamefont {H.}~\bibnamefont {Cho}}, \bibinfo {author} {\bibfnamefont {B.~S.}\ \bibnamefont {Prather}}, \bibinfo {author} {\bibfnamefont {R.}~\bibnamefont {Narayan}}, \bibinfo {author} {\bibfnamefont {P.}~\bibnamefont {Natarajan}}, \bibinfo {author} {\bibfnamefont {K.-Y.}\ \bibnamefont {Su}}, \bibinfo {author} {\bibfnamefont {A.}~\bibnamefont {Ricarte}},\ and\ \bibinfo {author} {\bibfnamefont {K.}~\bibnamefont {Chatterjee}},\ }\bibfield  {title} {\bibinfo {title} {{Bridging Scales in Black Hole Accretion and Feedback: Magnetized Bondi Accretion in 3D GRMHD}},\ }\href {https://doi.org/10.3847/2041-8213/ad1048} {\bibfield  {journal} {\bibinfo  {journal} {Astrophys. J. Lett.}\ }\textbf {\bibinfo {volume} {959}},\ \bibinfo {pages} {L22} (\bibinfo {year} {2023})},\ \Eprint {https://arxiv.org/abs/2310.19135} {arXiv:2310.19135 [astro-ph.HE]} \BibitemShut {NoStop}%
\bibitem [{\citenamefont {Cho}\ \emph {et~al.}(2024)\citenamefont {Cho}, \citenamefont {Prather}, \citenamefont {Su}, \citenamefont {Narayan},\ and\ \citenamefont {Natarajan}}]{Cho:2024wsp}%
  \BibitemOpen
  \bibfield  {author} {\bibinfo {author} {\bibfnamefont {H.}~\bibnamefont {Cho}}, \bibinfo {author} {\bibfnamefont {B.~S.}\ \bibnamefont {Prather}}, \bibinfo {author} {\bibfnamefont {K.-Y.}\ \bibnamefont {Su}}, \bibinfo {author} {\bibfnamefont {R.}~\bibnamefont {Narayan}},\ and\ \bibinfo {author} {\bibfnamefont {P.}~\bibnamefont {Natarajan}},\ }\bibfield  {title} {\bibinfo {title} {{Multizone Modeling of Black Hole Accretion and Feedback in 3D GRMHD: Bridging Vast Spatial and Temporal Scales}},\ }\href {https://doi.org/10.3847/1538-4357/ad9561} {\bibfield  {journal} {\bibinfo  {journal} {Astrophys. J.}\ }\textbf {\bibinfo {volume} {977}},\ \bibinfo {pages} {200} (\bibinfo {year} {2024})},\ \Eprint {https://arxiv.org/abs/2405.13887} {arXiv:2405.13887 [astro-ph.HE]} \BibitemShut {NoStop}%
\bibitem [{\citenamefont {Shi}\ \emph {et~al.}(2024)\citenamefont {Shi}, \citenamefont {Kremer},\ and\ \citenamefont {Hopkins}}]{Shi:2024wgh}%
  \BibitemOpen
  \bibfield  {author} {\bibinfo {author} {\bibfnamefont {Y.}~\bibnamefont {Shi}}, \bibinfo {author} {\bibfnamefont {K.}~\bibnamefont {Kremer}},\ and\ \bibinfo {author} {\bibfnamefont {P.~F.}\ \bibnamefont {Hopkins}},\ }\bibfield  {title} {\bibinfo {title} {{Feedback-regulated seed black hole growth in star-forming molecular clouds and galactic nuclei}},\ }\href {https://doi.org/10.1051/0004-6361/202450964} {\bibfield  {journal} {\bibinfo  {journal} {Astron. Astrophys.}\ }\textbf {\bibinfo {volume} {691}},\ \bibinfo {pages} {A24} (\bibinfo {year} {2024})},\ \Eprint {https://arxiv.org/abs/2405.12164} {arXiv:2405.12164 [astro-ph.GA]} \BibitemShut {NoStop}%
\bibitem [{\citenamefont {{Shi}}\ \emph {et~al.}(2024)\citenamefont {{Shi}}, \citenamefont {{Kremer}},\ and\ \citenamefont {{Hopkins}}}]{Shi2024}%
  \BibitemOpen
  \bibfield  {author} {\bibinfo {author} {\bibfnamefont {Y.}~\bibnamefont {{Shi}}}, \bibinfo {author} {\bibfnamefont {K.}~\bibnamefont {{Kremer}}},\ and\ \bibinfo {author} {\bibfnamefont {P.~F.}\ \bibnamefont {{Hopkins}}},\ }\bibfield  {title} {\bibinfo {title} {{From Seeds to Supermassive Black Holes: Capture, Growth, Migration, and Pairing in Dense Protobulge Environments}},\ }\href {https://doi.org/10.3847/2041-8213/ad5a95} {\bibfield  {journal} {\bibinfo  {journal} {Astrophys. J. Lett.}\ }\textbf {\bibinfo {volume} {969}},\ \bibinfo {eid} {L31} (\bibinfo {year} {2024})},\ \Eprint {https://arxiv.org/abs/2405.17338} {arXiv:2405.17338 [astro-ph.GA]} \BibitemShut {NoStop}%
\bibitem [{\citenamefont {Kaaz}\ \emph {et~al.}(2025)\citenamefont {Kaaz}, \citenamefont {Liska}, \citenamefont {Tchekhovskoy}, \citenamefont {Hopkins},\ and\ \citenamefont {Jacquemin-Ide}}]{Kaaz:2024jxl}%
  \BibitemOpen
  \bibfield  {author} {\bibinfo {author} {\bibfnamefont {N.}~\bibnamefont {Kaaz}}, \bibinfo {author} {\bibfnamefont {M.}~\bibnamefont {Liska}}, \bibinfo {author} {\bibfnamefont {A.}~\bibnamefont {Tchekhovskoy}}, \bibinfo {author} {\bibfnamefont {P.~F.}\ \bibnamefont {Hopkins}},\ and\ \bibinfo {author} {\bibfnamefont {J.}~\bibnamefont {Jacquemin-Ide}},\ }\bibfield  {title} {\bibinfo {title} {{H-AMR FORGE\textquoteright{}d in FIRE. I. Magnetic State Transitions, Jet Launching, and Radiative Emission in Super-Eddington, Highly Magnetized Quasar Disks Formed from Cosmological Initial Conditions}},\ }\href {https://doi.org/10.3847/1538-4357/ad9a86} {\bibfield  {journal} {\bibinfo  {journal} {Astrophys. J.}\ }\textbf {\bibinfo {volume} {979}},\ \bibinfo {pages} {248} (\bibinfo {year} {2025})},\ \Eprint {https://arxiv.org/abs/2410.01877} {arXiv:2410.01877 [astro-ph.HE]} \BibitemShut {NoStop}%
\bibitem [{\citenamefont {Gaburov}\ \emph {et~al.}(2012)\citenamefont {Gaburov}, \citenamefont {Johansen},\ and\ \citenamefont {Levin}}]{Gaburov:2012jd}%
  \BibitemOpen
  \bibfield  {author} {\bibinfo {author} {\bibfnamefont {E.}~\bibnamefont {Gaburov}}, \bibinfo {author} {\bibfnamefont {A.}~\bibnamefont {Johansen}},\ and\ \bibinfo {author} {\bibfnamefont {Y.}~\bibnamefont {Levin}},\ }\bibfield  {title} {\bibinfo {title} {{Magnetically-levitating disks around supermassive black holes}},\ }\href {https://doi.org/10.1088/0004-637X/758/2/103} {\bibfield  {journal} {\bibinfo  {journal} {Astrophys. J.}\ }\textbf {\bibinfo {volume} {758}},\ \bibinfo {pages} {103} (\bibinfo {year} {2012})},\ \Eprint {https://arxiv.org/abs/1201.4873} {arXiv:1201.4873 [astro-ph.GA]} \BibitemShut {NoStop}%
\bibitem [{\citenamefont {{Amaro-Seoane}}\ \emph {et~al.}(2023{\natexlab{a}})\citenamefont {{Amaro-Seoane}}, \citenamefont {{Andrews}}, \citenamefont {{Arca Sedda}}, \citenamefont {{Askar}}, \citenamefont {{Baghi}}, \citenamefont {{Balasov}}, \citenamefont {{Bartos}}, \citenamefont {{Bavera}}, \citenamefont {{Bellovary}}, \citenamefont {{Berry}}, \citenamefont {{Berti}}, \citenamefont {{Bianchi}}, \citenamefont {{Blecha}}, \citenamefont {{Blondin}}, \citenamefont {{Bogdanovi{\'c}}}, \citenamefont {{Boissier}}, \citenamefont {{Bonetti}}, \citenamefont {{Bonoli}}, \citenamefont {{Bortolas}}, \citenamefont {{Breivik}}, \citenamefont {{Capelo}}, \citenamefont {{Caramete}}, \citenamefont {{Cattorini}}, \citenamefont {{Charisi}}, \citenamefont {{Chaty}}, \citenamefont {{Chen}}, \citenamefont {{Chru{\'s}li{\'n}ska}}, \citenamefont {{Chua}}, \citenamefont {{Church}}, \citenamefont {{Colpi}}, \citenamefont {{D'Orazio}}, \citenamefont {{Danielski}}, \citenamefont {{Davies}}, \citenamefont {{Dayal}}, \citenamefont {{De Rosa}}, \citenamefont {{Derdzinski}}, \citenamefont {{Destounis}}, \citenamefont {{Dotti}}, \citenamefont {{Dutan}}, \citenamefont {{Dvorkin}}, \citenamefont {{Fabj}}, \citenamefont {{Foglizzo}}, \citenamefont {{Ford}}, \citenamefont {{Fouvry}}, \citenamefont {{Franchini}}, \citenamefont {{Fragos}}, \citenamefont {{Fryer}}, \citenamefont {{Gaspari}}, \citenamefont {{Gerosa}}, \citenamefont {{Graziani}}, \citenamefont {{Groot}}, \citenamefont {{Habouzit}}, \citenamefont {{Haggard}}, \citenamefont {{Haiman}}, \citenamefont {{Han}}, \citenamefont {{Istrate}}, \citenamefont {{Johansson}}, \citenamefont {{Khan}}, \citenamefont {{Kimpson}}, \citenamefont {{Kokkotas}}, \citenamefont {{Kong}}, \citenamefont {{Korol}}, \citenamefont {{Kremer}}, \citenamefont {{Kupfer}}, \citenamefont {{Lamberts}}, \citenamefont {{Larson}}, \citenamefont {{Lau}}, \citenamefont {{Liu}}, \citenamefont {{Lloyd-Ronning}}, \citenamefont {{Lodato}}, \citenamefont {{Lupi}}, \citenamefont {{Ma}}, \citenamefont {{Maccarone}}, \citenamefont {{Mandel}}, \citenamefont {{Mangiagli}}, \citenamefont {{Mapelli}}, \citenamefont {{Mathis}}, \citenamefont {{Mayer}}, \citenamefont {{McGee}}, \citenamefont {{McKernan}}, \citenamefont {{Miller}}, \citenamefont {{Mota}}, \citenamefont {{Mumpower}}, \citenamefont {{Nasim}}, \citenamefont {{Nelemans}}, \citenamefont {{Noble}}, \citenamefont {{Pacucci}}, \citenamefont {{Panessa}}, \citenamefont {{Paschalidis}}, \citenamefont {{Pfister}}, \citenamefont {{Porquet}}, \citenamefont {{Quenby}}, \citenamefont {{Ricarte}}, \citenamefont {{R{\"o}pke}}, \citenamefont {{Regan}}, \citenamefont {{Rosswog}}, \citenamefont {{Ruiter}}, \citenamefont {{Ruiz}}, \citenamefont {{Runnoe}}, \citenamefont {{Schneider}}, \citenamefont {{Schnittman}}, \citenamefont {{Secunda}}, \citenamefont {{Sesana}}, \citenamefont {{Seto}}, \citenamefont {{Shao}}, \citenamefont {{Shapiro}}, \citenamefont {{Sopuerta}}, \citenamefont {{Stone}}, \citenamefont {{Suvorov}}, \citenamefont {{Tamanini}}, \citenamefont {{Tamfal}}, \citenamefont {{Tauris}}, \citenamefont {{Temmink}}, \citenamefont {{Tomsick}}, \citenamefont {{Toonen}}, \citenamefont {{Torres-Orjuela}}, \citenamefont {{Toscani}}, \citenamefont {{Tsokaros}}, \citenamefont {{Unal}}, \citenamefont {{V{\'a}zquez-Aceves}}, \citenamefont {{Valiante}}, \citenamefont {{van Putten}}, \citenamefont {{van Roestel}}, \citenamefont {{Vignali}}, \citenamefont {{Volonteri}}, \citenamefont {{Wu}}, \citenamefont {{Younsi}}, \citenamefont {{Yu}}, \citenamefont {{Zane}}, \citenamefont {{Zwick}}, \citenamefont {{Antonini}}, \citenamefont {{Baibhav}}, \citenamefont {{Barausse}}, \citenamefont {{Bonilla Rivera}}, \citenamefont {{Branchesi}}, \citenamefont {{Branduardi-Raymont}}, \citenamefont {{Burdge}}, \citenamefont {{Chakraborty}}, \citenamefont {{Cuadra}}, \citenamefont {{Dage}}, \citenamefont {{Davis}}, \citenamefont {{de Mink}}, \citenamefont {{Decarli}}, \citenamefont {{Doneva}}, \citenamefont {{Escoffier}}, \citenamefont {{Gandhi}}, \citenamefont {{Haardt}}, \citenamefont {{Lousto}}, \citenamefont {{Nissanke}}, \citenamefont {{Nordhaus}}, \citenamefont {{O'Shaughnessy}}, \citenamefont {{Portegies Zwart}}, \citenamefont {{Pound}}, \citenamefont {{Schussler}}, \citenamefont {{Sergijenko}}, \citenamefont {{Spallicci}}, \citenamefont {{Vernieri}},\ and\ \citenamefont {{Vigna-G{\'o}mez}}}]{lisa2023}%
  \BibitemOpen
  \bibfield  {author} {\bibinfo {author} {\bibfnamefont {P.}~\bibnamefont {{Amaro-Seoane}}}, \bibinfo {author} {\bibfnamefont {J.}~\bibnamefont {{Andrews}}}, \bibinfo {author} {\bibfnamefont {M.}~\bibnamefont {{Arca Sedda}}}, \bibinfo {author} {\bibfnamefont {A.}~\bibnamefont {{Askar}}}, \bibinfo {author} {\bibfnamefont {Q.}~\bibnamefont {{Baghi}}}, \bibinfo {author} {\bibfnamefont {R.}~\bibnamefont {{Balasov}}}, \bibinfo {author} {\bibfnamefont {I.}~\bibnamefont {{Bartos}}}, \bibinfo {author} {\bibfnamefont {S.~S.}\ \bibnamefont {{Bavera}}}, \bibinfo {author} {\bibfnamefont {J.}~\bibnamefont {{Bellovary}}}, \bibinfo {author} {\bibfnamefont {C.~P.~L.}\ \bibnamefont {{Berry}}}, \bibinfo {author} {\bibfnamefont {E.}~\bibnamefont {{Berti}}}, \bibinfo {author} {\bibfnamefont {S.}~\bibnamefont {{Bianchi}}}, \bibinfo {author} {\bibfnamefont {L.}~\bibnamefont {{Blecha}}}, \bibinfo {author} {\bibfnamefont {S.}~\bibnamefont {{Blondin}}}, \bibinfo {author} {\bibfnamefont {T.}~\bibnamefont {{Bogdanovi{\'c}}}}, \bibinfo {author} {\bibfnamefont {S.}~\bibnamefont {{Boissier}}}, \bibinfo {author} {\bibfnamefont {M.}~\bibnamefont {{Bonetti}}}, \bibinfo {author} {\bibfnamefont {S.}~\bibnamefont {{Bonoli}}}, \bibinfo {author} {\bibfnamefont {E.}~\bibnamefont {{Bortolas}}}, \bibinfo {author} {\bibfnamefont {K.}~\bibnamefont {{Breivik}}}, \bibinfo {author} {\bibfnamefont {P.~R.}\ \bibnamefont {{Capelo}}}, \bibinfo {author} {\bibfnamefont {L.}~\bibnamefont {{Caramete}}}, \bibinfo {author} {\bibfnamefont {F.}~\bibnamefont {{Cattorini}}}, \bibinfo {author} {\bibfnamefont {M.}~\bibnamefont {{Charisi}}}, \bibinfo {author} {\bibfnamefont {S.}~\bibnamefont {{Chaty}}}, \bibinfo {author} {\bibfnamefont {X.}~\bibnamefont {{Chen}}}, \bibinfo {author} {\bibfnamefont {M.}~\bibnamefont {{Chru{\'s}li{\'n}ska}}}, \bibinfo {author} {\bibfnamefont {A.~J.~K.}\ \bibnamefont {{Chua}}}, \bibinfo {author} {\bibfnamefont {R.}~\bibnamefont {{Church}}}, \bibinfo {author} {\bibfnamefont {M.}~\bibnamefont {{Colpi}}}, \bibinfo {author} {\bibfnamefont {D.}~\bibnamefont {{D'Orazio}}}, \bibinfo {author} {\bibfnamefont {C.}~\bibnamefont {{Danielski}}}, \bibinfo {author} {\bibfnamefont {M.~B.}\ \bibnamefont {{Davies}}}, \bibinfo {author} {\bibfnamefont {P.}~\bibnamefont {{Dayal}}}, \bibinfo {author} {\bibfnamefont {A.}~\bibnamefont {{De Rosa}}}, \bibinfo {author} {\bibfnamefont {A.}~\bibnamefont {{Derdzinski}}}, \bibinfo {author} {\bibfnamefont {K.}~\bibnamefont {{Destounis}}}, \bibinfo {author} {\bibfnamefont {M.}~\bibnamefont {{Dotti}}}, \bibinfo {author} {\bibfnamefont {I.}~\bibnamefont {{Dutan}}}, \bibinfo {author} {\bibfnamefont {I.}~\bibnamefont {{Dvorkin}}}, \bibinfo {author} {\bibfnamefont {G.}~\bibnamefont {{Fabj}}}, \bibinfo {author} {\bibfnamefont {T.}~\bibnamefont {{Foglizzo}}}, \bibinfo {author} {\bibfnamefont {S.}~\bibnamefont {{Ford}}}, \bibinfo {author} {\bibfnamefont {J.-B.}\ \bibnamefont {{Fouvry}}}, \bibinfo {author} {\bibfnamefont {A.}~\bibnamefont {{Franchini}}}, \bibinfo {author} {\bibfnamefont {T.}~\bibnamefont {{Fragos}}}, \bibinfo {author} {\bibfnamefont {C.}~\bibnamefont {{Fryer}}}, \bibinfo {author} {\bibfnamefont {M.}~\bibnamefont {{Gaspari}}}, \bibinfo {author} {\bibfnamefont {D.}~\bibnamefont {{Gerosa}}}, \bibinfo {author} {\bibfnamefont {L.}~\bibnamefont {{Graziani}}}, \bibinfo {author} {\bibfnamefont {P.}~\bibnamefont {{Groot}}}, \bibinfo {author} {\bibfnamefont {M.}~\bibnamefont {{Habouzit}}}, \bibinfo {author} {\bibfnamefont {D.}~\bibnamefont {{Haggard}}}, \bibinfo {author} {\bibfnamefont {Z.}~\bibnamefont {{Haiman}}}, \bibinfo {author} {\bibfnamefont {W.-B.}\ \bibnamefont {{Han}}}, \bibinfo {author} {\bibfnamefont {A.}~\bibnamefont {{Istrate}}}, \bibinfo {author} {\bibfnamefont {P.~H.}\ \bibnamefont {{Johansson}}}, \bibinfo {author} {\bibfnamefont {F.~M.}\ \bibnamefont {{Khan}}}, \bibinfo {author} {\bibfnamefont {T.}~\bibnamefont {{Kimpson}}}, \bibinfo {author} {\bibfnamefont {K.}~\bibnamefont {{Kokkotas}}}, \bibinfo {author} {\bibfnamefont {A.}~\bibnamefont {{Kong}}}, \bibinfo {author} {\bibfnamefont {V.}~\bibnamefont {{Korol}}}, \bibinfo {author} {\bibfnamefont {K.}~\bibnamefont {{Kremer}}}, \bibinfo {author} {\bibfnamefont {T.}~\bibnamefont {{Kupfer}}}, \bibinfo {author} {\bibfnamefont {A.}~\bibnamefont {{Lamberts}}}, \bibinfo {author} {\bibfnamefont {S.}~\bibnamefont {{Larson}}}, \bibinfo {author} {\bibfnamefont {M.}~\bibnamefont {{Lau}}}, \bibinfo {author} {\bibfnamefont {D.}~\bibnamefont {{Liu}}}, \bibinfo {author} {\bibfnamefont {N.}~\bibnamefont {{Lloyd-Ronning}}}, \bibinfo {author} {\bibfnamefont {G.}~\bibnamefont {{Lodato}}}, \bibinfo {author} {\bibfnamefont {A.}~\bibnamefont {{Lupi}}}, \bibinfo {author} {\bibfnamefont {C.-P.}\ \bibnamefont {{Ma}}}, \bibinfo {author} {\bibfnamefont {T.}~\bibnamefont {{Maccarone}}}, \bibinfo {author} {\bibfnamefont {I.}~\bibnamefont {{Mandel}}}, \bibinfo {author} {\bibfnamefont {A.}~\bibnamefont {{Mangiagli}}}, \bibinfo {author} {\bibfnamefont {M.}~\bibnamefont {{Mapelli}}}, \bibinfo {author} {\bibfnamefont {S.}~\bibnamefont {{Mathis}}}, \bibinfo {author} {\bibfnamefont {L.}~\bibnamefont {{Mayer}}}, \bibinfo {author} {\bibfnamefont {S.}~\bibnamefont {{McGee}}}, \bibinfo {author} {\bibfnamefont {B.}~\bibnamefont {{McKernan}}}, \bibinfo {author} {\bibfnamefont {M.~C.}\ \bibnamefont {{Miller}}}, \bibinfo {author} {\bibfnamefont {D.~F.}\ \bibnamefont {{Mota}}}, \bibinfo {author} {\bibfnamefont {M.}~\bibnamefont {{Mumpower}}}, \bibinfo {author} {\bibfnamefont {S.~S.}\ \bibnamefont {{Nasim}}}, \bibinfo {author} {\bibfnamefont {G.}~\bibnamefont {{Nelemans}}}, \bibinfo {author} {\bibfnamefont {S.}~\bibnamefont {{Noble}}}, \bibinfo {author} {\bibfnamefont {F.}~\bibnamefont {{Pacucci}}}, \bibinfo {author} {\bibfnamefont {F.}~\bibnamefont {{Panessa}}}, \bibinfo {author} {\bibfnamefont {V.}~\bibnamefont {{Paschalidis}}}, \bibinfo {author} {\bibfnamefont {H.}~\bibnamefont {{Pfister}}}, \bibinfo {author} {\bibfnamefont {D.}~\bibnamefont {{Porquet}}}, \bibinfo {author} {\bibfnamefont {J.}~\bibnamefont {{Quenby}}}, \bibinfo {author} {\bibfnamefont {A.}~\bibnamefont {{Ricarte}}}, \bibinfo {author} {\bibfnamefont {F.~K.}\ \bibnamefont {{R{\"o}pke}}}, \bibinfo {author} {\bibfnamefont {J.}~\bibnamefont {{Regan}}}, \bibinfo {author} {\bibfnamefont {S.}~\bibnamefont {{Rosswog}}}, \bibinfo {author} {\bibfnamefont {A.}~\bibnamefont {{Ruiter}}}, \bibinfo {author} {\bibfnamefont {M.}~\bibnamefont {{Ruiz}}}, \bibinfo {author} {\bibfnamefont {J.}~\bibnamefont {{Runnoe}}}, \bibinfo {author} {\bibfnamefont {R.}~\bibnamefont {{Schneider}}}, \bibinfo {author} {\bibfnamefont {J.}~\bibnamefont {{Schnittman}}}, \bibinfo {author} {\bibfnamefont {A.}~\bibnamefont {{Secunda}}}, \bibinfo {author} {\bibfnamefont {A.}~\bibnamefont {{Sesana}}}, \bibinfo {author} {\bibfnamefont {N.}~\bibnamefont {{Seto}}}, \bibinfo {author} {\bibfnamefont {L.}~\bibnamefont {{Shao}}}, \bibinfo {author} {\bibfnamefont {S.}~\bibnamefont {{Shapiro}}}, \bibinfo {author} {\bibfnamefont {C.}~\bibnamefont {{Sopuerta}}}, \bibinfo {author} {\bibfnamefont {N.~C.}\ \bibnamefont {{Stone}}}, \bibinfo {author} {\bibfnamefont {A.}~\bibnamefont {{Suvorov}}}, \bibinfo {author} {\bibfnamefont {N.}~\bibnamefont {{Tamanini}}}, \bibinfo {author} {\bibfnamefont {T.}~\bibnamefont {{Tamfal}}}, \bibinfo {author} {\bibfnamefont {T.}~\bibnamefont {{Tauris}}}, \bibinfo {author} {\bibfnamefont {K.}~\bibnamefont {{Temmink}}}, \bibinfo {author} {\bibfnamefont {J.}~\bibnamefont {{Tomsick}}}, \bibinfo {author} {\bibfnamefont {S.}~\bibnamefont {{Toonen}}}, \bibinfo {author} {\bibfnamefont {A.}~\bibnamefont {{Torres-Orjuela}}}, \bibinfo {author} {\bibfnamefont {M.}~\bibnamefont {{Toscani}}}, \bibinfo {author} {\bibfnamefont {A.}~\bibnamefont {{Tsokaros}}}, \bibinfo {author} {\bibfnamefont {C.}~\bibnamefont {{Unal}}}, \bibinfo {author} {\bibfnamefont {V.}~\bibnamefont {{V{\'a}zquez-Aceves}}}, \bibinfo {author} {\bibfnamefont {R.}~\bibnamefont {{Valiante}}}, \bibinfo {author} {\bibfnamefont {M.}~\bibnamefont {{van Putten}}}, \bibinfo {author} {\bibfnamefont {J.}~\bibnamefont {{van Roestel}}}, \bibinfo {author} {\bibfnamefont {C.}~\bibnamefont {{Vignali}}}, \bibinfo {author} {\bibfnamefont {M.}~\bibnamefont {{Volonteri}}}, \bibinfo {author} {\bibfnamefont {K.}~\bibnamefont {{Wu}}}, \bibinfo {author} {\bibfnamefont {Z.}~\bibnamefont {{Younsi}}}, \bibinfo {author} {\bibfnamefont {S.}~\bibnamefont {{Yu}}}, \bibinfo {author} {\bibfnamefont {S.}~\bibnamefont {{Zane}}}, \bibinfo {author} {\bibfnamefont {L.}~\bibnamefont {{Zwick}}}, \bibinfo {author} {\bibfnamefont {F.}~\bibnamefont {{Antonini}}}, \bibinfo {author} {\bibfnamefont {V.}~\bibnamefont {{Baibhav}}}, \bibinfo {author} {\bibfnamefont {E.}~\bibnamefont {{Barausse}}}, \bibinfo {author} {\bibfnamefont {A.}~\bibnamefont {{Bonilla Rivera}}}, \bibinfo {author} {\bibfnamefont {M.}~\bibnamefont {{Branchesi}}}, \bibinfo {author} {\bibfnamefont {G.}~\bibnamefont {{Branduardi-Raymont}}}, \bibinfo {author} {\bibfnamefont {K.}~\bibnamefont {{Burdge}}}, \bibinfo {author} {\bibfnamefont {S.}~\bibnamefont {{Chakraborty}}}, \bibinfo {author} {\bibfnamefont {J.}~\bibnamefont {{Cuadra}}}, \bibinfo {author} {\bibfnamefont {K.}~\bibnamefont {{Dage}}}, \bibinfo {author} {\bibfnamefont {B.}~\bibnamefont {{Davis}}}, \bibinfo {author} {\bibfnamefont {S.~E.}\ \bibnamefont {{de Mink}}}, \bibinfo {author} {\bibfnamefont {R.}~\bibnamefont {{Decarli}}}, \bibinfo {author} {\bibfnamefont {D.}~\bibnamefont {{Doneva}}}, \bibinfo {author} {\bibfnamefont {S.}~\bibnamefont {{Escoffier}}}, \bibinfo {author} {\bibfnamefont {P.}~\bibnamefont {{Gandhi}}}, \bibinfo {author} {\bibfnamefont {F.}~\bibnamefont {{Haardt}}}, \bibinfo {author} {\bibfnamefont {C.~O.}\ \bibnamefont {{Lousto}}}, \bibinfo {author} {\bibfnamefont {S.}~\bibnamefont {{Nissanke}}}, \bibinfo {author} {\bibfnamefont {J.}~\bibnamefont {{Nordhaus}}}, \bibinfo {author} {\bibfnamefont {R.}~\bibnamefont {{O'Shaughnessy}}}, \bibinfo {author} {\bibfnamefont {S.}~\bibnamefont {{Portegies Zwart}}}, \bibinfo
  {author} {\bibfnamefont {A.}~\bibnamefont {{Pound}}}, \bibinfo {author} {\bibfnamefont {F.}~\bibnamefont {{Schussler}}}, \bibinfo {author} {\bibfnamefont {O.}~\bibnamefont {{Sergijenko}}}, \bibinfo {author} {\bibfnamefont {A.}~\bibnamefont {{Spallicci}}}, \bibinfo {author} {\bibfnamefont {D.}~\bibnamefont {{Vernieri}}},\ and\ \bibinfo {author} {\bibfnamefont {A.}~\bibnamefont {{Vigna-G{\'o}mez}}},\ }\bibfield  {title} {\bibinfo {title} {{Astrophysics with the Laser Interferometer Space Antenna}},\ }\href {https://doi.org/10.1007/s41114-022-00041-y} {\bibfield  {journal} {\bibinfo  {journal} {Living Reviews in Relativity}\ }\textbf {\bibinfo {volume} {26}},\ \bibinfo {eid} {2} (\bibinfo {year} {2023}{\natexlab{a}})},\ \Eprint {https://arxiv.org/abs/2203.06016} {arXiv:2203.06016 [gr-qc]} \BibitemShut {NoStop}%
\bibitem [{\citenamefont {{Li}}\ \emph {et~al.}(2024)\citenamefont {{Li}}, \citenamefont {{Liu}}, \citenamefont {{Torres-Orjuela}}, \citenamefont {{Chen}}, \citenamefont {{Inayoshi}}, \citenamefont {{Wang}}, \citenamefont {{Hu}}, \citenamefont {{Amaro-Seoane}}, \citenamefont {{Askar}}, \citenamefont {{Bambi}}, \citenamefont {{Capelo}}, \citenamefont {{Chen}}, \citenamefont {{Chua}}, \citenamefont {{Cond{\'e}s-Bre{\~n}a}}, \citenamefont {{Dai}}, \citenamefont {{Das}}, \citenamefont {{Derdzinski}}, \citenamefont {{Fan}}, \citenamefont {{Fujii}}, \citenamefont {{Gao}}, \citenamefont {{Garg}}, \citenamefont {{Ge}}, \citenamefont {{Giersz}}, \citenamefont {{Huang}}, \citenamefont {{Hypki}}, \citenamefont {{Liang}}, \citenamefont {{Liu}}, \citenamefont {{Liu}}, \citenamefont {{Liu}}, \citenamefont {{Liu}}, \citenamefont {{Mayer}}, \citenamefont {{Napolitano}}, \citenamefont {{Peng}}, \citenamefont {{Shao}}, \citenamefont {{Shashank}}, \citenamefont {{Shen}}, \citenamefont {{Tagawa}}, \citenamefont {{Tanikawa}}, \citenamefont {{Toscani}}, \citenamefont {{V{\'a}zquez-Aceves}}, \citenamefont {{Wang}}, \citenamefont {{Wang}}, \citenamefont {{Yi}}, \citenamefont {{Zhang}}, \citenamefont {{Zhang}}, \citenamefont {{Zhu}}, \citenamefont {{Zwick}}, \citenamefont {{Huang}}, \citenamefont {{Mei}}, \citenamefont {{Wang}}, \citenamefont {{Xie}}, \citenamefont {{Zhang}},\ and\ \citenamefont {{Luo}}}]{tianqin}%
  \BibitemOpen
  \bibfield  {author} {\bibinfo {author} {\bibfnamefont {E.-K.}\ \bibnamefont {{Li}}}, \bibinfo {author} {\bibfnamefont {S.}~\bibnamefont {{Liu}}}, \bibinfo {author} {\bibfnamefont {A.}~\bibnamefont {{Torres-Orjuela}}}, \bibinfo {author} {\bibfnamefont {X.}~\bibnamefont {{Chen}}}, \bibinfo {author} {\bibfnamefont {K.}~\bibnamefont {{Inayoshi}}}, \bibinfo {author} {\bibfnamefont {L.}~\bibnamefont {{Wang}}}, \bibinfo {author} {\bibfnamefont {Y.-M.}\ \bibnamefont {{Hu}}}, \bibinfo {author} {\bibfnamefont {P.}~\bibnamefont {{Amaro-Seoane}}}, \bibinfo {author} {\bibfnamefont {A.}~\bibnamefont {{Askar}}}, \bibinfo {author} {\bibfnamefont {C.}~\bibnamefont {{Bambi}}}, \bibinfo {author} {\bibfnamefont {P.~R.}\ \bibnamefont {{Capelo}}}, \bibinfo {author} {\bibfnamefont {H.-Y.}\ \bibnamefont {{Chen}}}, \bibinfo {author} {\bibfnamefont {A.~J.~K.}\ \bibnamefont {{Chua}}}, \bibinfo {author} {\bibfnamefont {E.}~\bibnamefont {{Cond{\'e}s-Bre{\~n}a}}}, \bibinfo {author} {\bibfnamefont {L.}~\bibnamefont {{Dai}}}, \bibinfo {author} {\bibfnamefont {D.}~\bibnamefont {{Das}}}, \bibinfo {author} {\bibfnamefont {A.}~\bibnamefont {{Derdzinski}}}, \bibinfo {author} {\bibfnamefont {H.-M.}\ \bibnamefont {{Fan}}}, \bibinfo {author} {\bibfnamefont {M.}~\bibnamefont {{Fujii}}}, \bibinfo {author} {\bibfnamefont {J.}~\bibnamefont {{Gao}}}, \bibinfo {author} {\bibfnamefont {M.}~\bibnamefont {{Garg}}}, \bibinfo {author} {\bibfnamefont {H.}~\bibnamefont {{Ge}}}, \bibinfo {author} {\bibfnamefont {M.}~\bibnamefont {{Giersz}}}, \bibinfo {author} {\bibfnamefont {S.-J.}\ \bibnamefont {{Huang}}}, \bibinfo {author} {\bibfnamefont {A.}~\bibnamefont {{Hypki}}}, \bibinfo {author} {\bibfnamefont {Z.-C.}\ \bibnamefont {{Liang}}}, \bibinfo {author} {\bibfnamefont {B.}~\bibnamefont {{Liu}}}, \bibinfo {author} {\bibfnamefont {D.}~\bibnamefont {{Liu}}}, \bibinfo {author} {\bibfnamefont {M.}~\bibnamefont {{Liu}}}, \bibinfo {author} {\bibfnamefont {Y.}~\bibnamefont {{Liu}}}, \bibinfo {author} {\bibfnamefont {L.}~\bibnamefont {{Mayer}}}, \bibinfo {author} {\bibfnamefont {N.~R.}\ \bibnamefont {{Napolitano}}}, \bibinfo {author} {\bibfnamefont {P.}~\bibnamefont {{Peng}}}, \bibinfo {author} {\bibfnamefont {Y.}~\bibnamefont {{Shao}}}, \bibinfo {author} {\bibfnamefont {S.}~\bibnamefont {{Shashank}}}, \bibinfo {author} {\bibfnamefont {R.}~\bibnamefont {{Shen}}}, \bibinfo {author} {\bibfnamefont {H.}~\bibnamefont {{Tagawa}}}, \bibinfo {author} {\bibfnamefont {A.}~\bibnamefont {{Tanikawa}}}, \bibinfo {author} {\bibfnamefont {M.}~\bibnamefont {{Toscani}}}, \bibinfo {author} {\bibfnamefont {V.}~\bibnamefont {{V{\'a}zquez-Aceves}}}, \bibinfo {author} {\bibfnamefont {H.-T.}\ \bibnamefont {{Wang}}}, \bibinfo {author} {\bibfnamefont {H.}~\bibnamefont {{Wang}}}, \bibinfo {author} {\bibfnamefont {S.-X.}\ \bibnamefont {{Yi}}}, \bibinfo {author} {\bibfnamefont {J.-d.}\ \bibnamefont {{Zhang}}}, \bibinfo {author} {\bibfnamefont {X.-T.}\ \bibnamefont {{Zhang}}}, \bibinfo {author} {\bibfnamefont {L.}~\bibnamefont {{Zhu}}}, \bibinfo {author} {\bibfnamefont {L.}~\bibnamefont {{Zwick}}}, \bibinfo {author} {\bibfnamefont {S.}~\bibnamefont {{Huang}}}, \bibinfo {author} {\bibfnamefont {J.}~\bibnamefont {{Mei}}}, \bibinfo {author} {\bibfnamefont {Y.}~\bibnamefont {{Wang}}}, \bibinfo {author} {\bibfnamefont {Y.}~\bibnamefont {{Xie}}}, \bibinfo {author} {\bibfnamefont {J.}~\bibnamefont {{Zhang}}},\ and\ \bibinfo {author} {\bibfnamefont {J.}~\bibnamefont {{Luo}}},\ }\bibfield  {title} {\bibinfo {title} {{Gravitational Wave Astronomy With TianQin}},\ }\href {https://doi.org/10.48550/arXiv.2409.19665} {\bibfield  {journal} {\bibinfo  {journal} {arXiv e-prints}\ ,\ \bibinfo {eid} {arXiv:2409.19665}} (\bibinfo {year} {2024})},\ \Eprint {https://arxiv.org/abs/2409.19665} {arXiv:2409.19665 [astro-ph.GA]} \BibitemShut {NoStop}%
\bibitem [{\citenamefont {{Agazie}}\ \emph {et~al.}(2023{\natexlab{a}})\citenamefont {{Agazie}}, \citenamefont {{Anumarlapudi}}, \citenamefont {{Archibald}}, \citenamefont {{Arzoumanian}}, \citenamefont {{Baker}}, \citenamefont {{B{\'e}csy}}, \citenamefont {{Blecha}}, \citenamefont {{Brazier}}, \citenamefont {{Brook}}, \citenamefont {{Burke-Spolaor}}, \citenamefont {{Burnette}}, \citenamefont {{Case}}, \citenamefont {{Charisi}}, \citenamefont {{Chatterjee}}, \citenamefont {{Chatziioannou}}, \citenamefont {{Cheeseboro}}, \citenamefont {{Chen}}, \citenamefont {{Cohen}}, \citenamefont {{Cordes}}, \citenamefont {{Cornish}}, \citenamefont {{Crawford}}, \citenamefont {{Cromartie}}, \citenamefont {{Crowter}}, \citenamefont {{Cutler}}, \citenamefont {{Decesar}}, \citenamefont {{Degan}}, \citenamefont {{Demorest}}, \citenamefont {{Deng}}, \citenamefont {{Dolch}}, \citenamefont {{Drachler}}, \citenamefont {{Ellis}}, \citenamefont {{Ferrara}}, \citenamefont {{Fiore}}, \citenamefont {{Fonseca}}, \citenamefont {{Freedman}}, \citenamefont {{Garver-Daniels}}, \citenamefont {{Gentile}}, \citenamefont {{Gersbach}}, \citenamefont {{Glaser}}, \citenamefont {{Good}}, \citenamefont {{G{\"u}ltekin}}, \citenamefont {{Hazboun}}, \citenamefont {{Hourihane}}, \citenamefont {{Islo}}, \citenamefont {{Jennings}}, \citenamefont {{Johnson}}, \citenamefont {{Jones}}, \citenamefont {{Kaiser}}, \citenamefont {{Kaplan}}, \citenamefont {{Kelley}}, \citenamefont {{Kerr}}, \citenamefont {{Key}}, \citenamefont {{Klein}}, \citenamefont {{Laal}}, \citenamefont {{Lam}}, \citenamefont {{Lamb}}, \citenamefont {{Lazio}}, \citenamefont {{Lewandowska}}, \citenamefont {{Littenberg}}, \citenamefont {{Liu}}, \citenamefont {{Lommen}}, \citenamefont {{Lorimer}}, \citenamefont {{Luo}}, \citenamefont {{Lynch}}, \citenamefont {{Ma}}, \citenamefont {{Madison}}, \citenamefont {{Mattson}}, \citenamefont {{McEwen}}, \citenamefont {{McKee}}, \citenamefont {{McLaughlin}}, \citenamefont {{McMann}}, \citenamefont {{Meyers}}, \citenamefont {{Meyers}}, \citenamefont {{Mingarelli}}, \citenamefont {{Mitridate}}, \citenamefont {{Natarajan}}, \citenamefont {{Ng}}, \citenamefont {{Nice}}, \citenamefont {{Ocker}}, \citenamefont {{Olum}}, \citenamefont {{Pennucci}}, \citenamefont {{Perera}}, \citenamefont {{Petrov}}, \citenamefont {{Pol}}, \citenamefont {{Radovan}}, \citenamefont {{Ransom}}, \citenamefont {{Ray}}, \citenamefont {{Romano}}, \citenamefont {{Sardesai}}, \citenamefont {{Schmiedekamp}}, \citenamefont {{Schmiedekamp}}, \citenamefont {{Schmitz}}, \citenamefont {{Schult}}, \citenamefont {{Shapiro-Albert}}, \citenamefont {{Siemens}}, \citenamefont {{Simon}}, \citenamefont {{Siwek}}, \citenamefont {{Stairs}}, \citenamefont {{Stinebring}}, \citenamefont {{Stovall}}, \citenamefont {{Sun}}, \citenamefont {{Susobhanan}}, \citenamefont {{Swiggum}}, \citenamefont {{Taylor}}, \citenamefont {{Taylor}}, \citenamefont {{Turner}}, \citenamefont {{Unal}}, \citenamefont {{Vallisneri}}, \citenamefont {{van Haasteren}}, \citenamefont {{Vigeland}}, \citenamefont {{Wahl}}, \citenamefont {{Wang}}, \citenamefont {{Witt}}, \citenamefont {{Young}},\ and\ \citenamefont {{Nanograv Collaboration}}}]{Nanograv2023a}%
  \BibitemOpen
  \bibfield  {author} {\bibinfo {author} {\bibfnamefont {G.}~\bibnamefont {{Agazie}}}, \bibinfo {author} {\bibfnamefont {A.}~\bibnamefont {{Anumarlapudi}}}, \bibinfo {author} {\bibfnamefont {A.~M.}\ \bibnamefont {{Archibald}}}, \bibinfo {author} {\bibfnamefont {Z.}~\bibnamefont {{Arzoumanian}}}, \bibinfo {author} {\bibfnamefont {P.~T.}\ \bibnamefont {{Baker}}}, \bibinfo {author} {\bibfnamefont {B.}~\bibnamefont {{B{\'e}csy}}}, \bibinfo {author} {\bibfnamefont {L.}~\bibnamefont {{Blecha}}}, \bibinfo {author} {\bibfnamefont {A.}~\bibnamefont {{Brazier}}}, \bibinfo {author} {\bibfnamefont {P.~R.}\ \bibnamefont {{Brook}}}, \bibinfo {author} {\bibfnamefont {S.}~\bibnamefont {{Burke-Spolaor}}}, \bibinfo {author} {\bibfnamefont {R.}~\bibnamefont {{Burnette}}}, \bibinfo {author} {\bibfnamefont {R.}~\bibnamefont {{Case}}}, \bibinfo {author} {\bibfnamefont {M.}~\bibnamefont {{Charisi}}}, \bibinfo {author} {\bibfnamefont {S.}~\bibnamefont {{Chatterjee}}}, \bibinfo {author} {\bibfnamefont {K.}~\bibnamefont {{Chatziioannou}}}, \bibinfo {author} {\bibfnamefont {B.~D.}\ \bibnamefont {{Cheeseboro}}}, \bibinfo {author} {\bibfnamefont {S.}~\bibnamefont {{Chen}}}, \bibinfo {author} {\bibfnamefont {T.}~\bibnamefont {{Cohen}}}, \bibinfo {author} {\bibfnamefont {J.~M.}\ \bibnamefont {{Cordes}}}, \bibinfo {author} {\bibfnamefont {N.~J.}\ \bibnamefont {{Cornish}}}, \bibinfo {author} {\bibfnamefont {F.}~\bibnamefont {{Crawford}}}, \bibinfo {author} {\bibfnamefont {H.~T.}\ \bibnamefont {{Cromartie}}}, \bibinfo {author} {\bibfnamefont {K.}~\bibnamefont {{Crowter}}}, \bibinfo {author} {\bibfnamefont {C.~J.}\ \bibnamefont {{Cutler}}}, \bibinfo {author} {\bibfnamefont {M.~E.}\ \bibnamefont {{Decesar}}}, \bibinfo {author} {\bibfnamefont {D.}~\bibnamefont {{Degan}}}, \bibinfo {author} {\bibfnamefont {P.~B.}\ \bibnamefont {{Demorest}}}, \bibinfo {author} {\bibfnamefont {H.}~\bibnamefont {{Deng}}}, \bibinfo {author} {\bibfnamefont {T.}~\bibnamefont {{Dolch}}}, \bibinfo {author} {\bibfnamefont {B.}~\bibnamefont {{Drachler}}}, \bibinfo {author} {\bibfnamefont {J.~A.}\ \bibnamefont {{Ellis}}}, \bibinfo {author} {\bibfnamefont {E.~C.}\ \bibnamefont {{Ferrara}}}, \bibinfo {author} {\bibfnamefont {W.}~\bibnamefont {{Fiore}}}, \bibinfo {author} {\bibfnamefont {E.}~\bibnamefont {{Fonseca}}}, \bibinfo {author} {\bibfnamefont {G.~E.}\ \bibnamefont {{Freedman}}}, \bibinfo {author} {\bibfnamefont {N.}~\bibnamefont {{Garver-Daniels}}}, \bibinfo {author} {\bibfnamefont {P.~A.}\ \bibnamefont {{Gentile}}}, \bibinfo {author} {\bibfnamefont {K.~A.}\ \bibnamefont {{Gersbach}}}, \bibinfo {author} {\bibfnamefont {J.}~\bibnamefont {{Glaser}}}, \bibinfo {author} {\bibfnamefont {D.~C.}\ \bibnamefont {{Good}}}, \bibinfo {author} {\bibfnamefont {K.}~\bibnamefont {{G{\"u}ltekin}}}, \bibinfo {author} {\bibfnamefont {J.~S.}\ \bibnamefont {{Hazboun}}}, \bibinfo {author} {\bibfnamefont {S.}~\bibnamefont {{Hourihane}}}, \bibinfo {author} {\bibfnamefont {K.}~\bibnamefont {{Islo}}}, \bibinfo {author} {\bibfnamefont {R.~J.}\ \bibnamefont {{Jennings}}}, \bibinfo {author} {\bibfnamefont {A.~D.}\ \bibnamefont {{Johnson}}}, \bibinfo {author} {\bibfnamefont {M.~L.}\ \bibnamefont {{Jones}}}, \bibinfo {author} {\bibfnamefont {A.~R.}\ \bibnamefont {{Kaiser}}}, \bibinfo {author} {\bibfnamefont {D.~L.}\ \bibnamefont {{Kaplan}}}, \bibinfo {author} {\bibfnamefont {L.~Z.}\ \bibnamefont {{Kelley}}}, \bibinfo {author} {\bibfnamefont {M.}~\bibnamefont {{Kerr}}}, \bibinfo {author} {\bibfnamefont {J.~S.}\ \bibnamefont {{Key}}}, \bibinfo {author} {\bibfnamefont {T.~C.}\ \bibnamefont {{Klein}}}, \bibinfo {author} {\bibfnamefont {N.}~\bibnamefont {{Laal}}}, \bibinfo {author} {\bibfnamefont {M.~T.}\ \bibnamefont {{Lam}}}, \bibinfo {author} {\bibfnamefont {W.~G.}\ \bibnamefont {{Lamb}}}, \bibinfo {author} {\bibfnamefont {T.~J.~W.}\ \bibnamefont {{Lazio}}}, \bibinfo {author} {\bibfnamefont {N.}~\bibnamefont {{Lewandowska}}}, \bibinfo {author} {\bibfnamefont {T.~B.}\ \bibnamefont {{Littenberg}}}, \bibinfo {author} {\bibfnamefont {T.}~\bibnamefont {{Liu}}}, \bibinfo {author} {\bibfnamefont {A.}~\bibnamefont {{Lommen}}}, \bibinfo {author} {\bibfnamefont {D.~R.}\ \bibnamefont {{Lorimer}}}, \bibinfo {author} {\bibfnamefont {J.}~\bibnamefont {{Luo}}}, \bibinfo {author} {\bibfnamefont {R.~S.}\ \bibnamefont {{Lynch}}}, \bibinfo {author} {\bibfnamefont {C.-P.}\ \bibnamefont {{Ma}}}, \bibinfo {author} {\bibfnamefont {D.~R.}\ \bibnamefont {{Madison}}}, \bibinfo {author} {\bibfnamefont {M.~A.}\ \bibnamefont {{Mattson}}}, \bibinfo {author} {\bibfnamefont {A.}~\bibnamefont {{McEwen}}}, \bibinfo {author} {\bibfnamefont {J.~W.}\ \bibnamefont {{McKee}}}, \bibinfo {author} {\bibfnamefont {M.~A.}\ \bibnamefont {{McLaughlin}}}, \bibinfo {author} {\bibfnamefont {N.}~\bibnamefont {{McMann}}}, \bibinfo {author} {\bibfnamefont {B.~W.}\ \bibnamefont {{Meyers}}}, \bibinfo {author} {\bibfnamefont {P.~M.}\ \bibnamefont {{Meyers}}}, \bibinfo {author} {\bibfnamefont {C.~M.~F.}\ \bibnamefont {{Mingarelli}}}, \bibinfo {author} {\bibfnamefont {A.}~\bibnamefont {{Mitridate}}}, \bibinfo {author} {\bibfnamefont {P.}~\bibnamefont {{Natarajan}}}, \bibinfo {author} {\bibfnamefont {C.}~\bibnamefont {{Ng}}}, \bibinfo {author} {\bibfnamefont {D.~J.}\ \bibnamefont {{Nice}}}, \bibinfo {author} {\bibfnamefont {S.~K.}\ \bibnamefont {{Ocker}}}, \bibinfo {author} {\bibfnamefont {K.~D.}\ \bibnamefont {{Olum}}}, \bibinfo {author} {\bibfnamefont {T.~T.}\ \bibnamefont {{Pennucci}}}, \bibinfo {author} {\bibfnamefont {B.~B.~P.}\ \bibnamefont {{Perera}}}, \bibinfo {author} {\bibfnamefont {P.}~\bibnamefont {{Petrov}}}, \bibinfo {author} {\bibfnamefont {N.~S.}\ \bibnamefont {{Pol}}}, \bibinfo {author} {\bibfnamefont {H.~A.}\ \bibnamefont {{Radovan}}}, \bibinfo {author} {\bibfnamefont {S.~M.}\ \bibnamefont {{Ransom}}}, \bibinfo {author} {\bibfnamefont {P.~S.}\ \bibnamefont {{Ray}}}, \bibinfo {author} {\bibfnamefont {J.~D.}\ \bibnamefont {{Romano}}}, \bibinfo {author} {\bibfnamefont {S.~C.}\ \bibnamefont {{Sardesai}}}, \bibinfo {author} {\bibfnamefont {A.}~\bibnamefont {{Schmiedekamp}}}, \bibinfo {author} {\bibfnamefont {C.}~\bibnamefont {{Schmiedekamp}}}, \bibinfo {author} {\bibfnamefont {K.}~\bibnamefont {{Schmitz}}}, \bibinfo {author} {\bibfnamefont {L.}~\bibnamefont {{Schult}}}, \bibinfo {author} {\bibfnamefont {B.~J.}\ \bibnamefont {{Shapiro-Albert}}}, \bibinfo {author} {\bibfnamefont {X.}~\bibnamefont {{Siemens}}}, \bibinfo {author} {\bibfnamefont {J.}~\bibnamefont {{Simon}}}, \bibinfo {author} {\bibfnamefont {M.~S.}\ \bibnamefont {{Siwek}}}, \bibinfo {author} {\bibfnamefont {I.~H.}\ \bibnamefont {{Stairs}}}, \bibinfo {author} {\bibfnamefont {D.~R.}\ \bibnamefont {{Stinebring}}}, \bibinfo {author} {\bibfnamefont {K.}~\bibnamefont {{Stovall}}}, \bibinfo {author} {\bibfnamefont {J.~P.}\ \bibnamefont {{Sun}}}, \bibinfo {author} {\bibfnamefont {A.}~\bibnamefont {{Susobhanan}}}, \bibinfo {author} {\bibfnamefont {J.~K.}\ \bibnamefont {{Swiggum}}}, \bibinfo {author} {\bibfnamefont {J.}~\bibnamefont {{Taylor}}}, \bibinfo {author} {\bibfnamefont {S.~R.}\ \bibnamefont {{Taylor}}}, \bibinfo {author} {\bibfnamefont {J.~E.}\ \bibnamefont {{Turner}}}, \bibinfo {author} {\bibfnamefont {C.}~\bibnamefont {{Unal}}}, \bibinfo {author} {\bibfnamefont {M.}~\bibnamefont {{Vallisneri}}}, \bibinfo {author} {\bibfnamefont {R.}~\bibnamefont {{van Haasteren}}}, \bibinfo {author} {\bibfnamefont {S.~J.}\ \bibnamefont {{Vigeland}}}, \bibinfo {author} {\bibfnamefont {H.~M.}\ \bibnamefont {{Wahl}}}, \bibinfo {author} {\bibfnamefont {Q.}~\bibnamefont {{Wang}}}, \bibinfo {author} {\bibfnamefont {C.~A.}\ \bibnamefont {{Witt}}}, \bibinfo {author} {\bibfnamefont {O.}~\bibnamefont {{Young}}},\ and\ \bibinfo {author} {\bibnamefont {{Nanograv Collaboration}}},\ }\bibfield  {title} {\bibinfo {title} {{The NANOGrav 15 yr Data Set: Evidence for a Gravitational-wave Background}},\ }\href {https://doi.org/10.3847/2041-8213/acdac6} {\bibfield  {journal} {\bibinfo  {journal} {Astrophys. J. Lett.}\ }\textbf {\bibinfo {volume} {951}},\ \bibinfo {eid} {L8} (\bibinfo {year} {2023}{\natexlab{a}})},\ \Eprint {https://arxiv.org/abs/2306.16213} {arXiv:2306.16213 [astro-ph.HE]} \BibitemShut {NoStop}%
\bibitem [{\citenamefont {{Agazie}}\ \emph {et~al.}(2023{\natexlab{b}})\citenamefont {{Agazie}}, \citenamefont {{Anumarlapudi}}, \citenamefont {{Archibald}}, \citenamefont {{Baker}}, \citenamefont {{B{\'e}csy}}, \citenamefont {{Blecha}}, \citenamefont {{Bonilla}}, \citenamefont {{Brazier}}, \citenamefont {{Brook}}, \citenamefont {{Burke-Spolaor}}, \citenamefont {{Burnette}}, \citenamefont {{Case}}, \citenamefont {{Casey-Clyde}}, \citenamefont {{Charisi}}, \citenamefont {{Chatterjee}}, \citenamefont {{Chatziioannou}}, \citenamefont {{Cheeseboro}}, \citenamefont {{Chen}}, \citenamefont {{Cohen}}, \citenamefont {{Cordes}}, \citenamefont {{Cornish}}, \citenamefont {{Crawford}}, \citenamefont {{Cromartie}}, \citenamefont {{Crowter}}, \citenamefont {{Cutler}}, \citenamefont {{D'Orazio}}, \citenamefont {{Decesar}}, \citenamefont {{Degan}}, \citenamefont {{Demorest}}, \citenamefont {{Deng}}, \citenamefont {{Dolch}}, \citenamefont {{Drachler}}, \citenamefont {{Ferrara}}, \citenamefont {{Fiore}}, \citenamefont {{Fonseca}}, \citenamefont {{Freedman}}, \citenamefont {{Gardiner}}, \citenamefont {{Garver-Daniels}}, \citenamefont {{Gentile}}, \citenamefont {{Gersbach}}, \citenamefont {{Glaser}}, \citenamefont {{Good}}, \citenamefont {{G{\"u}ltekin}}, \citenamefont {{Hazboun}}, \citenamefont {{Hourihane}}, \citenamefont {{Islo}}, \citenamefont {{Jennings}}, \citenamefont {{Johnson}}, \citenamefont {{Jones}}, \citenamefont {{Kaiser}}, \citenamefont {{Kaplan}}, \citenamefont {{Kelley}}, \citenamefont {{Kerr}}, \citenamefont {{Key}}, \citenamefont {{Laal}}, \citenamefont {{Lam}}, \citenamefont {{Lamb}}, \citenamefont {{Lazio}}, \citenamefont {{Lewandowska}}, \citenamefont {{Littenberg}}, \citenamefont {{Liu}}, \citenamefont {{Luo}}, \citenamefont {{Lynch}}, \citenamefont {{Ma}}, \citenamefont {{Madison}}, \citenamefont {{McEwen}}, \citenamefont {{McKee}}, \citenamefont {{McLaughlin}}, \citenamefont {{McMann}}, \citenamefont {{Meyers}}, \citenamefont {{Meyers}}, \citenamefont {{Mingarelli}}, \citenamefont {{Mitridate}}, \citenamefont {{Natarajan}}, \citenamefont {{Ng}}, \citenamefont {{Nice}}, \citenamefont {{Ocker}}, \citenamefont {{Olum}}, \citenamefont {{Pennucci}}, \citenamefont {{Perera}}, \citenamefont {{Petrov}}, \citenamefont {{Pol}}, \citenamefont {{Radovan}}, \citenamefont {{Ransom}}, \citenamefont {{Ray}}, \citenamefont {{Romano}}, \citenamefont {{Runnoe}}, \citenamefont {{Sardesai}}, \citenamefont {{Schmiedekamp}}, \citenamefont {{Schmiedekamp}}, \citenamefont {{Schmitz}}, \citenamefont {{Schult}}, \citenamefont {{Shapiro-Albert}}, \citenamefont {{Siemens}}, \citenamefont {{Simon}}, \citenamefont {{Siwek}}, \citenamefont {{Stairs}}, \citenamefont {{Stinebring}}, \citenamefont {{Stovall}}, \citenamefont {{Sun}}, \citenamefont {{Susobhanan}}, \citenamefont {{Swiggum}}, \citenamefont {{Taylor}}, \citenamefont {{Taylor}}, \citenamefont {{Turner}}, \citenamefont {{Unal}}, \citenamefont {{Vallisneri}}, \citenamefont {{Vigeland}}, \citenamefont {{Wachter}}, \citenamefont {{Wahl}}, \citenamefont {{Wang}}, \citenamefont {{Witt}}, \citenamefont {{Wright}}, \citenamefont {{Young}},\ and\ \citenamefont {{Nanograv Collaboration}}}]{Nanograv2023b}%
  \BibitemOpen
  \bibfield  {author} {\bibinfo {author} {\bibfnamefont {G.}~\bibnamefont {{Agazie}}}, \bibinfo {author} {\bibfnamefont {A.}~\bibnamefont {{Anumarlapudi}}}, \bibinfo {author} {\bibfnamefont {A.~M.}\ \bibnamefont {{Archibald}}}, \bibinfo {author} {\bibfnamefont {P.~T.}\ \bibnamefont {{Baker}}}, \bibinfo {author} {\bibfnamefont {B.}~\bibnamefont {{B{\'e}csy}}}, \bibinfo {author} {\bibfnamefont {L.}~\bibnamefont {{Blecha}}}, \bibinfo {author} {\bibfnamefont {A.}~\bibnamefont {{Bonilla}}}, \bibinfo {author} {\bibfnamefont {A.}~\bibnamefont {{Brazier}}}, \bibinfo {author} {\bibfnamefont {P.~R.}\ \bibnamefont {{Brook}}}, \bibinfo {author} {\bibfnamefont {S.}~\bibnamefont {{Burke-Spolaor}}}, \bibinfo {author} {\bibfnamefont {R.}~\bibnamefont {{Burnette}}}, \bibinfo {author} {\bibfnamefont {R.}~\bibnamefont {{Case}}}, \bibinfo {author} {\bibfnamefont {J.~A.}\ \bibnamefont {{Casey-Clyde}}}, \bibinfo {author} {\bibfnamefont {M.}~\bibnamefont {{Charisi}}}, \bibinfo {author} {\bibfnamefont {S.}~\bibnamefont {{Chatterjee}}}, \bibinfo {author} {\bibfnamefont {K.}~\bibnamefont {{Chatziioannou}}}, \bibinfo {author} {\bibfnamefont {B.~D.}\ \bibnamefont {{Cheeseboro}}}, \bibinfo {author} {\bibfnamefont {S.}~\bibnamefont {{Chen}}}, \bibinfo {author} {\bibfnamefont {T.}~\bibnamefont {{Cohen}}}, \bibinfo {author} {\bibfnamefont {J.~M.}\ \bibnamefont {{Cordes}}}, \bibinfo {author} {\bibfnamefont {N.~J.}\ \bibnamefont {{Cornish}}}, \bibinfo {author} {\bibfnamefont {F.}~\bibnamefont {{Crawford}}}, \bibinfo {author} {\bibfnamefont {H.~T.}\ \bibnamefont {{Cromartie}}}, \bibinfo {author} {\bibfnamefont {K.}~\bibnamefont {{Crowter}}}, \bibinfo {author} {\bibfnamefont {C.~J.}\ \bibnamefont {{Cutler}}}, \bibinfo {author} {\bibfnamefont {D.~J.}\ \bibnamefont {{D'Orazio}}}, \bibinfo {author} {\bibfnamefont {M.~E.}\ \bibnamefont {{Decesar}}}, \bibinfo {author} {\bibfnamefont {D.}~\bibnamefont {{Degan}}}, \bibinfo {author} {\bibfnamefont {P.~B.}\ \bibnamefont {{Demorest}}}, \bibinfo {author} {\bibfnamefont {H.}~\bibnamefont {{Deng}}}, \bibinfo {author} {\bibfnamefont {T.}~\bibnamefont {{Dolch}}}, \bibinfo {author} {\bibfnamefont {B.}~\bibnamefont {{Drachler}}}, \bibinfo {author} {\bibfnamefont {E.~C.}\ \bibnamefont {{Ferrara}}}, \bibinfo {author} {\bibfnamefont {W.}~\bibnamefont {{Fiore}}}, \bibinfo {author} {\bibfnamefont {E.}~\bibnamefont {{Fonseca}}}, \bibinfo {author} {\bibfnamefont {G.~E.}\ \bibnamefont {{Freedman}}}, \bibinfo {author} {\bibfnamefont {E.}~\bibnamefont {{Gardiner}}}, \bibinfo {author} {\bibfnamefont {N.}~\bibnamefont {{Garver-Daniels}}}, \bibinfo {author} {\bibfnamefont {P.~A.}\ \bibnamefont {{Gentile}}}, \bibinfo {author} {\bibfnamefont {K.~A.}\ \bibnamefont {{Gersbach}}}, \bibinfo {author} {\bibfnamefont {J.}~\bibnamefont {{Glaser}}}, \bibinfo {author} {\bibfnamefont {D.~C.}\ \bibnamefont {{Good}}}, \bibinfo {author} {\bibfnamefont {K.}~\bibnamefont {{G{\"u}ltekin}}}, \bibinfo {author} {\bibfnamefont {J.~S.}\ \bibnamefont {{Hazboun}}}, \bibinfo {author} {\bibfnamefont {S.}~\bibnamefont {{Hourihane}}}, \bibinfo {author} {\bibfnamefont {K.}~\bibnamefont {{Islo}}}, \bibinfo {author} {\bibfnamefont {R.~J.}\ \bibnamefont {{Jennings}}}, \bibinfo {author} {\bibfnamefont {A.}~\bibnamefont {{Johnson}}}, \bibinfo {author} {\bibfnamefont {M.~L.}\ \bibnamefont {{Jones}}}, \bibinfo {author} {\bibfnamefont {A.~R.}\ \bibnamefont {{Kaiser}}}, \bibinfo {author} {\bibfnamefont {D.~L.}\ \bibnamefont {{Kaplan}}}, \bibinfo {author} {\bibfnamefont {L.~Z.}\ \bibnamefont {{Kelley}}}, \bibinfo {author} {\bibfnamefont {M.}~\bibnamefont {{Kerr}}}, \bibinfo {author} {\bibfnamefont {J.~S.}\ \bibnamefont {{Key}}}, \bibinfo {author} {\bibfnamefont {N.}~\bibnamefont {{Laal}}}, \bibinfo {author} {\bibfnamefont {M.~T.}\ \bibnamefont {{Lam}}}, \bibinfo {author} {\bibfnamefont {W.~G.}\ \bibnamefont {{Lamb}}}, \bibinfo {author} {\bibfnamefont {T.~J.~W.}\ \bibnamefont {{Lazio}}}, \bibinfo {author} {\bibfnamefont {N.}~\bibnamefont {{Lewandowska}}}, \bibinfo {author} {\bibfnamefont {T.~B.}\ \bibnamefont {{Littenberg}}}, \bibinfo {author} {\bibfnamefont {T.}~\bibnamefont {{Liu}}}, \bibinfo {author} {\bibfnamefont {J.}~\bibnamefont {{Luo}}}, \bibinfo {author} {\bibfnamefont {R.~S.}\ \bibnamefont {{Lynch}}}, \bibinfo {author} {\bibfnamefont {C.-P.}\ \bibnamefont {{Ma}}}, \bibinfo {author} {\bibfnamefont {D.~R.}\ \bibnamefont {{Madison}}}, \bibinfo {author} {\bibfnamefont {A.}~\bibnamefont {{McEwen}}}, \bibinfo {author} {\bibfnamefont {J.~W.}\ \bibnamefont {{McKee}}}, \bibinfo {author} {\bibfnamefont {M.~A.}\ \bibnamefont {{McLaughlin}}}, \bibinfo {author} {\bibfnamefont {N.}~\bibnamefont {{McMann}}}, \bibinfo {author} {\bibfnamefont {B.~W.}\ \bibnamefont {{Meyers}}}, \bibinfo {author} {\bibfnamefont {P.~M.}\ \bibnamefont {{Meyers}}}, \bibinfo {author} {\bibfnamefont {C.~M.~F.}\ \bibnamefont {{Mingarelli}}}, \bibinfo {author} {\bibfnamefont {A.}~\bibnamefont {{Mitridate}}}, \bibinfo {author} {\bibfnamefont {P.}~\bibnamefont {{Natarajan}}}, \bibinfo {author} {\bibfnamefont {C.}~\bibnamefont {{Ng}}}, \bibinfo {author} {\bibfnamefont {D.~J.}\ \bibnamefont {{Nice}}}, \bibinfo {author} {\bibfnamefont {S.~K.}\ \bibnamefont {{Ocker}}}, \bibinfo {author} {\bibfnamefont {K.~D.}\ \bibnamefont {{Olum}}}, \bibinfo {author} {\bibfnamefont {T.~T.}\ \bibnamefont {{Pennucci}}}, \bibinfo {author} {\bibfnamefont {B.~B.~P.}\ \bibnamefont {{Perera}}}, \bibinfo {author} {\bibfnamefont {P.}~\bibnamefont {{Petrov}}}, \bibinfo {author} {\bibfnamefont {N.~S.}\ \bibnamefont {{Pol}}}, \bibinfo {author} {\bibfnamefont {H.~A.}\ \bibnamefont {{Radovan}}}, \bibinfo {author} {\bibfnamefont {S.~M.}\ \bibnamefont {{Ransom}}}, \bibinfo {author} {\bibfnamefont {P.~S.}\ \bibnamefont {{Ray}}}, \bibinfo {author} {\bibfnamefont {J.~D.}\ \bibnamefont {{Romano}}}, \bibinfo {author} {\bibfnamefont {J.~C.}\ \bibnamefont {{Runnoe}}}, \bibinfo {author} {\bibfnamefont {S.~C.}\ \bibnamefont {{Sardesai}}}, \bibinfo {author} {\bibfnamefont {A.}~\bibnamefont {{Schmiedekamp}}}, \bibinfo {author} {\bibfnamefont {C.}~\bibnamefont {{Schmiedekamp}}}, \bibinfo {author} {\bibfnamefont {K.}~\bibnamefont {{Schmitz}}}, \bibinfo {author} {\bibfnamefont {L.}~\bibnamefont {{Schult}}}, \bibinfo {author} {\bibfnamefont {B.~J.}\ \bibnamefont {{Shapiro-Albert}}}, \bibinfo {author} {\bibfnamefont {X.}~\bibnamefont {{Siemens}}}, \bibinfo {author} {\bibfnamefont {J.}~\bibnamefont {{Simon}}}, \bibinfo {author} {\bibfnamefont {M.~S.}\ \bibnamefont {{Siwek}}}, \bibinfo {author} {\bibfnamefont {I.~H.}\ \bibnamefont {{Stairs}}}, \bibinfo {author} {\bibfnamefont {D.~R.}\ \bibnamefont {{Stinebring}}}, \bibinfo {author} {\bibfnamefont {K.}~\bibnamefont {{Stovall}}}, \bibinfo {author} {\bibfnamefont {J.~P.}\ \bibnamefont {{Sun}}}, \bibinfo {author} {\bibfnamefont {A.}~\bibnamefont {{Susobhanan}}}, \bibinfo {author} {\bibfnamefont {J.~K.}\ \bibnamefont {{Swiggum}}}, \bibinfo {author} {\bibfnamefont {J.}~\bibnamefont {{Taylor}}}, \bibinfo {author} {\bibfnamefont {S.~R.}\ \bibnamefont {{Taylor}}}, \bibinfo {author} {\bibfnamefont {J.~E.}\ \bibnamefont {{Turner}}}, \bibinfo {author} {\bibfnamefont {C.}~\bibnamefont {{Unal}}}, \bibinfo {author} {\bibfnamefont {M.}~\bibnamefont {{Vallisneri}}}, \bibinfo {author} {\bibfnamefont {S.~J.}\ \bibnamefont {{Vigeland}}}, \bibinfo {author} {\bibfnamefont {J.~M.}\ \bibnamefont {{Wachter}}}, \bibinfo {author} {\bibfnamefont {H.~M.}\ \bibnamefont {{Wahl}}}, \bibinfo {author} {\bibfnamefont {Q.}~\bibnamefont {{Wang}}}, \bibinfo {author} {\bibfnamefont {C.~A.}\ \bibnamefont {{Witt}}}, \bibinfo {author} {\bibfnamefont {D.}~\bibnamefont {{Wright}}}, \bibinfo {author} {\bibfnamefont {O.}~\bibnamefont {{Young}}},\ and\ \bibinfo {author} {\bibnamefont {{Nanograv Collaboration}}},\ }\bibfield  {title} {\bibinfo {title} {{The NANOGrav 15 yr Data Set: Constraints on Supermassive Black Hole Binaries from the Gravitational-wave Background}},\ }\href {https://doi.org/10.3847/2041-8213/ace18b} {\bibfield  {journal} {\bibinfo  {journal} {Astrophys. J. Lett.}\ }\textbf {\bibinfo {volume} {952}},\ \bibinfo {eid} {L37} (\bibinfo {year} {2023}{\natexlab{b}})},\ \Eprint {https://arxiv.org/abs/2306.16220} {arXiv:2306.16220 [astro-ph.HE]} \BibitemShut {NoStop}%
\bibitem [{\citenamefont {Agazie}\ \emph {et~al.}(2024{\natexlab{a}})\citenamefont {Agazie} \emph {et~al.}}]{InternationalPulsarTimingArray:2023mzf}%
  \BibitemOpen
  \bibfield  {author} {\bibinfo {author} {\bibfnamefont {G.}~\bibnamefont {Agazie}} \emph {et~al.} (\bibinfo {collaboration} {International Pulsar Timing Array}),\ }\bibfield  {title} {\bibinfo {title} {{Comparing Recent Pulsar Timing Array Results on the Nanohertz Stochastic Gravitational-wave Background}},\ }\href {https://doi.org/10.3847/1538-4357/ad36be} {\bibfield  {journal} {\bibinfo  {journal} {Astrophys. J.}\ }\textbf {\bibinfo {volume} {966}},\ \bibinfo {pages} {105} (\bibinfo {year} {2024}{\natexlab{a}})},\ \Eprint {https://arxiv.org/abs/2309.00693} {arXiv:2309.00693 [astro-ph.HE]} \BibitemShut {NoStop}%
\bibitem [{\citenamefont {Agazie}\ \emph {et~al.}(2023{\natexlab{a}})\citenamefont {Agazie} \emph {et~al.}}]{NANOGrav:2023pdq}%
  \BibitemOpen
  \bibfield  {author} {\bibinfo {author} {\bibfnamefont {G.}~\bibnamefont {Agazie}} \emph {et~al.} (\bibinfo {collaboration} {NANOGrav}),\ }\bibfield  {title} {\bibinfo {title} {{The NANOGrav 15 yr Data Set: Bayesian Limits on Gravitational Waves from Individual Supermassive Black Hole Binaries}},\ }\href {https://doi.org/10.3847/2041-8213/ace18a} {\bibfield  {journal} {\bibinfo  {journal} {Astrophys. J. Lett.}\ }\textbf {\bibinfo {volume} {951}},\ \bibinfo {pages} {L50} (\bibinfo {year} {2023}{\natexlab{a}})},\ \Eprint {https://arxiv.org/abs/2306.16222} {arXiv:2306.16222 [astro-ph.HE]} \BibitemShut {NoStop}%
\bibitem [{\citenamefont {Agazie}\ \emph {et~al.}(2024{\natexlab{b}})\citenamefont {Agazie} \emph {et~al.}}]{NANOGrav:2023vfo}%
  \BibitemOpen
  \bibfield  {author} {\bibinfo {author} {\bibfnamefont {G.}~\bibnamefont {Agazie}} \emph {et~al.} (\bibinfo {collaboration} {NANOGrav}),\ }\bibfield  {title} {\bibinfo {title} {{The NANOGrav 12.5 yr Data Set: Search for Gravitational Wave Memory}},\ }\href {https://doi.org/10.3847/1538-4357/ad0726} {\bibfield  {journal} {\bibinfo  {journal} {Astrophys. J.}\ }\textbf {\bibinfo {volume} {963}},\ \bibinfo {pages} {61} (\bibinfo {year} {2024}{\natexlab{b}})},\ \Eprint {https://arxiv.org/abs/2307.13797} {arXiv:2307.13797 [gr-qc]} \BibitemShut {NoStop}%
\bibitem [{\citenamefont {Agazie}\ \emph {et~al.}(2023{\natexlab{b}})\citenamefont {Agazie} \emph {et~al.}}]{NANOGrav:2023hde}%
  \BibitemOpen
  \bibfield  {author} {\bibinfo {author} {\bibfnamefont {G.}~\bibnamefont {Agazie}} \emph {et~al.} (\bibinfo {collaboration} {NANOGrav}),\ }\bibfield  {title} {\bibinfo {title} {{The NANOGrav 15 yr Data Set: Observations and Timing of 68 Millisecond Pulsars}},\ }\href {https://doi.org/10.3847/2041-8213/acda9a} {\bibfield  {journal} {\bibinfo  {journal} {Astrophys. J. Lett.}\ }\textbf {\bibinfo {volume} {951}},\ \bibinfo {pages} {L9} (\bibinfo {year} {2023}{\natexlab{b}})},\ \Eprint {https://arxiv.org/abs/2306.16217} {arXiv:2306.16217 [astro-ph.HE]} \BibitemShut {NoStop}%
\bibitem [{\citenamefont {Agazie}\ \emph {et~al.}(2023{\natexlab{c}})\citenamefont {Agazie} \emph {et~al.}}]{NANOGrav:2023ctt}%
  \BibitemOpen
  \bibfield  {author} {\bibinfo {author} {\bibfnamefont {G.}~\bibnamefont {Agazie}} \emph {et~al.} (\bibinfo {collaboration} {NANOGrav}),\ }\bibfield  {title} {\bibinfo {title} {{The NANOGrav 15 yr Data Set: Detector Characterization and Noise Budget}},\ }\href {https://doi.org/10.3847/2041-8213/acda88} {\bibfield  {journal} {\bibinfo  {journal} {Astrophys. J. Lett.}\ }\textbf {\bibinfo {volume} {951}},\ \bibinfo {pages} {L10} (\bibinfo {year} {2023}{\natexlab{c}})},\ \Eprint {https://arxiv.org/abs/2306.16218} {arXiv:2306.16218 [astro-ph.HE]} \BibitemShut {NoStop}%
\bibitem [{\citenamefont {Agazie}\ \emph {et~al.}(2023{\natexlab{d}})\citenamefont {Agazie} \emph {et~al.}}]{NANOGrav:2023tcn}%
  \BibitemOpen
  \bibfield  {author} {\bibinfo {author} {\bibfnamefont {G.}~\bibnamefont {Agazie}} \emph {et~al.} (\bibinfo {collaboration} {NANOGrav}),\ }\bibfield  {title} {\bibinfo {title} {{The NANOGrav 15 yr Data Set: Search for Anisotropy in the Gravitational-wave Background}},\ }\href {https://doi.org/10.3847/2041-8213/acf4fd} {\bibfield  {journal} {\bibinfo  {journal} {Astrophys. J. Lett.}\ }\textbf {\bibinfo {volume} {956}},\ \bibinfo {pages} {L3} (\bibinfo {year} {2023}{\natexlab{d}})},\ \Eprint {https://arxiv.org/abs/2306.16221} {arXiv:2306.16221 [astro-ph.HE]} \BibitemShut {NoStop}%
\bibitem [{\citenamefont {Toomre}\ and\ \citenamefont {Toomre}(1972)}]{Toomre:1972vt}%
  \BibitemOpen
  \bibfield  {author} {\bibinfo {author} {\bibfnamefont {A.}~\bibnamefont {Toomre}}\ and\ \bibinfo {author} {\bibfnamefont {J.}~\bibnamefont {Toomre}},\ }\bibfield  {title} {\bibinfo {title} {{Galactic bridges and tails}},\ }\href {https://doi.org/10.1086/151823} {\bibfield  {journal} {\bibinfo  {journal} {Astrophys. J.}\ }\textbf {\bibinfo {volume} {178}},\ \bibinfo {pages} {623} (\bibinfo {year} {1972})}\BibitemShut {NoStop}%
\bibitem [{\citenamefont {Barnes}\ and\ \citenamefont {Hernquist}(1992)}]{Barnes:1992rm}%
  \BibitemOpen
  \bibfield  {author} {\bibinfo {author} {\bibfnamefont {J.~E.}\ \bibnamefont {Barnes}}\ and\ \bibinfo {author} {\bibfnamefont {L.~E.}\ \bibnamefont {Hernquist}},\ }\bibfield  {title} {\bibinfo {title} {{Dynamics of interacting galaxies}},\ }\href {https://doi.org/10.1146/annurev.aa.30.090192.003421} {\bibfield  {journal} {\bibinfo  {journal} {Ann. Rev. Astron. Astrophys.}\ }\textbf {\bibinfo {volume} {30}},\ \bibinfo {pages} {705} (\bibinfo {year} {1992})}\BibitemShut {NoStop}%
\bibitem [{\citenamefont {Boylan-Kolchin}\ \emph {et~al.}(2005)\citenamefont {Boylan-Kolchin}, \citenamefont {Ma},\ and\ \citenamefont {Quataert}}]{Boylan-Kolchin:2005cvz}%
  \BibitemOpen
  \bibfield  {author} {\bibinfo {author} {\bibfnamefont {M.}~\bibnamefont {Boylan-Kolchin}}, \bibinfo {author} {\bibfnamefont {C.-P.}\ \bibnamefont {Ma}},\ and\ \bibinfo {author} {\bibfnamefont {E.}~\bibnamefont {Quataert}},\ }\bibfield  {title} {\bibinfo {title} {{Dissipationless mergers of elliptical galaxies and the evolution of the fundamental plane}},\ }\href {https://doi.org/10.1111/j.1365-2966.2005.09278.x} {\bibfield  {journal} {\bibinfo  {journal} {Mon. Not. Roy. Astron. Soc.}\ }\textbf {\bibinfo {volume} {362}},\ \bibinfo {pages} {184} (\bibinfo {year} {2005})},\ \Eprint {https://arxiv.org/abs/astro-ph/0502495} {arXiv:astro-ph/0502495} \BibitemShut {NoStop}%
\bibitem [{\citenamefont {Cox}\ \emph {et~al.}(2006)\citenamefont {Cox}, \citenamefont {Dutta}, \citenamefont {Di~Matteo}, \citenamefont {Hernquist}, \citenamefont {Hopkins}, \citenamefont {Robertson},\ and\ \citenamefont {Springel}}]{Cox:2006hd}%
  \BibitemOpen
  \bibfield  {author} {\bibinfo {author} {\bibfnamefont {T.~J.}\ \bibnamefont {Cox}}, \bibinfo {author} {\bibfnamefont {S.~N.}\ \bibnamefont {Dutta}}, \bibinfo {author} {\bibfnamefont {T.}~\bibnamefont {Di~Matteo}}, \bibinfo {author} {\bibfnamefont {L.}~\bibnamefont {Hernquist}}, \bibinfo {author} {\bibfnamefont {P.~F.}\ \bibnamefont {Hopkins}}, \bibinfo {author} {\bibfnamefont {B.}~\bibnamefont {Robertson}},\ and\ \bibinfo {author} {\bibfnamefont {V.}~\bibnamefont {Springel}},\ }\bibfield  {title} {\bibinfo {title} {{Kinematic Structure of Merger Remnants}},\ }\href {https://doi.org/10.1086/507474} {\bibfield  {journal} {\bibinfo  {journal} {Astrophys. J.}\ }\textbf {\bibinfo {volume} {650}},\ \bibinfo {pages} {791} (\bibinfo {year} {2006})},\ \Eprint {https://arxiv.org/abs/astro-ph/0607446} {arXiv:astro-ph/0607446} \BibitemShut {NoStop}%
\bibitem [{\citenamefont {Naab}\ and\ \citenamefont {Trujillo}(2006)}]{Naab:2005wz}%
  \BibitemOpen
  \bibfield  {author} {\bibinfo {author} {\bibfnamefont {T.}~\bibnamefont {Naab}}\ and\ \bibinfo {author} {\bibfnamefont {I.}~\bibnamefont {Trujillo}},\ }\bibfield  {title} {\bibinfo {title} {{Surface density profiles of collisionless disc merger remnants}},\ }\href {https://doi.org/10.1111/j.1365-2966.2006.10252.x} {\bibfield  {journal} {\bibinfo  {journal} {Mon. Not. Roy. Astron. Soc.}\ }\textbf {\bibinfo {volume} {369}},\ \bibinfo {pages} {625} (\bibinfo {year} {2006})},\ \Eprint {https://arxiv.org/abs/astro-ph/0508362} {arXiv:astro-ph/0508362} \BibitemShut {NoStop}%
\bibitem [{\citenamefont {Conselice}(2014)}]{Conselice:2014joa}%
  \BibitemOpen
  \bibfield  {author} {\bibinfo {author} {\bibfnamefont {C.~J.}\ \bibnamefont {Conselice}},\ }\bibfield  {title} {\bibinfo {title} {{The Evolution of Galaxy Structure over Cosmic Time}},\ }\href {https://doi.org/10.1146/annurev-astro-081913-040037} {\bibfield  {journal} {\bibinfo  {journal} {Ann. Rev. Astron. Astrophys.}\ }\textbf {\bibinfo {volume} {52}},\ \bibinfo {pages} {291} (\bibinfo {year} {2014})},\ \Eprint {https://arxiv.org/abs/1403.2783} {arXiv:1403.2783 [astro-ph.GA]} \BibitemShut {NoStop}%
\bibitem [{\citenamefont {Fiacconi}\ \emph {et~al.}(2015)\citenamefont {Fiacconi}, \citenamefont {Feldmann},\ and\ \citenamefont {Mayer}}]{Fiacconi:2014lba}%
  \BibitemOpen
  \bibfield  {author} {\bibinfo {author} {\bibfnamefont {D.}~\bibnamefont {Fiacconi}}, \bibinfo {author} {\bibfnamefont {R.}~\bibnamefont {Feldmann}},\ and\ \bibinfo {author} {\bibfnamefont {L.}~\bibnamefont {Mayer}},\ }\bibfield  {title} {\bibinfo {title} {{The Argo simulation \textendash{} II. The early build-up of the Hubble sequence}},\ }\href {https://doi.org/10.1093/mnras/stu2228} {\bibfield  {journal} {\bibinfo  {journal} {Mon. Not. Roy. Astron. Soc.}\ }\textbf {\bibinfo {volume} {446}},\ \bibinfo {pages} {1957} (\bibinfo {year} {2015})},\ \Eprint {https://arxiv.org/abs/1410.6818} {arXiv:1410.6818 [astro-ph.GA]} \BibitemShut {NoStop}%
\bibitem [{\citenamefont {Talbot}\ \emph {et~al.}(2024)\citenamefont {Talbot}, \citenamefont {Sijacki},\ and\ \citenamefont {Bourne}}]{Talbot:2023kfn}%
  \BibitemOpen
  \bibfield  {author} {\bibinfo {author} {\bibfnamefont {R.~Y.}\ \bibnamefont {Talbot}}, \bibinfo {author} {\bibfnamefont {D.}~\bibnamefont {Sijacki}},\ and\ \bibinfo {author} {\bibfnamefont {M.~A.}\ \bibnamefont {Bourne}},\ }\bibfield  {title} {\bibinfo {title} {{Simulations of spin-driven AGN jets in gas-rich galaxy mergers}},\ }\href {https://doi.org/10.1093/mnras/stae392} {\bibfield  {journal} {\bibinfo  {journal} {Mon. Not. Roy. Astron. Soc.}\ }\textbf {\bibinfo {volume} {528}},\ \bibinfo {pages} {5432} (\bibinfo {year} {2024})},\ \Eprint {https://arxiv.org/abs/2306.07316} {arXiv:2306.07316 [astro-ph.GA]} \BibitemShut {NoStop}%
\bibitem [{\citenamefont {Peters}(1964)}]{Peters:1964zz}%
  \BibitemOpen
  \bibfield  {author} {\bibinfo {author} {\bibfnamefont {P.~C.}\ \bibnamefont {Peters}},\ }\bibfield  {title} {\bibinfo {title} {{Gravitational Radiation and the Motion of Two Point Masses}},\ }\href {https://doi.org/10.1103/PhysRev.136.B1224} {\bibfield  {journal} {\bibinfo  {journal} {Phys. Rev.}\ }\textbf {\bibinfo {volume} {136}},\ \bibinfo {pages} {B1224} (\bibinfo {year} {1964})}\BibitemShut {NoStop}%
\bibitem [{\citenamefont {Alonso-\'Alvarez}\ \emph {et~al.}(2024)\citenamefont {Alonso-\'Alvarez}, \citenamefont {Cline},\ and\ \citenamefont {Dewar}}]{Alonso-Alvarez:2024gdz}%
  \BibitemOpen
  \bibfield  {author} {\bibinfo {author} {\bibfnamefont {G.}~\bibnamefont {Alonso-\'Alvarez}}, \bibinfo {author} {\bibfnamefont {J.~M.}\ \bibnamefont {Cline}},\ and\ \bibinfo {author} {\bibfnamefont {C.}~\bibnamefont {Dewar}},\ }\bibfield  {title} {\bibinfo {title} {{Self-Interacting Dark Matter Solves the Final Parsec Problem of Supermassive Black Hole Mergers}},\ }\href {https://doi.org/10.1103/PhysRevLett.133.021401} {\bibfield  {journal} {\bibinfo  {journal} {Phys. Rev. Lett.}\ }\textbf {\bibinfo {volume} {133}},\ \bibinfo {pages} {021401} (\bibinfo {year} {2024})},\ \Eprint {https://arxiv.org/abs/2401.14450} {arXiv:2401.14450 [astro-ph.CO]} \BibitemShut {NoStop}%
\bibitem [{\citenamefont {Chen}\ \emph {et~al.}(2024)\citenamefont {Chen} \emph {et~al.}}]{NANOGrav:2024nmo}%
  \BibitemOpen
  \bibfield  {author} {\bibinfo {author} {\bibfnamefont {Y.}~\bibnamefont {Chen}} \emph {et~al.},\ }\bibfield  {title} {\bibinfo {title} {{Galaxy Tomography with the Gravitational Wave Background from Supermassive Black Hole Binaries}},\ }\href {https://doi.org/10.48550/arXiv.2411.05906} {\bibfield  {journal} {\bibinfo  {journal} {arXiv e-prints}\ ,\ \bibinfo {eid} {arXiv:2411.05906}} (\bibinfo {year} {2024})},\ \Eprint {https://arxiv.org/abs/2411.05906} {arXiv:2411.05906 [astro-ph.HE]} \BibitemShut {NoStop}%
\bibitem [{\citenamefont {Quinlan}(1996)}]{Quinlan:1996vp}%
  \BibitemOpen
  \bibfield  {author} {\bibinfo {author} {\bibfnamefont {G.~D.}\ \bibnamefont {Quinlan}},\ }\bibfield  {title} {\bibinfo {title} {{The dynamical evolution of massive black hole binaries - I. hardening in a fixed stellar background}},\ }\href {https://doi.org/10.1016/S1384-1076(96)00003-6} {\bibfield  {journal} {\bibinfo  {journal} {New Astron.}\ }\textbf {\bibinfo {volume} {1}},\ \bibinfo {pages} {35} (\bibinfo {year} {1996})},\ \Eprint {https://arxiv.org/abs/astro-ph/9601092} {arXiv:astro-ph/9601092} \BibitemShut {NoStop}%
\bibitem [{\citenamefont {Milosavljevic}\ and\ \citenamefont {Merritt}(2001)}]{Milosavljevic:2001vi}%
  \BibitemOpen
  \bibfield  {author} {\bibinfo {author} {\bibfnamefont {M.}~\bibnamefont {Milosavljevic}}\ and\ \bibinfo {author} {\bibfnamefont {D.}~\bibnamefont {Merritt}},\ }\bibfield  {title} {\bibinfo {title} {{Formation of galactic nuclei}},\ }\href {https://doi.org/10.1086/323830} {\bibfield  {journal} {\bibinfo  {journal} {Astrophys. J.}\ }\textbf {\bibinfo {volume} {563}},\ \bibinfo {pages} {34} (\bibinfo {year} {2001})},\ \Eprint {https://arxiv.org/abs/astro-ph/0103350} {arXiv:astro-ph/0103350} \BibitemShut {NoStop}%
\bibitem [{\citenamefont {{Milosavljevi{\'c}}}\ and\ \citenamefont {{Merritt}}(2003)}]{2003AIPC..686..201M}%
  \BibitemOpen
  \bibfield  {author} {\bibinfo {author} {\bibfnamefont {M.}~\bibnamefont {{Milosavljevi{\'c}}}}\ and\ \bibinfo {author} {\bibfnamefont {D.}~\bibnamefont {{Merritt}}},\ }\bibfield  {title} {\bibinfo {title} {{The Final Parsec Problem}},\ }in\ \href {https://doi.org/10.1063/1.1629432} {\emph {\bibinfo {booktitle} {The Astrophysics of Gravitational Wave Sources}}},\ \bibinfo {series} {American Institute of Physics Conference Series}, Vol.\ \bibinfo {volume} {686},\ \bibinfo {editor} {edited by\ \bibinfo {editor} {\bibfnamefont {J.~M.}\ \bibnamefont {{Centrella}}}}\ (\bibinfo {year} {2003})\ pp.\ \bibinfo {pages} {201--210},\ \Eprint {https://arxiv.org/abs/astro-ph/0212270} {arXiv:astro-ph/0212270 [astro-ph]} \BibitemShut {NoStop}%
\bibitem [{\citenamefont {Berczik}\ \emph {et~al.}(2006)\citenamefont {Berczik}, \citenamefont {Merritt}, \citenamefont {Spurzem},\ and\ \citenamefont {Bischof}}]{Berczik:2006tz}%
  \BibitemOpen
  \bibfield  {author} {\bibinfo {author} {\bibfnamefont {P.}~\bibnamefont {Berczik}}, \bibinfo {author} {\bibfnamefont {D.}~\bibnamefont {Merritt}}, \bibinfo {author} {\bibfnamefont {R.}~\bibnamefont {Spurzem}},\ and\ \bibinfo {author} {\bibfnamefont {H.-P.}\ \bibnamefont {Bischof}},\ }\bibfield  {title} {\bibinfo {title} {{Efficient merger of binary supermassive black holes in non-axisymmetric galaxies}},\ }\href {https://doi.org/10.1086/504426} {\bibfield  {journal} {\bibinfo  {journal} {Astrophys. J. Lett.}\ }\textbf {\bibinfo {volume} {642}},\ \bibinfo {pages} {L21} (\bibinfo {year} {2006})},\ \Eprint {https://arxiv.org/abs/astro-ph/0601698} {arXiv:astro-ph/0601698} \BibitemShut {NoStop}%
\bibitem [{\citenamefont {Sesana}\ \emph {et~al.}(2006)\citenamefont {Sesana}, \citenamefont {Haardt},\ and\ \citenamefont {Madau}}]{Sesana:2006xw}%
  \BibitemOpen
  \bibfield  {author} {\bibinfo {author} {\bibfnamefont {A.}~\bibnamefont {Sesana}}, \bibinfo {author} {\bibfnamefont {F.}~\bibnamefont {Haardt}},\ and\ \bibinfo {author} {\bibfnamefont {P.}~\bibnamefont {Madau}},\ }\bibfield  {title} {\bibinfo {title} {{Interaction of massive black hole binaries with their stellar environment. 1. Ejection of hypervelocity stars}},\ }\href {https://doi.org/10.1086/507596} {\bibfield  {journal} {\bibinfo  {journal} {Astrophys. J.}\ }\textbf {\bibinfo {volume} {651}},\ \bibinfo {pages} {392} (\bibinfo {year} {2006})},\ \Eprint {https://arxiv.org/abs/astro-ph/0604299} {arXiv:astro-ph/0604299} \BibitemShut {NoStop}%
\bibitem [{\citenamefont {Rantala}\ \emph {et~al.}(2017)\citenamefont {Rantala}, \citenamefont {Pihajoki}, \citenamefont {Johansson}, \citenamefont {Naab}, \citenamefont {Lah\'en},\ and\ \citenamefont {Sawala}}]{Rantala:2016rng}%
  \BibitemOpen
  \bibfield  {author} {\bibinfo {author} {\bibfnamefont {A.}~\bibnamefont {Rantala}}, \bibinfo {author} {\bibfnamefont {P.}~\bibnamefont {Pihajoki}}, \bibinfo {author} {\bibfnamefont {P.~H.}\ \bibnamefont {Johansson}}, \bibinfo {author} {\bibfnamefont {T.}~\bibnamefont {Naab}}, \bibinfo {author} {\bibfnamefont {N.}~\bibnamefont {Lah\'en}},\ and\ \bibinfo {author} {\bibfnamefont {T.}~\bibnamefont {Sawala}},\ }\bibfield  {title} {\bibinfo {title} {{Post-Newtonian dynamical modeling of supermassive black holes in galactic-scale simulations}},\ }\href {https://doi.org/10.3847/1538-4357/aa6d65} {\bibfield  {journal} {\bibinfo  {journal} {Astrophys. J.}\ }\textbf {\bibinfo {volume} {840}},\ \bibinfo {pages} {53} (\bibinfo {year} {2017})},\ \Eprint {https://arxiv.org/abs/1611.07028} {arXiv:1611.07028 [astro-ph.GA]} \BibitemShut {NoStop}%
\bibitem [{\citenamefont {Hoffman}\ and\ \citenamefont {Loeb}(2007)}]{Hoffman:2006iq}%
  \BibitemOpen
  \bibfield  {author} {\bibinfo {author} {\bibfnamefont {L.}~\bibnamefont {Hoffman}}\ and\ \bibinfo {author} {\bibfnamefont {A.}~\bibnamefont {Loeb}},\ }\bibfield  {title} {\bibinfo {title} {{Dynamics of triple black hole systems in hierarchically merging massive galaxies}},\ }\href {https://doi.org/10.1111/j.1365-2966.2007.11694.x} {\bibfield  {journal} {\bibinfo  {journal} {Mon. Not. Roy. Astron. Soc.}\ }\textbf {\bibinfo {volume} {377}},\ \bibinfo {pages} {957} (\bibinfo {year} {2007})},\ \Eprint {https://arxiv.org/abs/astro-ph/0612517} {arXiv:astro-ph/0612517} \BibitemShut {NoStop}%
\bibitem [{\citenamefont {Bonetti}\ \emph {et~al.}(2018)\citenamefont {Bonetti}, \citenamefont {Haardt}, \citenamefont {Sesana},\ and\ \citenamefont {Barausse}}]{Bonetti:2017dan}%
  \BibitemOpen
  \bibfield  {author} {\bibinfo {author} {\bibfnamefont {M.}~\bibnamefont {Bonetti}}, \bibinfo {author} {\bibfnamefont {F.}~\bibnamefont {Haardt}}, \bibinfo {author} {\bibfnamefont {A.}~\bibnamefont {Sesana}},\ and\ \bibinfo {author} {\bibfnamefont {E.}~\bibnamefont {Barausse}},\ }\bibfield  {title} {\bibinfo {title} {{Post-Newtonian evolution of massive black hole triplets in galactic nuclei \textendash{} II. Survey of the parameter space}},\ }\href {https://doi.org/10.1093/mnras/sty896} {\bibfield  {journal} {\bibinfo  {journal} {Mon. Not. Roy. Astron. Soc.}\ }\textbf {\bibinfo {volume} {477}},\ \bibinfo {pages} {3910} (\bibinfo {year} {2018})},\ \Eprint {https://arxiv.org/abs/1709.06088} {arXiv:1709.06088 [astro-ph.GA]} \BibitemShut {NoStop}%
\bibitem [{\citenamefont {Mannerkoski}\ \emph {et~al.}(2021)\citenamefont {Mannerkoski}, \citenamefont {Johansson}, \citenamefont {Rantala}, \citenamefont {Naab},\ and\ \citenamefont {Liao}}]{Mannerkoski:2021lal}%
  \BibitemOpen
  \bibfield  {author} {\bibinfo {author} {\bibfnamefont {M.}~\bibnamefont {Mannerkoski}}, \bibinfo {author} {\bibfnamefont {P.~H.}\ \bibnamefont {Johansson}}, \bibinfo {author} {\bibfnamefont {A.}~\bibnamefont {Rantala}}, \bibinfo {author} {\bibfnamefont {T.}~\bibnamefont {Naab}},\ and\ \bibinfo {author} {\bibfnamefont {S.}~\bibnamefont {Liao}},\ }\bibfield  {title} {\bibinfo {title} {{Resolving the Complex Evolution of a Supermassive Black Hole Triplet in a Cosmological Simulation}},\ }\href {https://doi.org/10.3847/2041-8213/abf9a5} {\bibfield  {journal} {\bibinfo  {journal} {Astrophys. J. Lett.}\ }\textbf {\bibinfo {volume} {912}},\ \bibinfo {pages} {L20} (\bibinfo {year} {2021})},\ \Eprint {https://arxiv.org/abs/2103.16254} {arXiv:2103.16254 [astro-ph.GA]} \BibitemShut {NoStop}%
\bibitem [{\citenamefont {{Lai}}\ and\ \citenamefont {{Mu{\~n}oz}}(2023)}]{Lai:2022ylu}%
  \BibitemOpen
  \bibfield  {author} {\bibinfo {author} {\bibfnamefont {D.}~\bibnamefont {{Lai}}}\ and\ \bibinfo {author} {\bibfnamefont {D.~J.}\ \bibnamefont {{Mu{\~n}oz}}},\ }\bibfield  {title} {\bibinfo {title} {{Circumbinary Accretion: From Binary Stars to Massive Binary Black Holes}},\ }\href {https://doi.org/10.1146/annurev-astro-052622-022933} {\bibfield  {journal} {\bibinfo  {journal} {Ann. Rev. Astron. Astrophys.}\ }\textbf {\bibinfo {volume} {61}},\ \bibinfo {pages} {517} (\bibinfo {year} {2023})},\ \Eprint {https://arxiv.org/abs/2211.00028} {arXiv:2211.00028 [astro-ph.HE]} \BibitemShut {NoStop}%
\bibitem [{\citenamefont {Mayer}\ \emph {et~al.}(2007)\citenamefont {Mayer}, \citenamefont {Kazantzidis}, \citenamefont {Madau}, \citenamefont {Colpi}, \citenamefont {Quinn},\ and\ \citenamefont {Wadsley}}]{Mayer:2007vk}%
  \BibitemOpen
  \bibfield  {author} {\bibinfo {author} {\bibfnamefont {L.}~\bibnamefont {Mayer}}, \bibinfo {author} {\bibfnamefont {S.}~\bibnamefont {Kazantzidis}}, \bibinfo {author} {\bibfnamefont {P.}~\bibnamefont {Madau}}, \bibinfo {author} {\bibfnamefont {M.}~\bibnamefont {Colpi}}, \bibinfo {author} {\bibfnamefont {T.~R.}\ \bibnamefont {Quinn}},\ and\ \bibinfo {author} {\bibfnamefont {J.}~\bibnamefont {Wadsley}},\ }\bibfield  {title} {\bibinfo {title} {{Rapid Formation of Supermassive Black Hole Binaries in Galaxy Mergers with Gas}},\ }\href {https://doi.org/10.1126/science.1141858} {\bibfield  {journal} {\bibinfo  {journal} {Science}\ }\textbf {\bibinfo {volume} {316}},\ \bibinfo {pages} {1874} (\bibinfo {year} {2007})},\ \Eprint {https://arxiv.org/abs/0706.1562} {arXiv:0706.1562 [astro-ph]} \BibitemShut {NoStop}%
\bibitem [{\citenamefont {Dunhill}\ \emph {et~al.}(2014)\citenamefont {Dunhill}, \citenamefont {Alexander}, \citenamefont {Nixon},\ and\ \citenamefont {King}}]{Dunhill:2014oka}%
  \BibitemOpen
  \bibfield  {author} {\bibinfo {author} {\bibfnamefont {A.}~\bibnamefont {Dunhill}}, \bibinfo {author} {\bibfnamefont {R.}~\bibnamefont {Alexander}}, \bibinfo {author} {\bibfnamefont {C.}~\bibnamefont {Nixon}},\ and\ \bibinfo {author} {\bibfnamefont {A.}~\bibnamefont {King}},\ }\bibfield  {title} {\bibinfo {title} {{Misaligned accretion on to supermassive black hole binaries}},\ }\href {https://doi.org/10.1093/mnras/stu1914} {\bibfield  {journal} {\bibinfo  {journal} {Mon. Not. Roy. Astron. Soc.}\ }\textbf {\bibinfo {volume} {445}},\ \bibinfo {pages} {2285} (\bibinfo {year} {2014})},\ \Eprint {https://arxiv.org/abs/1409.3842} {arXiv:1409.3842 [astro-ph.HE]} \BibitemShut {NoStop}%
\bibitem [{\citenamefont {Goicovic}\ \emph {et~al.}(2016)\citenamefont {Goicovic}, \citenamefont {Cuadra}, \citenamefont {Sesana}, \citenamefont {Stasyszyn}, \citenamefont {Amaro-Seoane},\ and\ \citenamefont {Tanaka}}]{Goicovic:2015kda}%
  \BibitemOpen
  \bibfield  {author} {\bibinfo {author} {\bibfnamefont {F.~G.}\ \bibnamefont {Goicovic}}, \bibinfo {author} {\bibfnamefont {J.}~\bibnamefont {Cuadra}}, \bibinfo {author} {\bibfnamefont {A.}~\bibnamefont {Sesana}}, \bibinfo {author} {\bibfnamefont {F.}~\bibnamefont {Stasyszyn}}, \bibinfo {author} {\bibfnamefont {P.}~\bibnamefont {Amaro-Seoane}},\ and\ \bibinfo {author} {\bibfnamefont {T.~L.}\ \bibnamefont {Tanaka}},\ }\bibfield  {title} {\bibinfo {title} {{Infalling clouds on to supermassive black hole binaries \textendash{} I. Formation of discs, accretion and gas dynamics}},\ }\href {https://doi.org/10.1093/mnras/stv2470} {\bibfield  {journal} {\bibinfo  {journal} {Mon. Not. Roy. Astron. Soc.}\ }\textbf {\bibinfo {volume} {455}},\ \bibinfo {pages} {1989} (\bibinfo {year} {2016})},\ \Eprint {https://arxiv.org/abs/1507.05596} {arXiv:1507.05596 [astro-ph.HE]} \BibitemShut {NoStop}%
\bibitem [{\citenamefont {Goicovic}\ \emph {et~al.}(2017)\citenamefont {Goicovic}, \citenamefont {Sesana}, \citenamefont {Cuadra},\ and\ \citenamefont {Stasyszyn}}]{Goicovic:2016dul}%
  \BibitemOpen
  \bibfield  {author} {\bibinfo {author} {\bibfnamefont {F.~G.}\ \bibnamefont {Goicovic}}, \bibinfo {author} {\bibfnamefont {A.}~\bibnamefont {Sesana}}, \bibinfo {author} {\bibfnamefont {J.}~\bibnamefont {Cuadra}},\ and\ \bibinfo {author} {\bibfnamefont {F.}~\bibnamefont {Stasyszyn}},\ }\bibfield  {title} {\bibinfo {title} {{Infalling clouds on to supermassive black hole binaries \textendash{} II. Binary evolution and the final parsec problem}},\ }\href {https://doi.org/10.1093/mnras/stx1996} {\bibfield  {journal} {\bibinfo  {journal} {Mon. Not. Roy. Astron. Soc.}\ }\textbf {\bibinfo {volume} {472}},\ \bibinfo {pages} {514} (\bibinfo {year} {2017})},\ \Eprint {https://arxiv.org/abs/1602.01966} {arXiv:1602.01966 [astro-ph.HE]} \BibitemShut {NoStop}%
\bibitem [{\citenamefont {Goicovic}\ \emph {et~al.}(2018)\citenamefont {Goicovic}, \citenamefont {Maureira-Fredes}, \citenamefont {Sesana}, \citenamefont {Amaro-Seoane},\ and\ \citenamefont {Cuadra}}]{Goicovic:2018xxi}%
  \BibitemOpen
  \bibfield  {author} {\bibinfo {author} {\bibfnamefont {F.~G.}\ \bibnamefont {Goicovic}}, \bibinfo {author} {\bibfnamefont {C.}~\bibnamefont {Maureira-Fredes}}, \bibinfo {author} {\bibfnamefont {A.}~\bibnamefont {Sesana}}, \bibinfo {author} {\bibfnamefont {P.}~\bibnamefont {Amaro-Seoane}},\ and\ \bibinfo {author} {\bibfnamefont {J.}~\bibnamefont {Cuadra}},\ }\bibfield  {title} {\bibinfo {title} {{Accretion of clumpy cold gas onto massive black hole binaries: a possible fast route to binary coalescence}},\ }\href {https://doi.org/10.1093/mnras/sty1709} {\bibfield  {journal} {\bibinfo  {journal} {Mon. Not. Roy. Astron. Soc.}\ }\textbf {\bibinfo {volume} {479}},\ \bibinfo {pages} {3438} (\bibinfo {year} {2018})},\ \Eprint {https://arxiv.org/abs/1801.04937} {arXiv:1801.04937 [astro-ph.HE]} \BibitemShut {NoStop}%
\bibitem [{\citenamefont {Barnes}\ and\ \citenamefont {Hernquist}(1991)}]{Barnes:1991zz}%
  \BibitemOpen
  \bibfield  {author} {\bibinfo {author} {\bibfnamefont {J.~E.}\ \bibnamefont {Barnes}}\ and\ \bibinfo {author} {\bibfnamefont {L.~E.}\ \bibnamefont {Hernquist}},\ }\bibfield  {title} {\bibinfo {title} {{Fueling starburst galaxies with gas-rich mergers}},\ }\href {https://doi.org/10.1086/185978} {\bibfield  {journal} {\bibinfo  {journal} {Astrophys. J. Lett.}\ }\textbf {\bibinfo {volume} {370}},\ \bibinfo {pages} {L65} (\bibinfo {year} {1991})}\BibitemShut {NoStop}%
\bibitem [{\citenamefont {Mihos}\ and\ \citenamefont {Hernquist}(1996)}]{Mihos:1995ri}%
  \BibitemOpen
  \bibfield  {author} {\bibinfo {author} {\bibfnamefont {J.~C.}\ \bibnamefont {Mihos}}\ and\ \bibinfo {author} {\bibfnamefont {L.}~\bibnamefont {Hernquist}},\ }\bibfield  {title} {\bibinfo {title} {{Gasdynamics and starbursts in major mergers}},\ }\href {https://doi.org/10.1086/177353} {\bibfield  {journal} {\bibinfo  {journal} {Astrophys. J.}\ }\textbf {\bibinfo {volume} {464}},\ \bibinfo {pages} {641} (\bibinfo {year} {1996})},\ \Eprint {https://arxiv.org/abs/astro-ph/9512099} {arXiv:astro-ph/9512099} \BibitemShut {NoStop}%
\bibitem [{\citenamefont {Barnes}(2002)}]{Barnes:2002sh}%
  \BibitemOpen
  \bibfield  {author} {\bibinfo {author} {\bibfnamefont {J.~E.}\ \bibnamefont {Barnes}},\ }\bibfield  {title} {\bibinfo {title} {{Formation of gas disks in merging galaxies}},\ }\href {https://doi.org/10.1046/j.1365-8711.2002.05335.x} {\bibfield  {journal} {\bibinfo  {journal} {Mon. Not. Roy. Astron. Soc.}\ }\textbf {\bibinfo {volume} {333}},\ \bibinfo {pages} {481} (\bibinfo {year} {2002})},\ \Eprint {https://arxiv.org/abs/astro-ph/0201250} {arXiv:astro-ph/0201250} \BibitemShut {NoStop}%
\bibitem [{\citenamefont {Di~Matteo}\ \emph {et~al.}(2005)\citenamefont {Di~Matteo}, \citenamefont {Springel},\ and\ \citenamefont {Hernquist}}]{DiMatteo:2005ttp}%
  \BibitemOpen
  \bibfield  {author} {\bibinfo {author} {\bibfnamefont {T.}~\bibnamefont {Di~Matteo}}, \bibinfo {author} {\bibfnamefont {V.}~\bibnamefont {Springel}},\ and\ \bibinfo {author} {\bibfnamefont {L.}~\bibnamefont {Hernquist}},\ }\bibfield  {title} {\bibinfo {title} {{Energy input from quasars regulates the growth and activity of black holes and their host galaxies}},\ }\href {https://doi.org/10.1038/nature03335} {\bibfield  {journal} {\bibinfo  {journal} {Nature}\ }\textbf {\bibinfo {volume} {433}},\ \bibinfo {pages} {604} (\bibinfo {year} {2005})},\ \Eprint {https://arxiv.org/abs/astro-ph/0502199} {arXiv:astro-ph/0502199} \BibitemShut {NoStop}%
\bibitem [{\citenamefont {Hopkins}\ \emph {et~al.}(2006)\citenamefont {Hopkins}, \citenamefont {Hernquist}, \citenamefont {Cox}, \citenamefont {Di~Matteo}, \citenamefont {Robertson},\ and\ \citenamefont {Springel}}]{Hopkins:2005fb}%
  \BibitemOpen
  \bibfield  {author} {\bibinfo {author} {\bibfnamefont {P.~F.}\ \bibnamefont {Hopkins}}, \bibinfo {author} {\bibfnamefont {L.}~\bibnamefont {Hernquist}}, \bibinfo {author} {\bibfnamefont {T.~J.}\ \bibnamefont {Cox}}, \bibinfo {author} {\bibfnamefont {T.}~\bibnamefont {Di~Matteo}}, \bibinfo {author} {\bibfnamefont {B.}~\bibnamefont {Robertson}},\ and\ \bibinfo {author} {\bibfnamefont {V.}~\bibnamefont {Springel}},\ }\bibfield  {title} {\bibinfo {title} {{A Unified, merger-driven model for the origin of starbursts, quasars, the cosmic x-ray background, supermassive black holes and galaxy spheroids}},\ }\href {https://doi.org/10.1086/499298} {\bibfield  {journal} {\bibinfo  {journal} {Astrophys. J. Suppl.}\ }\textbf {\bibinfo {volume} {163}},\ \bibinfo {pages} {1} (\bibinfo {year} {2006})},\ \Eprint {https://arxiv.org/abs/astro-ph/0506398} {arXiv:astro-ph/0506398} \BibitemShut {NoStop}%
\bibitem [{\citenamefont {Hopkins}\ \emph {et~al.}(2013)\citenamefont {Hopkins}, \citenamefont {Cox}, \citenamefont {Hernquist}, \citenamefont {Narayanan}, \citenamefont {Hayward},\ and\ \citenamefont {Murray}}]{Hopkins:2012fd}%
  \BibitemOpen
  \bibfield  {author} {\bibinfo {author} {\bibfnamefont {P.~F.}\ \bibnamefont {Hopkins}}, \bibinfo {author} {\bibfnamefont {T.~J.}\ \bibnamefont {Cox}}, \bibinfo {author} {\bibfnamefont {L.}~\bibnamefont {Hernquist}}, \bibinfo {author} {\bibfnamefont {D.}~\bibnamefont {Narayanan}}, \bibinfo {author} {\bibfnamefont {C.~C.}\ \bibnamefont {Hayward}},\ and\ \bibinfo {author} {\bibfnamefont {N.}~\bibnamefont {Murray}},\ }\bibfield  {title} {\bibinfo {title} {{Star Formation in Galaxy Mergers with Realistic Models of Stellar Feedback \& the Interstellar Medium}},\ }\href {https://doi.org/10.1093/mnras/stt017} {\bibfield  {journal} {\bibinfo  {journal} {Mon. Not. Roy. Astron. Soc.}\ }\textbf {\bibinfo {volume} {430}},\ \bibinfo {pages} {1901} (\bibinfo {year} {2013})},\ \Eprint {https://arxiv.org/abs/1206.0011} {arXiv:1206.0011 [astro-ph.CO]} \BibitemShut {NoStop}%
\bibitem [{\citenamefont {Cox}\ \emph {et~al.}(2008)\citenamefont {Cox}, \citenamefont {Jonsson}, \citenamefont {Somerville}, \citenamefont {Primack},\ and\ \citenamefont {Dekel}}]{Cox:2007mn}%
  \BibitemOpen
  \bibfield  {author} {\bibinfo {author} {\bibfnamefont {T.~J.}\ \bibnamefont {Cox}}, \bibinfo {author} {\bibfnamefont {P.~B.}\ \bibnamefont {Jonsson}}, \bibinfo {author} {\bibfnamefont {R.~S.}\ \bibnamefont {Somerville}}, \bibinfo {author} {\bibfnamefont {J.~R.}\ \bibnamefont {Primack}},\ and\ \bibinfo {author} {\bibfnamefont {A.}~\bibnamefont {Dekel}},\ }\bibfield  {title} {\bibinfo {title} {{The effect of galaxy mass ratio on merger-driven starbursts}},\ }\href {https://doi.org/10.1111/j.1365-2966.2007.12730.x} {\bibfield  {journal} {\bibinfo  {journal} {Mon. Not. Roy. Astron. Soc.}\ }\textbf {\bibinfo {volume} {384}},\ \bibinfo {pages} {386} (\bibinfo {year} {2008})},\ \Eprint {https://arxiv.org/abs/0709.3511} {arXiv:0709.3511 [astro-ph]} \BibitemShut {NoStop}%
\bibitem [{\citenamefont {Johansson}\ \emph {et~al.}(2009)\citenamefont {Johansson}, \citenamefont {Naab},\ and\ \citenamefont {Burkert}}]{Johansson:2008ib}%
  \BibitemOpen
  \bibfield  {author} {\bibinfo {author} {\bibfnamefont {P.~H.}\ \bibnamefont {Johansson}}, \bibinfo {author} {\bibfnamefont {T.}~\bibnamefont {Naab}},\ and\ \bibinfo {author} {\bibfnamefont {A.}~\bibnamefont {Burkert}},\ }\bibfield  {title} {\bibinfo {title} {{Equal- and unequal-mass mergers of disk and elliptical galaxies with black holes: The M\_BH-sigma and M\_BH-M\_* relations}},\ }\href {https://doi.org/10.1088/0004-637X/690/1/802} {\bibfield  {journal} {\bibinfo  {journal} {Astrophys. J.}\ }\textbf {\bibinfo {volume} {690}},\ \bibinfo {pages} {802} (\bibinfo {year} {2009})},\ \Eprint {https://arxiv.org/abs/0802.0210} {arXiv:0802.0210 [astro-ph]} \BibitemShut {NoStop}%
\bibitem [{\citenamefont {Capelo}\ \emph {et~al.}(2015)\citenamefont {Capelo}, \citenamefont {Volonteri}, \citenamefont {Dotti}, \citenamefont {Bellovary}, \citenamefont {Mayer},\ and\ \citenamefont {Governato}}]{Capelo:2014gqa}%
  \BibitemOpen
  \bibfield  {author} {\bibinfo {author} {\bibfnamefont {P.~R.}\ \bibnamefont {Capelo}}, \bibinfo {author} {\bibfnamefont {M.}~\bibnamefont {Volonteri}}, \bibinfo {author} {\bibfnamefont {M.}~\bibnamefont {Dotti}}, \bibinfo {author} {\bibfnamefont {J.~M.}\ \bibnamefont {Bellovary}}, \bibinfo {author} {\bibfnamefont {L.}~\bibnamefont {Mayer}},\ and\ \bibinfo {author} {\bibfnamefont {F.}~\bibnamefont {Governato}},\ }\bibfield  {title} {\bibinfo {title} {{Growth and activity of black holes in galaxy mergers with varying mass ratios}},\ }\href {https://doi.org/10.1093/mnras/stu2500} {\bibfield  {journal} {\bibinfo  {journal} {Mon. Not. Roy. Astron. Soc.}\ }\textbf {\bibinfo {volume} {447}},\ \bibinfo {pages} {2123} (\bibinfo {year} {2015})},\ \Eprint {https://arxiv.org/abs/1409.0004} {arXiv:1409.0004 [astro-ph.GA]} \BibitemShut {NoStop}%
\bibitem [{\citenamefont {Capelo}\ and\ \citenamefont {Dotti}(2017)}]{Capelo:2016vif}%
  \BibitemOpen
  \bibfield  {author} {\bibinfo {author} {\bibfnamefont {P.~R.}\ \bibnamefont {Capelo}}\ and\ \bibinfo {author} {\bibfnamefont {M.}~\bibnamefont {Dotti}},\ }\bibfield  {title} {\bibinfo {title} {{Shocks and angular momentum flips: a different path to feeding the nuclear regions of merging galaxies}},\ }\href {https://doi.org/10.1093/mnras/stw2872} {\bibfield  {journal} {\bibinfo  {journal} {Mon. Not. Roy. Astron. Soc.}\ }\textbf {\bibinfo {volume} {465}},\ \bibinfo {pages} {2643} (\bibinfo {year} {2017})},\ \Eprint {https://arxiv.org/abs/1610.08507} {arXiv:1610.08507 [astro-ph.GA]} \BibitemShut {NoStop}%
\bibitem [{\citenamefont {{Begelman}}\ \emph {et~al.}(1980)\citenamefont {{Begelman}}, \citenamefont {{Blandford}},\ and\ \citenamefont {{Rees}}}]{1980Natur.287..307B}%
  \BibitemOpen
  \bibfield  {author} {\bibinfo {author} {\bibfnamefont {M.~C.}\ \bibnamefont {{Begelman}}}, \bibinfo {author} {\bibfnamefont {R.~D.}\ \bibnamefont {{Blandford}}},\ and\ \bibinfo {author} {\bibfnamefont {M.~J.}\ \bibnamefont {{Rees}}},\ }\bibfield  {title} {\bibinfo {title} {{Massive black hole binaries in active galactic nuclei}},\ }\href {https://doi.org/10.1038/287307a0} {\bibfield  {journal} {\bibinfo  {journal} {\nat}\ }\textbf {\bibinfo {volume} {287}},\ \bibinfo {pages} {307} (\bibinfo {year} {1980})}\BibitemShut {NoStop}%
\bibitem [{\citenamefont {{Escala}}\ \emph {et~al.}(2005)\citenamefont {{Escala}}, \citenamefont {{Larson}}, \citenamefont {{Coppi}},\ and\ \citenamefont {{Mardones}}}]{2005ApJ...630..152E}%
  \BibitemOpen
  \bibfield  {author} {\bibinfo {author} {\bibfnamefont {A.}~\bibnamefont {{Escala}}}, \bibinfo {author} {\bibfnamefont {R.~B.}\ \bibnamefont {{Larson}}}, \bibinfo {author} {\bibfnamefont {P.~S.}\ \bibnamefont {{Coppi}}},\ and\ \bibinfo {author} {\bibfnamefont {D.}~\bibnamefont {{Mardones}}},\ }\bibfield  {title} {\bibinfo {title} {{The Role of Gas in the Merging of Massive Black Holes in Galactic Nuclei. II. Black Hole Merging in a Nuclear Gas Disk}},\ }\href {https://doi.org/10.1086/431747} {\bibfield  {journal} {\bibinfo  {journal} {Astrophys. J.}\ }\textbf {\bibinfo {volume} {630}},\ \bibinfo {pages} {152} (\bibinfo {year} {2005})},\ \Eprint {https://arxiv.org/abs/astro-ph/0406304} {arXiv:astro-ph/0406304 [astro-ph]} \BibitemShut {NoStop}%
\bibitem [{\citenamefont {Cuadra}\ \emph {et~al.}(2009)\citenamefont {Cuadra}, \citenamefont {Armitage}, \citenamefont {Alexander},\ and\ \citenamefont {Begelman}}]{Cuadra:2008xn}%
  \BibitemOpen
  \bibfield  {author} {\bibinfo {author} {\bibfnamefont {J.}~\bibnamefont {Cuadra}}, \bibinfo {author} {\bibfnamefont {P.~J.}\ \bibnamefont {Armitage}}, \bibinfo {author} {\bibfnamefont {R.~D.}\ \bibnamefont {Alexander}},\ and\ \bibinfo {author} {\bibfnamefont {M.~C.}\ \bibnamefont {Begelman}},\ }\bibfield  {title} {\bibinfo {title} {{Massive black hole binary mergers within sub-pc scale gas discs}},\ }\href {https://doi.org/10.1111/j.1365-2966.2008.14147.x} {\bibfield  {journal} {\bibinfo  {journal} {Mon. Not. Roy. Astron. Soc.}\ }\textbf {\bibinfo {volume} {393}},\ \bibinfo {pages} {1423} (\bibinfo {year} {2009})},\ \Eprint {https://arxiv.org/abs/0809.0311} {arXiv:0809.0311 [astro-ph]} \BibitemShut {NoStop}%
\bibitem [{\citenamefont {Nixon}\ \emph {et~al.}(2011{\natexlab{a}})\citenamefont {Nixon}, \citenamefont {Cossins}, \citenamefont {King},\ and\ \citenamefont {Pringle}}]{Nixon:2010by}%
  \BibitemOpen
  \bibfield  {author} {\bibinfo {author} {\bibfnamefont {C.~J.}\ \bibnamefont {Nixon}}, \bibinfo {author} {\bibfnamefont {P.~J.}\ \bibnamefont {Cossins}}, \bibinfo {author} {\bibfnamefont {A.~R.}\ \bibnamefont {King}},\ and\ \bibinfo {author} {\bibfnamefont {J.~E.}\ \bibnamefont {Pringle}},\ }\bibfield  {title} {\bibinfo {title} {{Retrograde Accretion and Merging Supermassive Black Holes}},\ }\href {https://doi.org/10.1111/j.1365-2966.2010.17952.x} {\bibfield  {journal} {\bibinfo  {journal} {Mon. Not. Roy. Astron. Soc.}\ }\textbf {\bibinfo {volume} {412}},\ \bibinfo {pages} {1591} (\bibinfo {year} {2011}{\natexlab{a}})},\ \Eprint {https://arxiv.org/abs/1011.1914} {arXiv:1011.1914 [astro-ph.HE]} \BibitemShut {NoStop}%
\bibitem [{\citenamefont {Roedig}\ \emph {et~al.}(2012)\citenamefont {Roedig}, \citenamefont {Sesana}, \citenamefont {Dotti}, \citenamefont {Cuadra}, \citenamefont {Amaro-Seoane},\ and\ \citenamefont {Haardt}}]{Roedig:2012nc}%
  \BibitemOpen
  \bibfield  {author} {\bibinfo {author} {\bibfnamefont {C.}~\bibnamefont {Roedig}}, \bibinfo {author} {\bibfnamefont {A.}~\bibnamefont {Sesana}}, \bibinfo {author} {\bibfnamefont {M.}~\bibnamefont {Dotti}}, \bibinfo {author} {\bibfnamefont {J.}~\bibnamefont {Cuadra}}, \bibinfo {author} {\bibfnamefont {P.}~\bibnamefont {Amaro-Seoane}},\ and\ \bibinfo {author} {\bibfnamefont {F.}~\bibnamefont {Haardt}},\ }\bibfield  {title} {\bibinfo {title} {{Evolution of binary black holes in self gravitating discs: dissecting the torques}},\ }\href {https://doi.org/10.1051/0004-6361/201219986} {\bibfield  {journal} {\bibinfo  {journal} {Astron. Astrophys.}\ }\textbf {\bibinfo {volume} {545}},\ \bibinfo {pages} {A127} (\bibinfo {year} {2012})},\ \Eprint {https://arxiv.org/abs/1202.6063} {arXiv:1202.6063 [astro-ph.CO]} \BibitemShut {NoStop}%
\bibitem [{\citenamefont {D'Orazio}\ \emph {et~al.}(2013)\citenamefont {D'Orazio}, \citenamefont {Haiman},\ and\ \citenamefont {MacFadyen}}]{DOrazio:2012vqt}%
  \BibitemOpen
  \bibfield  {author} {\bibinfo {author} {\bibfnamefont {D.~J.}\ \bibnamefont {D'Orazio}}, \bibinfo {author} {\bibfnamefont {Z.}~\bibnamefont {Haiman}},\ and\ \bibinfo {author} {\bibfnamefont {A.}~\bibnamefont {MacFadyen}},\ }\bibfield  {title} {\bibinfo {title} {{Accretion into the Central Cavity of a Circumbinary Disk}},\ }\href {https://doi.org/10.1093/mnras/stt1787} {\bibfield  {journal} {\bibinfo  {journal} {Mon. Not. Roy. Astron. Soc.}\ }\textbf {\bibinfo {volume} {436}},\ \bibinfo {pages} {2997} (\bibinfo {year} {2013})},\ \Eprint {https://arxiv.org/abs/1210.0536} {arXiv:1210.0536 [astro-ph.GA]} \BibitemShut {NoStop}%
\bibitem [{\citenamefont {Nixon}\ \emph {et~al.}(2013)\citenamefont {Nixon}, \citenamefont {King},\ and\ \citenamefont {Price}}]{Nixon:2013qfa}%
  \BibitemOpen
  \bibfield  {author} {\bibinfo {author} {\bibfnamefont {C.}~\bibnamefont {Nixon}}, \bibinfo {author} {\bibfnamefont {A.}~\bibnamefont {King}},\ and\ \bibinfo {author} {\bibfnamefont {D.}~\bibnamefont {Price}},\ }\bibfield  {title} {\bibinfo {title} {{Tearing up the disc: misaligned accretion on to a binary}},\ }\href {https://doi.org/10.1093/mnras/stt1136} {\bibfield  {journal} {\bibinfo  {journal} {Mon. Not. Roy. Astron. Soc.}\ }\textbf {\bibinfo {volume} {434}},\ \bibinfo {pages} {1946} (\bibinfo {year} {2013})},\ \Eprint {https://arxiv.org/abs/1307.0010} {arXiv:1307.0010 [astro-ph.HE]} \BibitemShut {NoStop}%
\bibitem [{\citenamefont {Farris}\ \emph {et~al.}(2014)\citenamefont {Farris}, \citenamefont {Duffell}, \citenamefont {MacFadyen},\ and\ \citenamefont {Haiman}}]{Farris:2013uqa}%
  \BibitemOpen
  \bibfield  {author} {\bibinfo {author} {\bibfnamefont {B.~D.}\ \bibnamefont {Farris}}, \bibinfo {author} {\bibfnamefont {P.}~\bibnamefont {Duffell}}, \bibinfo {author} {\bibfnamefont {A.~I.}\ \bibnamefont {MacFadyen}},\ and\ \bibinfo {author} {\bibfnamefont {Z.}~\bibnamefont {Haiman}},\ }\bibfield  {title} {\bibinfo {title} {{Binary Black Hole Accretion From a Circumbinary Disk: Gas Dynamics Inside the Central Cavity}},\ }\href {https://doi.org/10.1088/0004-637X/783/2/134} {\bibfield  {journal} {\bibinfo  {journal} {Astrophys. J.}\ }\textbf {\bibinfo {volume} {783}},\ \bibinfo {pages} {134} (\bibinfo {year} {2014})},\ \Eprint {https://arxiv.org/abs/1310.0492} {arXiv:1310.0492 [astro-ph.HE]} \BibitemShut {NoStop}%
\bibitem [{\citenamefont {{Miranda}}\ \emph {et~al.}(2017)\citenamefont {{Miranda}}, \citenamefont {{Mu{\~n}oz}},\ and\ \citenamefont {{Lai}}}]{Miranda2017}%
  \BibitemOpen
  \bibfield  {author} {\bibinfo {author} {\bibfnamefont {R.}~\bibnamefont {{Miranda}}}, \bibinfo {author} {\bibfnamefont {D.~J.}\ \bibnamefont {{Mu{\~n}oz}}},\ and\ \bibinfo {author} {\bibfnamefont {D.}~\bibnamefont {{Lai}}},\ }\bibfield  {title} {\bibinfo {title} {{Viscous hydrodynamics simulations of circumbinary accretion discs: variability, quasi-steady state and angular momentum transfer}},\ }\href {https://doi.org/10.1093/mnras/stw3189} {\bibfield  {journal} {\bibinfo  {journal} {Mon. Not. Roy. Astron. Soc.}\ }\textbf {\bibinfo {volume} {466}},\ \bibinfo {pages} {1170} (\bibinfo {year} {2017})},\ \Eprint {https://arxiv.org/abs/1610.07263} {arXiv:1610.07263 [astro-ph.SR]} \BibitemShut {NoStop}%
\bibitem [{\citenamefont {Moody}\ \emph {et~al.}(2019)\citenamefont {Moody}, \citenamefont {Shi},\ and\ \citenamefont {Stone}}]{Moody:2019nes}%
  \BibitemOpen
  \bibfield  {author} {\bibinfo {author} {\bibfnamefont {M.~S.~L.}\ \bibnamefont {Moody}}, \bibinfo {author} {\bibfnamefont {J.-M.}\ \bibnamefont {Shi}},\ and\ \bibinfo {author} {\bibfnamefont {J.~M.}\ \bibnamefont {Stone}},\ }\bibfield  {title} {\bibinfo {title} {{Hydrodynamic Torques in Circumbinary Accretion Disks}},\ }\href {https://doi.org/10.3847/1538-4357/ab09ee} {\bibfield  {journal} {\bibinfo  {journal} {Astrophys. J.}\ }\textbf {\bibinfo {volume} {875}},\ \bibinfo {pages} {66} (\bibinfo {year} {2019})},\ \Eprint {https://arxiv.org/abs/1903.00008} {arXiv:1903.00008 [astro-ph.HE]} \BibitemShut {NoStop}%
\bibitem [{\citenamefont {Mu\~noz}\ \emph {et~al.}(2019)\citenamefont {Mu\~noz}, \citenamefont {Miranda},\ and\ \citenamefont {Lai}}]{Munoz:2018tnj}%
  \BibitemOpen
  \bibfield  {author} {\bibinfo {author} {\bibfnamefont {D.~J.}\ \bibnamefont {Mu\~noz}}, \bibinfo {author} {\bibfnamefont {R.}~\bibnamefont {Miranda}},\ and\ \bibinfo {author} {\bibfnamefont {D.}~\bibnamefont {Lai}},\ }\bibfield  {title} {\bibinfo {title} {{Hydrodynamics of circumbinary accretion: Angular momentum transfer and binary orbital evolution}},\ }\href {https://doi.org/10.3847/1538-4357/aaf867} {\bibfield  {journal} {\bibinfo  {journal} {Astrophys. J.}\ }\textbf {\bibinfo {volume} {871}},\ \bibinfo {pages} {84} (\bibinfo {year} {2019})},\ \Eprint {https://arxiv.org/abs/1810.04676} {arXiv:1810.04676 [astro-ph.HE]} \BibitemShut {NoStop}%
\bibitem [{\citenamefont {Artymowicz}\ and\ \citenamefont {Lubow}(1994)}]{Artymowicz:1994bw}%
  \BibitemOpen
  \bibfield  {author} {\bibinfo {author} {\bibfnamefont {P.}~\bibnamefont {Artymowicz}}\ and\ \bibinfo {author} {\bibfnamefont {S.~H.}\ \bibnamefont {Lubow}},\ }\bibfield  {title} {\bibinfo {title} {{Dynamics of binary-disk interaction. 1: Resonances and disk gap sizes}},\ }\href {https://doi.org/10.1086/173679} {\bibfield  {journal} {\bibinfo  {journal} {Astrophys. J.}\ }\textbf {\bibinfo {volume} {421}},\ \bibinfo {pages} {651} (\bibinfo {year} {1994})}\BibitemShut {NoStop}%
\bibitem [{\citenamefont {{Duffell}}\ \emph {et~al.}(2020)\citenamefont {{Duffell}}, \citenamefont {{D'Orazio}}, \citenamefont {{Derdzinski}}, \citenamefont {{Haiman}}, \citenamefont {{MacFadyen}}, \citenamefont {{Rosen}},\ and\ \citenamefont {{Zrake}}}]{Duffell2020}%
  \BibitemOpen
  \bibfield  {author} {\bibinfo {author} {\bibfnamefont {P.~C.}\ \bibnamefont {{Duffell}}}, \bibinfo {author} {\bibfnamefont {D.}~\bibnamefont {{D'Orazio}}}, \bibinfo {author} {\bibfnamefont {A.}~\bibnamefont {{Derdzinski}}}, \bibinfo {author} {\bibfnamefont {Z.}~\bibnamefont {{Haiman}}}, \bibinfo {author} {\bibfnamefont {A.}~\bibnamefont {{MacFadyen}}}, \bibinfo {author} {\bibfnamefont {A.~L.}\ \bibnamefont {{Rosen}}},\ and\ \bibinfo {author} {\bibfnamefont {J.}~\bibnamefont {{Zrake}}},\ }\bibfield  {title} {\bibinfo {title} {{Circumbinary Disks: Accretion and Torque as a Function of Mass Ratio and Disk Viscosity}},\ }\href {https://doi.org/10.3847/1538-4357/abab95} {\bibfield  {journal} {\bibinfo  {journal} {Astrophys. J.}\ }\textbf {\bibinfo {volume} {901}},\ \bibinfo {eid} {25} (\bibinfo {year} {2020})},\ \Eprint {https://arxiv.org/abs/1911.05506} {arXiv:1911.05506 [astro-ph.SR]} \BibitemShut {NoStop}%
\bibitem [{\citenamefont {Derdzinski}\ \emph {et~al.}(2021)\citenamefont {Derdzinski}, \citenamefont {D'Orazio}, \citenamefont {Duffell}, \citenamefont {Haiman},\ and\ \citenamefont {MacFadyen}}]{Derdzinski:2020wlw}%
  \BibitemOpen
  \bibfield  {author} {\bibinfo {author} {\bibfnamefont {A.}~\bibnamefont {Derdzinski}}, \bibinfo {author} {\bibfnamefont {D.}~\bibnamefont {D'Orazio}}, \bibinfo {author} {\bibfnamefont {P.}~\bibnamefont {Duffell}}, \bibinfo {author} {\bibfnamefont {Z.}~\bibnamefont {Haiman}},\ and\ \bibinfo {author} {\bibfnamefont {A.}~\bibnamefont {MacFadyen}},\ }\bibfield  {title} {\bibinfo {title} {{Evolution of gas disc\textendash{}embedded intermediate mass ratio inspirals in the $LISA$ band}},\ }\href {https://doi.org/10.1093/mnras/staa3976} {\bibfield  {journal} {\bibinfo  {journal} {Mon. Not. Roy. Astron. Soc.}\ }\textbf {\bibinfo {volume} {501}},\ \bibinfo {pages} {3540} (\bibinfo {year} {2021})},\ \Eprint {https://arxiv.org/abs/2005.11333} {arXiv:2005.11333 [astro-ph.HE]} \BibitemShut {NoStop}%
\bibitem [{\citenamefont {Siwek}\ \emph {et~al.}(2022)\citenamefont {Siwek}, \citenamefont {Weinberger}, \citenamefont {Mu\~noz},\ and\ \citenamefont {Hernquist}}]{Siwek:2022xhf}%
  \BibitemOpen
  \bibfield  {author} {\bibinfo {author} {\bibfnamefont {M.}~\bibnamefont {Siwek}}, \bibinfo {author} {\bibfnamefont {R.}~\bibnamefont {Weinberger}}, \bibinfo {author} {\bibfnamefont {D.~J.}\ \bibnamefont {Mu\~noz}},\ and\ \bibinfo {author} {\bibfnamefont {L.}~\bibnamefont {Hernquist}},\ }\bibfield  {title} {\bibinfo {title} {{Preferential accretion and circumbinary disc precession in eccentric binary systems}},\ }\href {https://doi.org/10.1093/mnras/stac3263} {\bibfield  {journal} {\bibinfo  {journal} {Mon. Not. Roy. Astron. Soc.}\ }\textbf {\bibinfo {volume} {518}},\ \bibinfo {pages} {5059} (\bibinfo {year} {2022})},\ \Eprint {https://arxiv.org/abs/2203.02514} {arXiv:2203.02514 [astro-ph.HE]} \BibitemShut {NoStop}%
\bibitem [{\citenamefont {Siwek}\ \emph {et~al.}(2023)\citenamefont {Siwek}, \citenamefont {Weinberger},\ and\ \citenamefont {Hernquist}}]{Siwek:2023rlk}%
  \BibitemOpen
  \bibfield  {author} {\bibinfo {author} {\bibfnamefont {M.}~\bibnamefont {Siwek}}, \bibinfo {author} {\bibfnamefont {R.}~\bibnamefont {Weinberger}},\ and\ \bibinfo {author} {\bibfnamefont {L.}~\bibnamefont {Hernquist}},\ }\bibfield  {title} {\bibinfo {title} {{Orbital evolution of binaries in circumbinary discs}},\ }\href {https://doi.org/10.1093/mnras/stad1131} {\bibfield  {journal} {\bibinfo  {journal} {Mon. Not. Roy. Astron. Soc.}\ }\textbf {\bibinfo {volume} {522}},\ \bibinfo {pages} {2707} (\bibinfo {year} {2023})},\ \Eprint {https://arxiv.org/abs/2302.01785} {arXiv:2302.01785 [astro-ph.HE]} \BibitemShut {NoStop}%
\bibitem [{\citenamefont {{D'Orazio}}\ and\ \citenamefont {{Duffell}}(2021)}]{DOrazio2021}%
  \BibitemOpen
  \bibfield  {author} {\bibinfo {author} {\bibfnamefont {D.~J.}\ \bibnamefont {{D'Orazio}}}\ and\ \bibinfo {author} {\bibfnamefont {P.~C.}\ \bibnamefont {{Duffell}}},\ }\bibfield  {title} {\bibinfo {title} {{Orbital Evolution of Equal-mass Eccentric Binaries due to a Gas Disk: Eccentric Inspirals and Circular Outspirals}},\ }\href {https://doi.org/10.3847/2041-8213/ac0621} {\bibfield  {journal} {\bibinfo  {journal} {Astrophys. J. Lett.}\ }\textbf {\bibinfo {volume} {914}},\ \bibinfo {eid} {L21} (\bibinfo {year} {2021})},\ \Eprint {https://arxiv.org/abs/2103.09251} {arXiv:2103.09251 [astro-ph.HE]} \BibitemShut {NoStop}%
\bibitem [{\citenamefont {{Siwek}}\ \emph {et~al.}(2023{\natexlab{a}})\citenamefont {{Siwek}}, \citenamefont {{Weinberger}},\ and\ \citenamefont {{Hernquist}}}]{Siwek2023b}%
  \BibitemOpen
  \bibfield  {author} {\bibinfo {author} {\bibfnamefont {M.}~\bibnamefont {{Siwek}}}, \bibinfo {author} {\bibfnamefont {R.}~\bibnamefont {{Weinberger}}},\ and\ \bibinfo {author} {\bibfnamefont {L.}~\bibnamefont {{Hernquist}}},\ }\bibfield  {title} {\bibinfo {title} {{Orbital evolution of binaries in circumbinary discs}},\ }\href {https://doi.org/10.1093/mnras/stad1131} {\bibfield  {journal} {\bibinfo  {journal} {Mon. Not. Roy. Astron. Soc.}\ }\textbf {\bibinfo {volume} {522}},\ \bibinfo {pages} {2707} (\bibinfo {year} {2023}{\natexlab{a}})},\ \Eprint {https://arxiv.org/abs/2302.01785} {arXiv:2302.01785 [astro-ph.HE]} \BibitemShut {NoStop}%
\bibitem [{\citenamefont {{Siwek}}\ \emph {et~al.}(2023{\natexlab{b}})\citenamefont {{Siwek}}, \citenamefont {{Weinberger}}, \citenamefont {{Mu{\~n}oz}},\ and\ \citenamefont {{Hernquist}}}]{Siwek2023a}%
  \BibitemOpen
  \bibfield  {author} {\bibinfo {author} {\bibfnamefont {M.}~\bibnamefont {{Siwek}}}, \bibinfo {author} {\bibfnamefont {R.}~\bibnamefont {{Weinberger}}}, \bibinfo {author} {\bibfnamefont {D.~J.}\ \bibnamefont {{Mu{\~n}oz}}},\ and\ \bibinfo {author} {\bibfnamefont {L.}~\bibnamefont {{Hernquist}}},\ }\bibfield  {title} {\bibinfo {title} {{Preferential accretion and circumbinary disc precession in eccentric binary systems}},\ }\href {https://doi.org/10.1093/mnras/stac3263} {\bibfield  {journal} {\bibinfo  {journal} {Mon. Not. Roy. Astron. Soc.}\ }\textbf {\bibinfo {volume} {518}},\ \bibinfo {pages} {5059} (\bibinfo {year} {2023}{\natexlab{b}})},\ \Eprint {https://arxiv.org/abs/2203.02514} {arXiv:2203.02514 [astro-ph.HE]} \BibitemShut {NoStop}%
\bibitem [{\citenamefont {{Tiede}}\ \emph {et~al.}(2020)\citenamefont {{Tiede}}, \citenamefont {{Zrake}}, \citenamefont {{MacFadyen}},\ and\ \citenamefont {{Haiman}}}]{Tiede2020}%
  \BibitemOpen
  \bibfield  {author} {\bibinfo {author} {\bibfnamefont {C.}~\bibnamefont {{Tiede}}}, \bibinfo {author} {\bibfnamefont {J.}~\bibnamefont {{Zrake}}}, \bibinfo {author} {\bibfnamefont {A.}~\bibnamefont {{MacFadyen}}},\ and\ \bibinfo {author} {\bibfnamefont {Z.}~\bibnamefont {{Haiman}}},\ }\bibfield  {title} {\bibinfo {title} {{Gas-driven Inspiral of Binaries in Thin Accretion Disks}},\ }\href {https://doi.org/10.3847/1538-4357/aba432} {\bibfield  {journal} {\bibinfo  {journal} {Astrophys. J.}\ }\textbf {\bibinfo {volume} {900}},\ \bibinfo {eid} {43} (\bibinfo {year} {2020})},\ \Eprint {https://arxiv.org/abs/2005.09555} {arXiv:2005.09555 [astro-ph.GA]} \BibitemShut {NoStop}%
\bibitem [{\citenamefont {{Dittmann}}\ and\ \citenamefont {{Ryan}}(2024)}]{Dittmann2023b}%
  \BibitemOpen
  \bibfield  {author} {\bibinfo {author} {\bibfnamefont {A.~J.}\ \bibnamefont {{Dittmann}}}\ and\ \bibinfo {author} {\bibfnamefont {G.}~\bibnamefont {{Ryan}}},\ }\bibfield  {title} {\bibinfo {title} {{The Evolution of Accreting Binaries: From Brown Dwarfs to Supermassive Black Holes}},\ }\href {https://doi.org/10.3847/1538-4357/ad2f1e} {\bibfield  {journal} {\bibinfo  {journal} {Astrophys. J.}\ }\textbf {\bibinfo {volume} {967}},\ \bibinfo {eid} {12} (\bibinfo {year} {2024})},\ \Eprint {https://arxiv.org/abs/2310.07758} {arXiv:2310.07758 [astro-ph.GA]} \BibitemShut {NoStop}%
\bibitem [{\citenamefont {{Dittmann}}\ and\ \citenamefont {{Ryan}}(2022)}]{Dittmann2022}%
  \BibitemOpen
  \bibfield  {author} {\bibinfo {author} {\bibfnamefont {A.~J.}\ \bibnamefont {{Dittmann}}}\ and\ \bibinfo {author} {\bibfnamefont {G.}~\bibnamefont {{Ryan}}},\ }\bibfield  {title} {\bibinfo {title} {{A Survey of Disc Thickness and Viscosity in Circumbinary Accretion: Binary Evolution, Variability, and Disc Morphology}},\ }\href@noop {} {\bibfield  {journal} {\bibinfo  {journal} {arXiv e-prints}\ ,\ \bibinfo {eid} {arXiv:2201.07816}} (\bibinfo {year} {2022})},\ \Eprint {https://arxiv.org/abs/2201.07816} {arXiv:2201.07816 [astro-ph.HE]} \BibitemShut {NoStop}%
\bibitem [{\citenamefont {{Sudarshan}}\ \emph {et~al.}(2022)\citenamefont {{Sudarshan}}, \citenamefont {{Penzlin}}, \citenamefont {{Ziampras}}, \citenamefont {{Kley}},\ and\ \citenamefont {{Nelson}}}]{Sudarshan2022}%
  \BibitemOpen
  \bibfield  {author} {\bibinfo {author} {\bibfnamefont {P.}~\bibnamefont {{Sudarshan}}}, \bibinfo {author} {\bibfnamefont {A.~B.~T.}\ \bibnamefont {{Penzlin}}}, \bibinfo {author} {\bibfnamefont {A.}~\bibnamefont {{Ziampras}}}, \bibinfo {author} {\bibfnamefont {W.}~\bibnamefont {{Kley}}},\ and\ \bibinfo {author} {\bibfnamefont {R.~P.}\ \bibnamefont {{Nelson}}},\ }\bibfield  {title} {\bibinfo {title} {{How cooling influences circumbinary discs}},\ }\href@noop {} {\bibfield  {journal} {\bibinfo  {journal} {arXiv e-prints}\ ,\ \bibinfo {eid} {arXiv:2206.07749}} (\bibinfo {year} {2022})},\ \Eprint {https://arxiv.org/abs/2206.07749} {arXiv:2206.07749 [astro-ph.EP]} \BibitemShut {NoStop}%
\bibitem [{\citenamefont {{Wang}}\ \emph {et~al.}(2023{\natexlab{a}})\citenamefont {{Wang}}, \citenamefont {{Bai}},\ and\ \citenamefont {{Lai}}}]{Wang2022}%
  \BibitemOpen
  \bibfield  {author} {\bibinfo {author} {\bibfnamefont {H.-Y.}\ \bibnamefont {{Wang}}}, \bibinfo {author} {\bibfnamefont {X.-N.}\ \bibnamefont {{Bai}}},\ and\ \bibinfo {author} {\bibfnamefont {D.}~\bibnamefont {{Lai}}},\ }\bibfield  {title} {\bibinfo {title} {{On the Role of Dynamical Cooling in the Dynamics of Circumbinary Disks}},\ }\href {https://doi.org/10.3847/1538-4357/acac77} {\bibfield  {journal} {\bibinfo  {journal} {Astrophys. J.}\ }\textbf {\bibinfo {volume} {943}},\ \bibinfo {eid} {175} (\bibinfo {year} {2023}{\natexlab{a}})},\ \Eprint {https://arxiv.org/abs/2212.04199} {arXiv:2212.04199 [astro-ph.HE]} \BibitemShut {NoStop}%
\bibitem [{\citenamefont {{Wang}}\ \emph {et~al.}(2023{\natexlab{b}})\citenamefont {{Wang}}, \citenamefont {{Bai}}, \citenamefont {{Lai}},\ and\ \citenamefont {{Lin}}}]{Wang2023}%
  \BibitemOpen
  \bibfield  {author} {\bibinfo {author} {\bibfnamefont {H.-Y.}\ \bibnamefont {{Wang}}}, \bibinfo {author} {\bibfnamefont {X.-N.}\ \bibnamefont {{Bai}}}, \bibinfo {author} {\bibfnamefont {D.}~\bibnamefont {{Lai}}},\ and\ \bibinfo {author} {\bibfnamefont {D.~N.~C.}\ \bibnamefont {{Lin}}},\ }\bibfield  {title} {\bibinfo {title} {{Hydrodynamical simulations of circumbinary accretion: balance between heating and cooling}},\ }\href {https://doi.org/10.1093/mnras/stad2884} {\bibfield  {journal} {\bibinfo  {journal} {Mon. Not. Roy. Astron. Soc.}\ }\textbf {\bibinfo {volume} {526}},\ \bibinfo {pages} {3570} (\bibinfo {year} {2023}{\natexlab{b}})},\ \Eprint {https://arxiv.org/abs/2212.07416} {arXiv:2212.07416 [astro-ph.HE]} \BibitemShut {NoStop}%
\bibitem [{\citenamefont {{Tiwari}}\ \emph {et~al.}(2025)\citenamefont {{Tiwari}}, \citenamefont {{Chan}}, \citenamefont {{Bogdanovi{\'c}}}, \citenamefont {{Jiang}}, \citenamefont {{Davis}},\ and\ \citenamefont {{Ferrel}}}]{Tiwari:2025imm}%
  \BibitemOpen
  \bibfield  {author} {\bibinfo {author} {\bibfnamefont {V.}~\bibnamefont {{Tiwari}}}, \bibinfo {author} {\bibfnamefont {C.-H.}\ \bibnamefont {{Chan}}}, \bibinfo {author} {\bibfnamefont {T.}~\bibnamefont {{Bogdanovi{\'c}}}}, \bibinfo {author} {\bibfnamefont {Y.-F.}\ \bibnamefont {{Jiang}}}, \bibinfo {author} {\bibfnamefont {S.~W.}\ \bibnamefont {{Davis}}},\ and\ \bibinfo {author} {\bibfnamefont {S.}~\bibnamefont {{Ferrel}}},\ }\bibfield  {title} {\bibinfo {title} {{Radiation Magnetohydrodynamic Simulation of sub-Eddington Circumbinary Disk around an Equal-mass Massive Black Hole Binary}},\ }\href {https://doi.org/10.48550/arXiv.2502.18584} {\bibfield  {journal} {\bibinfo  {journal} {arXiv e-prints}\ ,\ \bibinfo {eid} {arXiv:2502.18584}} (\bibinfo {year} {2025})},\ \Eprint {https://arxiv.org/abs/2502.18584} {arXiv:2502.18584 [astro-ph.HE]} \BibitemShut {NoStop}%
\bibitem [{\citenamefont {Shi}\ \emph {et~al.}(2012)\citenamefont {Shi}, \citenamefont {Krolik}, \citenamefont {Lubow},\ and\ \citenamefont {Hawley}}]{Shi:2011us}%
  \BibitemOpen
  \bibfield  {author} {\bibinfo {author} {\bibfnamefont {J.-M.}\ \bibnamefont {Shi}}, \bibinfo {author} {\bibfnamefont {J.~H.}\ \bibnamefont {Krolik}}, \bibinfo {author} {\bibfnamefont {S.~H.}\ \bibnamefont {Lubow}},\ and\ \bibinfo {author} {\bibfnamefont {J.~F.}\ \bibnamefont {Hawley}},\ }\bibfield  {title} {\bibinfo {title} {{Three Dimensional MHD Simulation of Circumbinary Accretion Disks: Disk Structures and Angular Momentum Transport}},\ }\href {https://doi.org/10.1088/0004-637X/749/2/118} {\bibfield  {journal} {\bibinfo  {journal} {Astrophys. J.}\ }\textbf {\bibinfo {volume} {749}},\ \bibinfo {pages} {118} (\bibinfo {year} {2012})},\ \Eprint {https://arxiv.org/abs/1110.4866} {arXiv:1110.4866 [astro-ph.HE]} \BibitemShut {NoStop}%
\bibitem [{\citenamefont {Noble}\ \emph {et~al.}(2012)\citenamefont {Noble}, \citenamefont {Mundim}, \citenamefont {Nakano}, \citenamefont {Krolik}, \citenamefont {Campanelli}, \citenamefont {Zlochower},\ and\ \citenamefont {Yunes}}]{Noble:2012xz}%
  \BibitemOpen
  \bibfield  {author} {\bibinfo {author} {\bibfnamefont {S.~C.}\ \bibnamefont {Noble}}, \bibinfo {author} {\bibfnamefont {B.~C.}\ \bibnamefont {Mundim}}, \bibinfo {author} {\bibfnamefont {H.}~\bibnamefont {Nakano}}, \bibinfo {author} {\bibfnamefont {J.~H.}\ \bibnamefont {Krolik}}, \bibinfo {author} {\bibfnamefont {M.}~\bibnamefont {Campanelli}}, \bibinfo {author} {\bibfnamefont {Y.}~\bibnamefont {Zlochower}},\ and\ \bibinfo {author} {\bibfnamefont {N.}~\bibnamefont {Yunes}},\ }\bibfield  {title} {\bibinfo {title} {{Circumbinary MHD Accretion into Inspiraling Binary Black Holes}},\ }\href {https://doi.org/10.1088/0004-637X/755/1/51} {\bibfield  {journal} {\bibinfo  {journal} {Astrophys. J.}\ }\textbf {\bibinfo {volume} {755}},\ \bibinfo {pages} {51} (\bibinfo {year} {2012})},\ \Eprint {https://arxiv.org/abs/1204.1073} {arXiv:1204.1073 [astro-ph.HE]} \BibitemShut {NoStop}%
\bibitem [{\citenamefont {Noble}\ \emph {et~al.}(2021)\citenamefont {Noble}, \citenamefont {Krolik}, \citenamefont {Campanelli}, \citenamefont {Zlochower}, \citenamefont {Mundim}, \citenamefont {Nakano},\ and\ \citenamefont {Zilh\~ao}}]{Noble:2021vfg}%
  \BibitemOpen
  \bibfield  {author} {\bibinfo {author} {\bibfnamefont {S.~C.}\ \bibnamefont {Noble}}, \bibinfo {author} {\bibfnamefont {J.~H.}\ \bibnamefont {Krolik}}, \bibinfo {author} {\bibfnamefont {M.}~\bibnamefont {Campanelli}}, \bibinfo {author} {\bibfnamefont {Y.}~\bibnamefont {Zlochower}}, \bibinfo {author} {\bibfnamefont {B.~C.}\ \bibnamefont {Mundim}}, \bibinfo {author} {\bibfnamefont {H.}~\bibnamefont {Nakano}},\ and\ \bibinfo {author} {\bibfnamefont {M.}~\bibnamefont {Zilh\~ao}},\ }\bibfield  {title} {\bibinfo {title} {{Mass-ratio and Magnetic Flux Dependence of Modulated Accretion from Circumbinary Disks}},\ }\href {https://doi.org/10.3847/1538-4357/ac2229} {\bibfield  {journal} {\bibinfo  {journal} {Astrophys. J.}\ }\textbf {\bibinfo {volume} {922}},\ \bibinfo {pages} {175} (\bibinfo {year} {2021})},\ \Eprint {https://arxiv.org/abs/2103.12100} {arXiv:2103.12100 [astro-ph.HE]} \BibitemShut {NoStop}%
\bibitem [{\citenamefont {Most}\ and\ \citenamefont {Wang}(2024)}]{Most:2024qus}%
  \BibitemOpen
  \bibfield  {author} {\bibinfo {author} {\bibfnamefont {E.~R.}\ \bibnamefont {Most}}\ and\ \bibinfo {author} {\bibfnamefont {H.-Y.}\ \bibnamefont {Wang}},\ }\bibfield  {title} {\bibinfo {title} {{Magnetically Arrested Circumbinary Accretion Flows}},\ }\href {https://doi.org/10.3847/2041-8213/ad7713} {\bibfield  {journal} {\bibinfo  {journal} {Astrophys. J. Lett.}\ }\textbf {\bibinfo {volume} {973}},\ \bibinfo {pages} {L19} (\bibinfo {year} {2024})},\ \Eprint {https://arxiv.org/abs/2408.00757} {arXiv:2408.00757 [astro-ph.HE]} \BibitemShut {NoStop}%
\bibitem [{\citenamefont {{Most}}\ and\ \citenamefont {{Wang}}(2024)}]{Most:2024onq}%
  \BibitemOpen
  \bibfield  {author} {\bibinfo {author} {\bibfnamefont {E.~R.}\ \bibnamefont {{Most}}}\ and\ \bibinfo {author} {\bibfnamefont {H.-Y.}\ \bibnamefont {{Wang}}},\ }\bibfield  {title} {\bibinfo {title} {{Decoupling of a supermassive black hole binary from its magnetically arrested circumbinary accretion disk}},\ }\href {https://doi.org/10.48550/arXiv.2410.23264} {\bibfield  {journal} {\bibinfo  {journal} {arXiv e-prints}\ ,\ \bibinfo {eid} {arXiv:2410.23264}} (\bibinfo {year} {2024})},\ \Eprint {https://arxiv.org/abs/2410.23264} {arXiv:2410.23264 [astro-ph.HE]} \BibitemShut {NoStop}%
\bibitem [{\citenamefont {{Ennoggi}}\ \emph {et~al.}(2025)\citenamefont {{Ennoggi}}, \citenamefont {{Campanelli}}, \citenamefont {{Zlochower}}, \citenamefont {{Noble}}, \citenamefont {{Krolik}}, \citenamefont {{Cattorini}}, \citenamefont {{Kalinani}}, \citenamefont {{Mewes}}, \citenamefont {{Chabanov}}, \citenamefont {{Ji}},\ and\ \citenamefont {{de Simone}}}]{Ennoggi:2025nht}%
  \BibitemOpen
  \bibfield  {author} {\bibinfo {author} {\bibfnamefont {L.}~\bibnamefont {{Ennoggi}}}, \bibinfo {author} {\bibfnamefont {M.}~\bibnamefont {{Campanelli}}}, \bibinfo {author} {\bibfnamefont {Y.}~\bibnamefont {{Zlochower}}}, \bibinfo {author} {\bibfnamefont {S.~C.}\ \bibnamefont {{Noble}}}, \bibinfo {author} {\bibfnamefont {J.}~\bibnamefont {{Krolik}}}, \bibinfo {author} {\bibfnamefont {F.}~\bibnamefont {{Cattorini}}}, \bibinfo {author} {\bibfnamefont {J.~V.}\ \bibnamefont {{Kalinani}}}, \bibinfo {author} {\bibfnamefont {V.}~\bibnamefont {{Mewes}}}, \bibinfo {author} {\bibfnamefont {M.}~\bibnamefont {{Chabanov}}}, \bibinfo {author} {\bibfnamefont {L.}~\bibnamefont {{Ji}}},\ and\ \bibinfo {author} {\bibfnamefont {M.~C.}\ \bibnamefont {{de Simone}}},\ }\bibfield  {title} {\bibinfo {title} {{Relativistic Gas Accretion onto Supermassive Black Hole Binaries from Inspiral through Merger}},\ }\href {https://doi.org/10.48550/arXiv.2502.06389} {\bibfield  {journal} {\bibinfo  {journal} {arXiv e-prints}\ ,\ \bibinfo {eid} {arXiv:2502.06389}} (\bibinfo {year} {2025})},\ \Eprint {https://arxiv.org/abs/2502.06389} {arXiv:2502.06389 [astro-ph.HE]} \BibitemShut {NoStop}%
\bibitem [{\citenamefont {{Narayan}}\ \emph {et~al.}(2003)\citenamefont {{Narayan}}, \citenamefont {{Igumenshchev}},\ and\ \citenamefont {{Abramowicz}}}]{Narayan2003}%
  \BibitemOpen
  \bibfield  {author} {\bibinfo {author} {\bibfnamefont {R.}~\bibnamefont {{Narayan}}}, \bibinfo {author} {\bibfnamefont {I.~V.}\ \bibnamefont {{Igumenshchev}}},\ and\ \bibinfo {author} {\bibfnamefont {M.~A.}\ \bibnamefont {{Abramowicz}}},\ }\bibfield  {title} {\bibinfo {title} {{Magnetically Arrested Disk: an Energetically Efficient Accretion Flow}},\ }\href {https://doi.org/10.1093/pasj/55.6.L69} {\bibfield  {journal} {\bibinfo  {journal} {PASJ}\ }\textbf {\bibinfo {volume} {55}},\ \bibinfo {pages} {L69} (\bibinfo {year} {2003})},\ \Eprint {https://arxiv.org/abs/astro-ph/0305029} {arXiv:astro-ph/0305029 [astro-ph]} \BibitemShut {NoStop}%
\bibitem [{\citenamefont {Tchekhovskoy}\ \emph {et~al.}(2011)\citenamefont {Tchekhovskoy}, \citenamefont {Narayan},\ and\ \citenamefont {McKinney}}]{Tchekhovskoy:2011zx}%
  \BibitemOpen
  \bibfield  {author} {\bibinfo {author} {\bibfnamefont {A.}~\bibnamefont {Tchekhovskoy}}, \bibinfo {author} {\bibfnamefont {R.}~\bibnamefont {Narayan}},\ and\ \bibinfo {author} {\bibfnamefont {J.~C.}\ \bibnamefont {McKinney}},\ }\bibfield  {title} {\bibinfo {title} {{Efficient Generation of Jets from Magnetically Arrested Accretion on a Rapidly Spinning Black Hole}},\ }\href {https://doi.org/10.1111/j.1745-3933.2011.01147.x} {\bibfield  {journal} {\bibinfo  {journal} {Mon. Not. Roy. Astron. Soc.}\ }\textbf {\bibinfo {volume} {418}},\ \bibinfo {pages} {L79} (\bibinfo {year} {2011})},\ \Eprint {https://arxiv.org/abs/1108.0412} {arXiv:1108.0412 [astro-ph.HE]} \BibitemShut {NoStop}%
\bibitem [{\citenamefont {Ripperda}\ \emph {et~al.}(2022)\citenamefont {Ripperda}, \citenamefont {Liska}, \citenamefont {Chatterjee}, \citenamefont {Musoke}, \citenamefont {Philippov}, \citenamefont {Markoff}, \citenamefont {Tchekhovskoy},\ and\ \citenamefont {Younsi}}]{Ripperda:2021zpn}%
  \BibitemOpen
  \bibfield  {author} {\bibinfo {author} {\bibfnamefont {B.}~\bibnamefont {Ripperda}}, \bibinfo {author} {\bibfnamefont {M.}~\bibnamefont {Liska}}, \bibinfo {author} {\bibfnamefont {K.}~\bibnamefont {Chatterjee}}, \bibinfo {author} {\bibfnamefont {G.}~\bibnamefont {Musoke}}, \bibinfo {author} {\bibfnamefont {A.~A.}\ \bibnamefont {Philippov}}, \bibinfo {author} {\bibfnamefont {S.~B.}\ \bibnamefont {Markoff}}, \bibinfo {author} {\bibfnamefont {A.}~\bibnamefont {Tchekhovskoy}},\ and\ \bibinfo {author} {\bibfnamefont {Z.}~\bibnamefont {Younsi}},\ }\bibfield  {title} {\bibinfo {title} {{Black Hole Flares: Ejection of Accreted Magnetic Flux through 3D Plasmoid-mediated Reconnection}},\ }\href {https://doi.org/10.3847/2041-8213/ac46a1} {\bibfield  {journal} {\bibinfo  {journal} {Astrophys. J. Lett.}\ }\textbf {\bibinfo {volume} {924}},\ \bibinfo {pages} {L32} (\bibinfo {year} {2022})},\ \Eprint {https://arxiv.org/abs/2109.15115} {arXiv:2109.15115 [astro-ph.HE]} \BibitemShut {NoStop}%
\bibitem [{\citenamefont {Liska}\ \emph {et~al.}(2022)\citenamefont {Liska}, \citenamefont {Musoke}, \citenamefont {Tchekhovskoy}, \citenamefont {Porth},\ and\ \citenamefont {Beloborodov}}]{Liska:2022jdy}%
  \BibitemOpen
  \bibfield  {author} {\bibinfo {author} {\bibfnamefont {M.~T.~P.}\ \bibnamefont {Liska}}, \bibinfo {author} {\bibfnamefont {G.}~\bibnamefont {Musoke}}, \bibinfo {author} {\bibfnamefont {A.}~\bibnamefont {Tchekhovskoy}}, \bibinfo {author} {\bibfnamefont {O.}~\bibnamefont {Porth}},\ and\ \bibinfo {author} {\bibfnamefont {A.~M.}\ \bibnamefont {Beloborodov}},\ }\bibfield  {title} {\bibinfo {title} {{Formation of Magnetically Truncated Accretion Disks in 3D Radiation-transport Two-temperature GRMHD Simulations}},\ }\href {https://doi.org/10.3847/2041-8213/ac84db} {\bibfield  {journal} {\bibinfo  {journal} {Astrophys. J. Lett.}\ }\textbf {\bibinfo {volume} {935}},\ \bibinfo {pages} {L1} (\bibinfo {year} {2022})},\ \Eprint {https://arxiv.org/abs/2201.03526} {arXiv:2201.03526 [astro-ph.HE]} \BibitemShut {NoStop}%
\bibitem [{\citenamefont {Fragner}\ and\ \citenamefont {Nelson}(2010)}]{Fragner:2009mk}%
  \BibitemOpen
  \bibfield  {author} {\bibinfo {author} {\bibfnamefont {M.}~\bibnamefont {Fragner}}\ and\ \bibinfo {author} {\bibfnamefont {R.}~\bibnamefont {Nelson}},\ }\bibfield  {title} {\bibinfo {title} {{Evolution of warped and twisted accretion discs in close binary systems}},\ }\href {https://doi.org/10.1051/0004-6361/200913088} {\bibfield  {journal} {\bibinfo  {journal} {Astron. Astrophys.}\ }\textbf {\bibinfo {volume} {511}},\ \bibinfo {pages} {A77} (\bibinfo {year} {2010})},\ \Eprint {https://arxiv.org/abs/0912.3220} {arXiv:0912.3220 [astro-ph.EP]} \BibitemShut {NoStop}%
\bibitem [{\citenamefont {Ressler}\ \emph {et~al.}(2023)\citenamefont {Ressler}, \citenamefont {White},\ and\ \citenamefont {Quataert}}]{Ressler:2023ptc}%
  \BibitemOpen
  \bibfield  {author} {\bibinfo {author} {\bibfnamefont {S.~M.}\ \bibnamefont {Ressler}}, \bibinfo {author} {\bibfnamefont {C.~J.}\ \bibnamefont {White}},\ and\ \bibinfo {author} {\bibfnamefont {E.}~\bibnamefont {Quataert}},\ }\bibfield  {title} {\bibinfo {title} {{Wind-fed GRMHD simulations of Sagittarius~A*: tilt and alignment of jets and accretion discs, electron thermodynamics, and multiscale modelling of the rotation measure}},\ }\href {https://doi.org/10.1093/mnras/stad837} {\bibfield  {journal} {\bibinfo  {journal} {Mon. Not. Roy. Astron. Soc.}\ }\textbf {\bibinfo {volume} {521}},\ \bibinfo {pages} {4277} (\bibinfo {year} {2023})},\ \Eprint {https://arxiv.org/abs/2303.15503} {arXiv:2303.15503 [astro-ph.HE]} \BibitemShut {NoStop}%
\bibitem [{\citenamefont {{Stone}}\ \emph {et~al.}(2024)\citenamefont {{Stone}}, \citenamefont {{Mullen}}, \citenamefont {{Fielding}}, \citenamefont {{Grete}}, \citenamefont {{Guo}}, \citenamefont {{Kempski}}, \citenamefont {{Most}}, \citenamefont {{White}},\ and\ \citenamefont {{Wong}}}]{2024arXiv240916053S}%
  \BibitemOpen
  \bibfield  {author} {\bibinfo {author} {\bibfnamefont {J.~M.}\ \bibnamefont {{Stone}}}, \bibinfo {author} {\bibfnamefont {P.~D.}\ \bibnamefont {{Mullen}}}, \bibinfo {author} {\bibfnamefont {D.}~\bibnamefont {{Fielding}}}, \bibinfo {author} {\bibfnamefont {P.}~\bibnamefont {{Grete}}}, \bibinfo {author} {\bibfnamefont {M.}~\bibnamefont {{Guo}}}, \bibinfo {author} {\bibfnamefont {P.}~\bibnamefont {{Kempski}}}, \bibinfo {author} {\bibfnamefont {E.~R.}\ \bibnamefont {{Most}}}, \bibinfo {author} {\bibfnamefont {C.~J.}\ \bibnamefont {{White}}},\ and\ \bibinfo {author} {\bibfnamefont {G.~N.}\ \bibnamefont {{Wong}}},\ }\bibfield  {title} {\bibinfo {title} {{AthenaK: A Performance-Portable Version of the Athena++ AMR Framework}},\ }\href {https://doi.org/10.48550/arXiv.2409.16053} {\bibfield  {journal} {\bibinfo  {journal} {arXiv e-prints}\ ,\ \bibinfo {eid} {arXiv:2409.16053}} (\bibinfo {year} {2024})},\ \Eprint {https://arxiv.org/abs/2409.16053} {arXiv:2409.16053 [astro-ph.IM]} \BibitemShut {NoStop}%
\bibitem [{\citenamefont {{Stone}}\ \emph {et~al.}(2020)\citenamefont {{Stone}}, \citenamefont {{Tomida}}, \citenamefont {{White}},\ and\ \citenamefont {{Felker}}}]{2020ApJS..249....4S}%
  \BibitemOpen
  \bibfield  {author} {\bibinfo {author} {\bibfnamefont {J.~M.}\ \bibnamefont {{Stone}}}, \bibinfo {author} {\bibfnamefont {K.}~\bibnamefont {{Tomida}}}, \bibinfo {author} {\bibfnamefont {C.~J.}\ \bibnamefont {{White}}},\ and\ \bibinfo {author} {\bibfnamefont {K.~G.}\ \bibnamefont {{Felker}}},\ }\bibfield  {title} {\bibinfo {title} {{The Athena++ Adaptive Mesh Refinement Framework: Design and Magnetohydrodynamic Solvers}},\ }\href {https://doi.org/10.3847/1538-4365/ab929b} {\bibfield  {journal} {\bibinfo  {journal} {Astrophys. J. Supp.}\ }\textbf {\bibinfo {volume} {249}},\ \bibinfo {eid} {4} (\bibinfo {year} {2020})},\ \Eprint {https://arxiv.org/abs/2005.06651} {arXiv:2005.06651 [astro-ph.IM]} \BibitemShut {NoStop}%
\bibitem [{\citenamefont {{Navarro}}\ \emph {et~al.}(1997)\citenamefont {{Navarro}}, \citenamefont {{Frenk}},\ and\ \citenamefont {{White}}}]{1997ApJ...490..493N}%
  \BibitemOpen
  \bibfield  {author} {\bibinfo {author} {\bibfnamefont {J.~F.}\ \bibnamefont {{Navarro}}}, \bibinfo {author} {\bibfnamefont {C.~S.}\ \bibnamefont {{Frenk}}},\ and\ \bibinfo {author} {\bibfnamefont {S.~D.~M.}\ \bibnamefont {{White}}},\ }\bibfield  {title} {\bibinfo {title} {{A Universal Density Profile from Hierarchical Clustering}},\ }\href {https://doi.org/10.1086/304888} {\bibfield  {journal} {\bibinfo  {journal} {Astrophys. J.}\ }\textbf {\bibinfo {volume} {490}},\ \bibinfo {pages} {493} (\bibinfo {year} {1997})},\ \Eprint {https://arxiv.org/abs/astro-ph/9611107} {arXiv:astro-ph/9611107 [astro-ph]} \BibitemShut {NoStop}%
\bibitem [{\citenamefont {Akiyama}\ \emph {et~al.}(2019)\citenamefont {Akiyama} \emph {et~al.}}]{EventHorizonTelescope:2019pgp}%
  \BibitemOpen
  \bibfield  {author} {\bibinfo {author} {\bibfnamefont {K.}~\bibnamefont {Akiyama}} \emph {et~al.} (\bibinfo {collaboration} {Event Horizon Telescope}),\ }\bibfield  {title} {\bibinfo {title} {{First M87 Event Horizon Telescope Results. V. Physical Origin of the Asymmetric Ring}},\ }\href {https://doi.org/10.3847/2041-8213/ab0f43} {\bibfield  {journal} {\bibinfo  {journal} {Astrophys. J. Lett.}\ }\textbf {\bibinfo {volume} {875}},\ \bibinfo {pages} {L5} (\bibinfo {year} {2019})},\ \Eprint {https://arxiv.org/abs/1906.11242} {arXiv:1906.11242 [astro-ph.GA]} \BibitemShut {NoStop}%
\bibitem [{\citenamefont {{Li}}\ and\ \citenamefont {{Bryan}}(2012)}]{2012ApJ...747...26L}%
  \BibitemOpen
  \bibfield  {author} {\bibinfo {author} {\bibfnamefont {Y.}~\bibnamefont {{Li}}}\ and\ \bibinfo {author} {\bibfnamefont {G.~L.}\ \bibnamefont {{Bryan}}},\ }\bibfield  {title} {\bibinfo {title} {{Simulating the Cooling Flow of Cool-core Clusters}},\ }\href {https://doi.org/10.1088/0004-637X/747/1/26} {\bibfield  {journal} {\bibinfo  {journal} {Astrophys. J.}\ }\textbf {\bibinfo {volume} {747}},\ \bibinfo {eid} {26} (\bibinfo {year} {2012})},\ \Eprint {https://arxiv.org/abs/1112.2701} {arXiv:1112.2701 [astro-ph.CO]} \BibitemShut {NoStop}%
\bibitem [{\citenamefont {{Wang}}\ \emph {et~al.}(2020)\citenamefont {{Wang}}, \citenamefont {{Ruszkowski}},\ and\ \citenamefont {{Yang}}}]{2020MNRAS.493.4065W}%
  \BibitemOpen
  \bibfield  {author} {\bibinfo {author} {\bibfnamefont {C.}~\bibnamefont {{Wang}}}, \bibinfo {author} {\bibfnamefont {M.}~\bibnamefont {{Ruszkowski}}},\ and\ \bibinfo {author} {\bibfnamefont {H.~Y.~K.}\ \bibnamefont {{Yang}}},\ }\bibfield  {title} {\bibinfo {title} {{Chaotic cold accretion in giant elliptical galaxies heated by AGN cosmic rays}},\ }\href {https://doi.org/10.1093/mnras/staa550} {\bibfield  {journal} {\bibinfo  {journal} {"Mon. Not. Roy. Astron. Soc.",}\ }\textbf {\bibinfo {volume} {493}},\ \bibinfo {pages} {4065} (\bibinfo {year} {2020})},\ \Eprint {https://arxiv.org/abs/1910.03608} {arXiv:1910.03608 [astro-ph.GA]} \BibitemShut {NoStop}%
\bibitem [{\citenamefont {Aly}\ \emph {et~al.}(2015)\citenamefont {Aly}, \citenamefont {Dehnen}, \citenamefont {Nixon},\ and\ \citenamefont {King}}]{Aly:2015vqa}%
  \BibitemOpen
  \bibfield  {author} {\bibinfo {author} {\bibfnamefont {H.}~\bibnamefont {Aly}}, \bibinfo {author} {\bibfnamefont {W.}~\bibnamefont {Dehnen}}, \bibinfo {author} {\bibfnamefont {C.}~\bibnamefont {Nixon}},\ and\ \bibinfo {author} {\bibfnamefont {A.}~\bibnamefont {King}},\ }\bibfield  {title} {\bibinfo {title} {{Misaligned gas discs around eccentric black-hole binaries and implications for the final-parsec problem}},\ }\href {https://doi.org/10.1093/mnras/stv128} {\bibfield  {journal} {\bibinfo  {journal} {Mon. Not. Roy. Astron. Soc.}\ }\textbf {\bibinfo {volume} {449}},\ \bibinfo {pages} {65} (\bibinfo {year} {2015})},\ \Eprint {https://arxiv.org/abs/1501.04623} {arXiv:1501.04623 [astro-ph.HE]} \BibitemShut {NoStop}%
\bibitem [{\citenamefont {{Martin}}\ and\ \citenamefont {{Lubow}}(2017)}]{Rebecca:2017ApJ...835L..28M}%
  \BibitemOpen
  \bibfield  {author} {\bibinfo {author} {\bibfnamefont {R.~G.}\ \bibnamefont {{Martin}}}\ and\ \bibinfo {author} {\bibfnamefont {S.~H.}\ \bibnamefont {{Lubow}}},\ }\bibfield  {title} {\bibinfo {title} {{Polar Alignment of a Protoplanetary Disk around an Eccentric Binary}},\ }\href {https://doi.org/10.3847/2041-8213/835/2/L28} {\bibfield  {journal} {\bibinfo  {journal} {Astrophys. J. Lett.}\ }\textbf {\bibinfo {volume} {835}},\ \bibinfo {eid} {L28} (\bibinfo {year} {2017})},\ \Eprint {https://arxiv.org/abs/1702.00545} {arXiv:1702.00545 [astro-ph.EP]} \BibitemShut {NoStop}%
\bibitem [{\citenamefont {Lepp}\ \emph {et~al.}(2022)\citenamefont {Lepp}, \citenamefont {Martin},\ and\ \citenamefont {Childs}}]{Lepp:2022gni}%
  \BibitemOpen
  \bibfield  {author} {\bibinfo {author} {\bibfnamefont {S.}~\bibnamefont {Lepp}}, \bibinfo {author} {\bibfnamefont {R.~G.}\ \bibnamefont {Martin}},\ and\ \bibinfo {author} {\bibfnamefont {A.~C.}\ \bibnamefont {Childs}},\ }\bibfield  {title} {\bibinfo {title} {{A Radial Limit on Polar Circumbinary Orbits from General Relativity}},\ }\href {https://doi.org/10.3847/2041-8213/ac61e1} {\bibfield  {journal} {\bibinfo  {journal} {Astrophys. J. Lett.}\ }\textbf {\bibinfo {volume} {929}},\ \bibinfo {pages} {L5} (\bibinfo {year} {2022})},\ \Eprint {https://arxiv.org/abs/2203.15160} {arXiv:2203.15160 [astro-ph.EP]} \BibitemShut {NoStop}%
\bibitem [{\citenamefont {Childs}\ \emph {et~al.}(2024)\citenamefont {Childs}, \citenamefont {Martin}, \citenamefont {Nixon}, \citenamefont {Geller}, \citenamefont {Lubow}, \citenamefont {Zhu},\ and\ \citenamefont {Lepp}}]{Childs:2023zsf}%
  \BibitemOpen
  \bibfield  {author} {\bibinfo {author} {\bibfnamefont {A.~C.}\ \bibnamefont {Childs}}, \bibinfo {author} {\bibfnamefont {R.~G.}\ \bibnamefont {Martin}}, \bibinfo {author} {\bibfnamefont {C.~J.}\ \bibnamefont {Nixon}}, \bibinfo {author} {\bibfnamefont {A.~M.}\ \bibnamefont {Geller}}, \bibinfo {author} {\bibfnamefont {S.~H.}\ \bibnamefont {Lubow}}, \bibinfo {author} {\bibfnamefont {Z.}~\bibnamefont {Zhu}},\ and\ \bibinfo {author} {\bibfnamefont {S.}~\bibnamefont {Lepp}},\ }\bibfield  {title} {\bibinfo {title} {{Relativistic Effects on Circumbinary Disk Evolution: Breaking the Polar Alignment around Eccentric Black Hole Binary Systems}},\ }\href {https://doi.org/10.3847/1538-4357/ad1a11} {\bibfield  {journal} {\bibinfo  {journal} {Astrophys. J.}\ }\textbf {\bibinfo {volume} {962}},\ \bibinfo {pages} {77} (\bibinfo {year} {2024})},\ \Eprint {https://arxiv.org/abs/2312.09495} {arXiv:2312.09495 [astro-ph.HE]} \BibitemShut {NoStop}%
\bibitem [{\citenamefont {{Squire}}\ \emph {et~al.}(2024)\citenamefont {{Squire}}, \citenamefont {{Quataert}},\ and\ \citenamefont {{Hopkins}}}]{Squire:2024yhe}%
  \BibitemOpen
  \bibfield  {author} {\bibinfo {author} {\bibfnamefont {J.}~\bibnamefont {{Squire}}}, \bibinfo {author} {\bibfnamefont {E.}~\bibnamefont {{Quataert}}},\ and\ \bibinfo {author} {\bibfnamefont {P.~F.}\ \bibnamefont {{Hopkins}}},\ }\bibfield  {title} {\bibinfo {title} {{Rapid, strongly magnetized accretion in the zero-net-vertical-flux shearing box}},\ }\href {https://doi.org/10.48550/arXiv.2409.05467} {\bibfield  {journal} {\bibinfo  {journal} {arXiv e-prints}\ ,\ \bibinfo {eid} {arXiv:2409.05467}} (\bibinfo {year} {2024})},\ \Eprint {https://arxiv.org/abs/2409.05467} {arXiv:2409.05467 [astro-ph.HE]} \BibitemShut {NoStop}%
\bibitem [{\citenamefont {{Miranda}}\ and\ \citenamefont {{Lai}}(2015)}]{2015MNRAS.452.2396M}%
  \BibitemOpen
  \bibfield  {author} {\bibinfo {author} {\bibfnamefont {R.}~\bibnamefont {{Miranda}}}\ and\ \bibinfo {author} {\bibfnamefont {D.}~\bibnamefont {{Lai}}},\ }\bibfield  {title} {\bibinfo {title} {{Tidal truncation of inclined circumstellar and circumbinary discs in young stellar binaries}},\ }\href {https://doi.org/10.1093/mnras/stv1450} {\bibfield  {journal} {\bibinfo  {journal} {Mon. Not. Roy. Astron. Soc.}\ }\textbf {\bibinfo {volume} {452}},\ \bibinfo {pages} {2396} (\bibinfo {year} {2015})},\ \Eprint {https://arxiv.org/abs/1504.02917} {arXiv:1504.02917 [astro-ph.EP]} \BibitemShut {NoStop}%
\bibitem [{\citenamefont {{Bate}}(2018)}]{2018MNRAS.475.5618B}%
  \BibitemOpen
  \bibfield  {author} {\bibinfo {author} {\bibfnamefont {M.~R.}\ \bibnamefont {{Bate}}},\ }\bibfield  {title} {\bibinfo {title} {{On the diversity and statistical properties of protostellar discs}},\ }\href {https://doi.org/10.1093/mnras/sty169} {\bibfield  {journal} {\bibinfo  {journal} {Mon. Not. Roy. Astron. Soc.}\ }\textbf {\bibinfo {volume} {475}},\ \bibinfo {pages} {5618} (\bibinfo {year} {2018})},\ \Eprint {https://arxiv.org/abs/1801.07721} {arXiv:1801.07721 [astro-ph.SR]} \BibitemShut {NoStop}%
\bibitem [{\citenamefont {Etienne}\ \emph {et~al.}(2012)\citenamefont {Etienne}, \citenamefont {Liu}, \citenamefont {Paschalidis},\ and\ \citenamefont {Shapiro}}]{Etienne:2011ea}%
  \BibitemOpen
  \bibfield  {author} {\bibinfo {author} {\bibfnamefont {Z.~B.}\ \bibnamefont {Etienne}}, \bibinfo {author} {\bibfnamefont {Y.~T.}\ \bibnamefont {Liu}}, \bibinfo {author} {\bibfnamefont {V.}~\bibnamefont {Paschalidis}},\ and\ \bibinfo {author} {\bibfnamefont {S.~L.}\ \bibnamefont {Shapiro}},\ }\bibfield  {title} {\bibinfo {title} {{General relativistic simulations of black hole-neutron star mergers: Effects of magnetic fields}},\ }\href {https://doi.org/10.1103/PhysRevD.85.064029} {\bibfield  {journal} {\bibinfo  {journal} {Phys. Rev. D}\ }\textbf {\bibinfo {volume} {85}},\ \bibinfo {pages} {064029} (\bibinfo {year} {2012})},\ \Eprint {https://arxiv.org/abs/1112.0568} {arXiv:1112.0568 [astro-ph.HE]} \BibitemShut {NoStop}%
\bibitem [{\citenamefont {Most}\ \emph {et~al.}(2021)\citenamefont {Most}, \citenamefont {Papenfort}, \citenamefont {Tootle},\ and\ \citenamefont {Rezzolla}}]{Most:2021ytn}%
  \BibitemOpen
  \bibfield  {author} {\bibinfo {author} {\bibfnamefont {E.~R.}\ \bibnamefont {Most}}, \bibinfo {author} {\bibfnamefont {L.~J.}\ \bibnamefont {Papenfort}}, \bibinfo {author} {\bibfnamefont {S.~D.}\ \bibnamefont {Tootle}},\ and\ \bibinfo {author} {\bibfnamefont {L.}~\bibnamefont {Rezzolla}},\ }\bibfield  {title} {\bibinfo {title} {{On accretion discs formed in MHD simulations of black hole\textendash{}neutron star mergers with accurate microphysics}},\ }\href {https://doi.org/10.1093/mnras/stab1824} {\bibfield  {journal} {\bibinfo  {journal} {Mon. Not. Roy. Astron. Soc.}\ }\textbf {\bibinfo {volume} {506}},\ \bibinfo {pages} {3511} (\bibinfo {year} {2021})},\ \Eprint {https://arxiv.org/abs/2106.06391} {arXiv:2106.06391 [astro-ph.HE]} \BibitemShut {NoStop}%
\bibitem [{\citenamefont {Izquierdo}\ \emph {et~al.}(2024)\citenamefont {Izquierdo}, \citenamefont {Bezares}, \citenamefont {Liebling},\ and\ \citenamefont {Palenzuela}}]{Izquierdo:2024rbb}%
  \BibitemOpen
  \bibfield  {author} {\bibinfo {author} {\bibfnamefont {M.~R.}\ \bibnamefont {Izquierdo}}, \bibinfo {author} {\bibfnamefont {M.}~\bibnamefont {Bezares}}, \bibinfo {author} {\bibfnamefont {S.}~\bibnamefont {Liebling}},\ and\ \bibinfo {author} {\bibfnamefont {C.}~\bibnamefont {Palenzuela}},\ }\bibfield  {title} {\bibinfo {title} {{Large eddy simulations of magnetized mergers of black holes and neutron stars}},\ }\href {https://doi.org/10.1103/PhysRevD.110.083017} {\bibfield  {journal} {\bibinfo  {journal} {Phys. Rev. D}\ }\textbf {\bibinfo {volume} {110}},\ \bibinfo {pages} {083017} (\bibinfo {year} {2024})},\ \Eprint {https://arxiv.org/abs/2403.09770} {arXiv:2403.09770 [astro-ph.HE]} \BibitemShut {NoStop}%
\bibitem [{\citenamefont {Ressler}\ \emph {et~al.}(2020)\citenamefont {Ressler}, \citenamefont {White}, \citenamefont {Quataert},\ and\ \citenamefont {Stone}}]{Ressler:2020voz}%
  \BibitemOpen
  \bibfield  {author} {\bibinfo {author} {\bibfnamefont {S.~M.}\ \bibnamefont {Ressler}}, \bibinfo {author} {\bibfnamefont {C.~J.}\ \bibnamefont {White}}, \bibinfo {author} {\bibfnamefont {E.}~\bibnamefont {Quataert}},\ and\ \bibinfo {author} {\bibfnamefont {J.~M.}\ \bibnamefont {Stone}},\ }\bibfield  {title} {\bibinfo {title} {{Ab Initio Horizon-Scale Simulations of Magnetically Arrested Accretion in Sagittarius A* Fed by Stellar Winds}},\ }\href {https://doi.org/10.3847/2041-8213/ab9532} {\bibfield  {journal} {\bibinfo  {journal} {Astrophys. J. Lett.}\ }\textbf {\bibinfo {volume} {896}},\ \bibinfo {pages} {L6} (\bibinfo {year} {2020})},\ \Eprint {https://arxiv.org/abs/2006.00005} {arXiv:2006.00005 [astro-ph.HE]} \BibitemShut {NoStop}%
\bibitem [{\citenamefont {Combi}\ \emph {et~al.}(2022)\citenamefont {Combi}, \citenamefont {Armengol}, \citenamefont {Campanelli}, \citenamefont {Noble}, \citenamefont {Avara}, \citenamefont {Krolik},\ and\ \citenamefont {Bowen}}]{Combi:2021dks}%
  \BibitemOpen
  \bibfield  {author} {\bibinfo {author} {\bibfnamefont {L.}~\bibnamefont {Combi}}, \bibinfo {author} {\bibfnamefont {F.~G.~L.}\ \bibnamefont {Armengol}}, \bibinfo {author} {\bibfnamefont {M.}~\bibnamefont {Campanelli}}, \bibinfo {author} {\bibfnamefont {S.~C.}\ \bibnamefont {Noble}}, \bibinfo {author} {\bibfnamefont {M.}~\bibnamefont {Avara}}, \bibinfo {author} {\bibfnamefont {J.~H.}\ \bibnamefont {Krolik}},\ and\ \bibinfo {author} {\bibfnamefont {D.}~\bibnamefont {Bowen}},\ }\bibfield  {title} {\bibinfo {title} {{Minidisk Accretion onto Spinning Black Hole Binaries: Quasi-periodicities and Outflows}},\ }\href {https://doi.org/10.3847/1538-4357/ac532a} {\bibfield  {journal} {\bibinfo  {journal} {Astrophys. J.}\ }\textbf {\bibinfo {volume} {928}},\ \bibinfo {pages} {187} (\bibinfo {year} {2022})},\ \Eprint {https://arxiv.org/abs/2109.01307} {arXiv:2109.01307 [astro-ph.HE]} \BibitemShut {NoStop}%
\bibitem [{\citenamefont {Igumenshchev}(2008)}]{Igumenshchev:2007bh}%
  \BibitemOpen
  \bibfield  {author} {\bibinfo {author} {\bibfnamefont {I.~V.}\ \bibnamefont {Igumenshchev}},\ }\bibfield  {title} {\bibinfo {title} {{Magnetically Arrested Disks and Origin of Poynting Jets: Numerical Study}},\ }\href {https://doi.org/10.1086/529025} {\bibfield  {journal} {\bibinfo  {journal} {Astrophys. J.}\ }\textbf {\bibinfo {volume} {677}},\ \bibinfo {pages} {317} (\bibinfo {year} {2008})},\ \Eprint {https://arxiv.org/abs/0711.4391} {arXiv:0711.4391 [astro-ph]} \BibitemShut {NoStop}%
\bibitem [{\citenamefont {Spruit}\ \emph {et~al.}(1995)\citenamefont {Spruit}, \citenamefont {Stehle},\ and\ \citenamefont {Papaloizou}}]{Spruit:1995fr}%
  \BibitemOpen
  \bibfield  {author} {\bibinfo {author} {\bibfnamefont {H.~C.}\ \bibnamefont {Spruit}}, \bibinfo {author} {\bibfnamefont {R.}~\bibnamefont {Stehle}},\ and\ \bibinfo {author} {\bibfnamefont {J.~C.~B.}\ \bibnamefont {Papaloizou}},\ }\bibfield  {title} {\bibinfo {title} {{Interchange instability in an accretion disc with a poloidal magnetic field}},\ }\href {https://doi.org/10.1093/mnras/275.4.1223} {\bibfield  {journal} {\bibinfo  {journal} {Mon. Not. Roy. Astron. Soc.}\ }\textbf {\bibinfo {volume} {275}},\ \bibinfo {pages} {1223} (\bibinfo {year} {1995})},\ \Eprint {https://arxiv.org/abs/astro-ph/9504043} {arXiv:astro-ph/9504043} \BibitemShut {NoStop}%
\bibitem [{\citenamefont {Kulkarni}\ and\ \citenamefont {Romanova}(2008)}]{Kulkarni:2008vk}%
  \BibitemOpen
  \bibfield  {author} {\bibinfo {author} {\bibfnamefont {A.~K.}\ \bibnamefont {Kulkarni}}\ and\ \bibinfo {author} {\bibfnamefont {M.~M.}\ \bibnamefont {Romanova}},\ }\bibfield  {title} {\bibinfo {title} {{Accretion to Magnetized Stars through the Rayleigh-Taylor Instability: Global Three-Dimensional Simulations}},\ }\href {https://doi.org/10.1111/j.1365-2966.2008.13094.x} {\bibfield  {journal} {\bibinfo  {journal} {Mon. Not. Roy. Astron. Soc.}\ }\textbf {\bibinfo {volume} {386}},\ \bibinfo {pages} {673} (\bibinfo {year} {2008})},\ \Eprint {https://arxiv.org/abs/0802.1759} {arXiv:0802.1759 [astro-ph]} \BibitemShut {NoStop}%
\bibitem [{\citenamefont {Begelman}\ \emph {et~al.}(2022)\citenamefont {Begelman}, \citenamefont {Scepi},\ and\ \citenamefont {Dexter}}]{Begelman:2021ufo}%
  \BibitemOpen
  \bibfield  {author} {\bibinfo {author} {\bibfnamefont {M.~C.}\ \bibnamefont {Begelman}}, \bibinfo {author} {\bibfnamefont {N.}~\bibnamefont {Scepi}},\ and\ \bibinfo {author} {\bibfnamefont {J.}~\bibnamefont {Dexter}},\ }\bibfield  {title} {\bibinfo {title} {{What really makes an accretion disc MAD}},\ }\href {https://doi.org/10.1093/mnras/stab3790} {\bibfield  {journal} {\bibinfo  {journal} {Mon. Not. Roy. Astron. Soc.}\ }\textbf {\bibinfo {volume} {511}},\ \bibinfo {pages} {2040} (\bibinfo {year} {2022})},\ \Eprint {https://arxiv.org/abs/2111.02439} {arXiv:2111.02439 [astro-ph.HE]} \BibitemShut {NoStop}%
\bibitem [{\citenamefont {Kaaz}\ \emph {et~al.}(2023)\citenamefont {Kaaz}, \citenamefont {Liska}, \citenamefont {Jacquemin-Ide}, \citenamefont {Andalman}, \citenamefont {Musoke}, \citenamefont {Tchekhovskoy},\ and\ \citenamefont {Porth}}]{Kaaz:2022fbg}%
  \BibitemOpen
  \bibfield  {author} {\bibinfo {author} {\bibfnamefont {N.}~\bibnamefont {Kaaz}}, \bibinfo {author} {\bibfnamefont {M.~T.~P.}\ \bibnamefont {Liska}}, \bibinfo {author} {\bibfnamefont {J.}~\bibnamefont {Jacquemin-Ide}}, \bibinfo {author} {\bibfnamefont {Z.~L.}\ \bibnamefont {Andalman}}, \bibinfo {author} {\bibfnamefont {G.}~\bibnamefont {Musoke}}, \bibinfo {author} {\bibfnamefont {A.}~\bibnamefont {Tchekhovskoy}},\ and\ \bibinfo {author} {\bibfnamefont {O.}~\bibnamefont {Porth}},\ }\bibfield  {title} {\bibinfo {title} {{Nozzle Shocks, Disk Tearing, and Streamers Drive Rapid Accretion in 3D GRMHD Simulations of Warped Thin Disks}},\ }\href {https://doi.org/10.3847/1538-4357/ace051} {\bibfield  {journal} {\bibinfo  {journal} {Astrophys. J.}\ }\textbf {\bibinfo {volume} {955}},\ \bibinfo {pages} {72} (\bibinfo {year} {2023})},\ \Eprint {https://arxiv.org/abs/2210.10053} {arXiv:2210.10053 [astro-ph.HE]} \BibitemShut {NoStop}%
\bibitem [{\citenamefont {White}\ \emph {et~al.}(2019)\citenamefont {White}, \citenamefont {Quataert},\ and\ \citenamefont {Blaes}}]{White:2019udt}%
  \BibitemOpen
  \bibfield  {author} {\bibinfo {author} {\bibfnamefont {C.~J.}\ \bibnamefont {White}}, \bibinfo {author} {\bibfnamefont {E.}~\bibnamefont {Quataert}},\ and\ \bibinfo {author} {\bibfnamefont {O.}~\bibnamefont {Blaes}},\ }\bibfield  {title} {\bibinfo {title} {{Tilted Disks around Black Holes: A Numerical Parameter Survey for Spin and Inclination Angle}},\ }\href {https://doi.org/10.3847/1538-4357/ab089e} {\bibfield  {journal} {\bibinfo  {journal} {Astrophys. J.}\ }\textbf {\bibinfo {volume} {878}},\ \bibinfo {pages} {51} (\bibinfo {year} {2019})},\ \Eprint {https://arxiv.org/abs/1902.09662} {arXiv:1902.09662 [astro-ph.HE]} \BibitemShut {NoStop}%
\bibitem [{\citenamefont {{Fragile}}\ \emph {et~al.}(2007)\citenamefont {{Fragile}}, \citenamefont {{Blaes}}, \citenamefont {{Anninos}},\ and\ \citenamefont {{Salmonson}}}]{2007ApJ...668..417F}%
  \BibitemOpen
  \bibfield  {author} {\bibinfo {author} {\bibfnamefont {P.~C.}\ \bibnamefont {{Fragile}}}, \bibinfo {author} {\bibfnamefont {O.~M.}\ \bibnamefont {{Blaes}}}, \bibinfo {author} {\bibfnamefont {P.}~\bibnamefont {{Anninos}}},\ and\ \bibinfo {author} {\bibfnamefont {J.~D.}\ \bibnamefont {{Salmonson}}},\ }\bibfield  {title} {\bibinfo {title} {{Global General Relativistic Magnetohydrodynamic Simulation of a Tilted Black Hole Accretion Disk}},\ }\href {https://doi.org/10.1086/521092} {\bibfield  {journal} {\bibinfo  {journal} {Astrophys. J.}\ }\textbf {\bibinfo {volume} {668}},\ \bibinfo {pages} {417} (\bibinfo {year} {2007})},\ \Eprint {https://arxiv.org/abs/0706.4303} {arXiv:0706.4303 [astro-ph]} \BibitemShut {NoStop}%
\bibitem [{\citenamefont {Palenzuela}\ \emph {et~al.}(2010)\citenamefont {Palenzuela}, \citenamefont {Lehner},\ and\ \citenamefont {Liebling}}]{Palenzuela:2010nf}%
  \BibitemOpen
  \bibfield  {author} {\bibinfo {author} {\bibfnamefont {C.}~\bibnamefont {Palenzuela}}, \bibinfo {author} {\bibfnamefont {L.}~\bibnamefont {Lehner}},\ and\ \bibinfo {author} {\bibfnamefont {S.~L.}\ \bibnamefont {Liebling}},\ }\bibfield  {title} {\bibinfo {title} {{Dual Jets from Binary Black Holes}},\ }\href {https://doi.org/10.1126/science.1191766} {\bibfield  {journal} {\bibinfo  {journal} {Science}\ }\textbf {\bibinfo {volume} {329}},\ \bibinfo {pages} {927} (\bibinfo {year} {2010})},\ \Eprint {https://arxiv.org/abs/1005.1067} {arXiv:1005.1067 [astro-ph.HE]} \BibitemShut {NoStop}%
\bibitem [{\citenamefont {Ressler}\ \emph {et~al.}(2025)\citenamefont {Ressler}, \citenamefont {Combi}, \citenamefont {Ripperda},\ and\ \citenamefont {Most}}]{Ressler:2024tan}%
  \BibitemOpen
  \bibfield  {author} {\bibinfo {author} {\bibfnamefont {S.~M.}\ \bibnamefont {Ressler}}, \bibinfo {author} {\bibfnamefont {L.}~\bibnamefont {Combi}}, \bibinfo {author} {\bibfnamefont {B.}~\bibnamefont {Ripperda}},\ and\ \bibinfo {author} {\bibfnamefont {E.~R.}\ \bibnamefont {Most}},\ }\bibfield  {title} {\bibinfo {title} {{Dual Jet Interaction, Magnetically Arrested Flows, and Flares in Accreting Binary Black Holes}},\ }\href {https://doi.org/10.3847/2041-8213/ad9eb5} {\bibfield  {journal} {\bibinfo  {journal} {Astrophys. J. Lett.}\ }\textbf {\bibinfo {volume} {979}},\ \bibinfo {pages} {L24} (\bibinfo {year} {2025})},\ \Eprint {https://arxiv.org/abs/2410.10944} {arXiv:2410.10944 [astro-ph.HE]} \BibitemShut {NoStop}%
\bibitem [{\citenamefont {{Lin}}\ and\ \citenamefont {{Papaloizou}}(1979)}]{lin1979}%
  \BibitemOpen
  \bibfield  {author} {\bibinfo {author} {\bibfnamefont {D.~N.~C.}\ \bibnamefont {{Lin}}}\ and\ \bibinfo {author} {\bibfnamefont {J.}~\bibnamefont {{Papaloizou}}},\ }\bibfield  {title} {\bibinfo {title} {{Tidal torques on accretion discs in binary systems with extreme mass ratios.}},\ }\href {https://doi.org/10.1093/mnras/186.4.799} {\bibfield  {journal} {\bibinfo  {journal} {Mon. Not. Roy. Astron. Soc.}\ }\textbf {\bibinfo {volume} {186}},\ \bibinfo {pages} {799} (\bibinfo {year} {1979})}\BibitemShut {NoStop}%
\bibitem [{\citenamefont {{Lin}}\ and\ \citenamefont {{Papaloizou}}(1993)}]{lin1993}%
  \BibitemOpen
  \bibfield  {author} {\bibinfo {author} {\bibfnamefont {D.~N.~C.}\ \bibnamefont {{Lin}}}\ and\ \bibinfo {author} {\bibfnamefont {J.~C.~B.}\ \bibnamefont {{Papaloizou}}},\ }\bibfield  {title} {\bibinfo {title} {{On the Tidal Interaction Between Protostellar Disks and Companions}},\ }in\ \href@noop {} {\emph {\bibinfo {booktitle} {Protostars and Planets III}}},\ \bibinfo {editor} {edited by\ \bibinfo {editor} {\bibfnamefont {E.~H.}\ \bibnamefont {{Levy}}}\ and\ \bibinfo {editor} {\bibfnamefont {J.~I.}\ \bibnamefont {{Lunine}}}}\ (\bibinfo {year} {1993})\ p.\ \bibinfo {pages} {749}\BibitemShut {NoStop}%
\bibitem [{\citenamefont {Artymowicz}\ and\ \citenamefont {Lubow}(1996)}]{Artymowicz:1996zz}%
  \BibitemOpen
  \bibfield  {author} {\bibinfo {author} {\bibfnamefont {P.}~\bibnamefont {Artymowicz}}\ and\ \bibinfo {author} {\bibfnamefont {S.~H.}\ \bibnamefont {Lubow}},\ }\bibfield  {title} {\bibinfo {title} {{Mass Flow through Gaps in Circumbinary Disks}},\ }\href {https://doi.org/10.1086/310200} {\bibfield  {journal} {\bibinfo  {journal} {Astrophys. J. Lett.}\ }\textbf {\bibinfo {volume} {467}},\ \bibinfo {pages} {L77} (\bibinfo {year} {1996})}\BibitemShut {NoStop}%
\bibitem [{\citenamefont {Nixon}\ \emph {et~al.}(2011{\natexlab{b}})\citenamefont {Nixon}, \citenamefont {King},\ and\ \citenamefont {Pringle}}]{Nixon:2011tn}%
  \BibitemOpen
  \bibfield  {author} {\bibinfo {author} {\bibfnamefont {C.}~\bibnamefont {Nixon}}, \bibinfo {author} {\bibfnamefont {A.}~\bibnamefont {King}},\ and\ \bibinfo {author} {\bibfnamefont {J.}~\bibnamefont {Pringle}},\ }\bibfield  {title} {\bibinfo {title} {{The final parsec problem: aligning a binary with an external accretion disc}},\ }\href {https://doi.org/10.1111/j.1745-3933.2011.01121.x} {\bibfield  {journal} {\bibinfo  {journal} {Mon. Not. Roy. Astron. Soc.}\ }\textbf {\bibinfo {volume} {417}},\ \bibinfo {pages} {66} (\bibinfo {year} {2011}{\natexlab{b}})},\ \Eprint {https://arxiv.org/abs/1107.5056} {arXiv:1107.5056 [astro-ph.GA]} \BibitemShut {NoStop}%
\bibitem [{\citenamefont {Martin}\ \emph {et~al.}(2023)\citenamefont {Martin}, \citenamefont {Lepp}, \citenamefont {Zhang}, \citenamefont {Nixon},\ and\ \citenamefont {Childs}}]{Martin:2023dfp}%
  \BibitemOpen
  \bibfield  {author} {\bibinfo {author} {\bibfnamefont {R.~G.}\ \bibnamefont {Martin}}, \bibinfo {author} {\bibfnamefont {S.}~\bibnamefont {Lepp}}, \bibinfo {author} {\bibfnamefont {B.}~\bibnamefont {Zhang}}, \bibinfo {author} {\bibfnamefont {C.~J.}\ \bibnamefont {Nixon}},\ and\ \bibinfo {author} {\bibfnamefont {A.~C.}\ \bibnamefont {Childs}},\ }\bibfield  {title} {\bibinfo {title} {{Mergers of black hole binaries driven by misaligned circumbinary discs}},\ }\href {https://doi.org/10.1093/mnrasl/slad174} {\bibfield  {journal} {\bibinfo  {journal} {Mon. Not. Roy. Astron. Soc.}\ }\textbf {\bibinfo {volume} {528}},\ \bibinfo {pages} {L161} (\bibinfo {year} {2023})},\ \Eprint {https://arxiv.org/abs/2311.10160} {arXiv:2311.10160 [astro-ph.HE]} \BibitemShut {NoStop}%
\bibitem [{\citenamefont {{Hopkins}}(2024)}]{Hopkins2024}%
  \BibitemOpen
  \bibfield  {author} {\bibinfo {author} {\bibfnamefont {P.~F.}\ \bibnamefont {{Hopkins}}},\ }\bibfield  {title} {\bibinfo {title} {{Multi-Phase Thermal Structure \& The Origin of the Broad-Line Region, Torus, and Corona in Magnetically-Dominated Accretion Disks}},\ }\href {https://doi.org/10.48550/arXiv.2407.00160} {\bibfield  {journal} {\bibinfo  {journal} {arXiv e-prints}\ ,\ \bibinfo {eid} {arXiv:2407.00160}} (\bibinfo {year} {2024})},\ \Eprint {https://arxiv.org/abs/2407.00160} {arXiv:2407.00160 [astro-ph.GA]} \BibitemShut {NoStop}%
\bibitem [{\citenamefont {{Kelley}}\ \emph {et~al.}(2017)\citenamefont {{Kelley}}, \citenamefont {{Blecha}},\ and\ \citenamefont {{Hernquist}}}]{2017MNRAS.464.3131K}%
  \BibitemOpen
  \bibfield  {author} {\bibinfo {author} {\bibfnamefont {L.~Z.}\ \bibnamefont {{Kelley}}}, \bibinfo {author} {\bibfnamefont {L.}~\bibnamefont {{Blecha}}},\ and\ \bibinfo {author} {\bibfnamefont {L.}~\bibnamefont {{Hernquist}}},\ }\bibfield  {title} {\bibinfo {title} {{Massive black hole binary mergers in dynamical galactic environments}},\ }\href {https://doi.org/10.1093/mnras/stw2452} {\bibfield  {journal} {\bibinfo  {journal} {"Mon. Not. Roy. Astron. Soc.",}\ }\textbf {\bibinfo {volume} {464}},\ \bibinfo {pages} {3131} (\bibinfo {year} {2017})},\ \Eprint {https://arxiv.org/abs/1606.01900} {arXiv:1606.01900 [astro-ph.HE]} \BibitemShut {NoStop}%
\bibitem [{\citenamefont {{Amaro-Seoane}}\ \emph {et~al.}(2023{\natexlab{b}})\citenamefont {{Amaro-Seoane}}, \citenamefont {{Andrews}}, \citenamefont {{Arca Sedda}}, \citenamefont {{Askar}}, \citenamefont {{Baghi}}, \citenamefont {{Balasov}}, \citenamefont {{Bartos}}, \citenamefont {{Bavera}}, \citenamefont {{Bellovary}}, \citenamefont {{Berry}}, \citenamefont {{Berti}}, \citenamefont {{Bianchi}}, \citenamefont {{Blecha}}, \citenamefont {{Blondin}}, \citenamefont {{Bogdanovi{\'c}}}, \citenamefont {{Boissier}}, \citenamefont {{Bonetti}}, \citenamefont {{Bonoli}}, \citenamefont {{Bortolas}}, \citenamefont {{Breivik}}, \citenamefont {{Capelo}}, \citenamefont {{Caramete}}, \citenamefont {{Cattorini}}, \citenamefont {{Charisi}}, \citenamefont {{Chaty}}, \citenamefont {{Chen}}, \citenamefont {{Chru{\'s}li{\'n}ska}}, \citenamefont {{Chua}}, \citenamefont {{Church}}, \citenamefont {{Colpi}}, \citenamefont {{D'Orazio}}, \citenamefont {{Danielski}}, \citenamefont {{Davies}}, \citenamefont {{Dayal}}, \citenamefont {{De Rosa}}, \citenamefont {{Derdzinski}}, \citenamefont {{Destounis}}, \citenamefont {{Dotti}}, \citenamefont {{Du{\c{t}}an}}, \citenamefont {{Dvorkin}}, \citenamefont {{Fabj}}, \citenamefont {{Foglizzo}}, \citenamefont {{Ford}}, \citenamefont {{Fouvry}}, \citenamefont {{Franchini}}, \citenamefont {{Fragos}}, \citenamefont {{Fryer}}, \citenamefont {{Gaspari}}, \citenamefont {{Gerosa}}, \citenamefont {{Graziani}}, \citenamefont {{Groot}}, \citenamefont {{Habouzit}}, \citenamefont {{Haggard}}, \citenamefont {{Haiman}}, \citenamefont {{Han}}, \citenamefont {{Istrate}}, \citenamefont {{Johansson}}, \citenamefont {{Khan}}, \citenamefont {{Kimpson}}, \citenamefont {{Kokkotas}}, \citenamefont {{Kong}}, \citenamefont {{Korol}}, \citenamefont {{Kremer}}, \citenamefont {{Kupfer}}, \citenamefont {{Lamberts}}, \citenamefont {{Larson}}, \citenamefont {{Lau}}, \citenamefont {{Liu}}, \citenamefont {{Lloyd-Ronning}}, \citenamefont {{Lodato}}, \citenamefont {{Lupi}}, \citenamefont {{Ma}}, \citenamefont {{Maccarone}}, \citenamefont {{Mandel}}, \citenamefont {{Mangiagli}}, \citenamefont {{Mapelli}}, \citenamefont {{Mathis}}, \citenamefont {{Mayer}}, \citenamefont {{McGee}}, \citenamefont {{McKernan}}, \citenamefont {{Miller}}, \citenamefont {{Mota}}, \citenamefont {{Mumpower}}, \citenamefont {{Nasim}}, \citenamefont {{Nelemans}}, \citenamefont {{Noble}}, \citenamefont {{Pacucci}}, \citenamefont {{Panessa}}, \citenamefont {{Paschalidis}}, \citenamefont {{Pfister}}, \citenamefont {{Porquet}}, \citenamefont {{Quenby}}, \citenamefont {{Ricarte}}, \citenamefont {{R{\"o}pke}}, \citenamefont {{Regan}}, \citenamefont {{Rosswog}}, \citenamefont {{Ruiter}}, \citenamefont {{Ruiz}}, \citenamefont {{Runnoe}}, \citenamefont {{Schneider}}, \citenamefont {{Schnittman}}, \citenamefont {{Secunda}}, \citenamefont {{Sesana}}, \citenamefont {{Seto}}, \citenamefont {{Shao}}, \citenamefont {{Shapiro}}, \citenamefont {{Sopuerta}}, \citenamefont {{Stone}}, \citenamefont {{Suvorov}}, \citenamefont {{Tamanini}}, \citenamefont {{Tamfal}}, \citenamefont {{Tauris}}, \citenamefont {{Temmink}}, \citenamefont {{Tomsick}}, \citenamefont {{Toonen}}, \citenamefont {{Torres-Orjuela}}, \citenamefont {{Toscani}}, \citenamefont {{Tsokaros}}, \citenamefont {{Unal}}, \citenamefont {{V{\'a}zquez-Aceves}}, \citenamefont {{Valiante}}, \citenamefont {{van Putten}}, \citenamefont {{van Roestel}}, \citenamefont {{Vignali}}, \citenamefont {{Volonteri}}, \citenamefont {{Wu}}, \citenamefont {{Younsi}}, \citenamefont {{Yu}}, \citenamefont {{Zane}}, \citenamefont {{Zwick}}, \citenamefont {{Antonini}}, \citenamefont {{Baibhav}}, \citenamefont {{Barausse}}, \citenamefont {{Bonilla Rivera}}, \citenamefont {{Branchesi}}, \citenamefont {{Branduardi-Raymont}}, \citenamefont {{Burdge}}, \citenamefont {{Chakraborty}}, \citenamefont {{Cuadra}}, \citenamefont {{Dage}}, \citenamefont {{Davis}}, \citenamefont {{de Mink}}, \citenamefont {{Decarli}}, \citenamefont {{Doneva}}, \citenamefont {{Escoffier}}, \citenamefont {{Gandhi}}, \citenamefont {{Haardt}}, \citenamefont {{Lousto}}, \citenamefont {{Nissanke}}, \citenamefont {{Nordhaus}}, \citenamefont {{O'Shaughnessy}}, \citenamefont {{Portegies Zwart}}, \citenamefont {{Pound}}, \citenamefont {{Schussler}}, \citenamefont {{Sergijenko}}, \citenamefont {{Spallicci}}, \citenamefont {{Vernieri}},\ and\ \citenamefont {{Vigna-G{\'o}mez}}}]{2023LRR....26....2A}%
  \BibitemOpen
  \bibfield  {author} {\bibinfo {author} {\bibfnamefont {P.}~\bibnamefont {{Amaro-Seoane}}}, \bibinfo {author} {\bibfnamefont {J.}~\bibnamefont {{Andrews}}}, \bibinfo {author} {\bibfnamefont {M.}~\bibnamefont {{Arca Sedda}}}, \bibinfo {author} {\bibfnamefont {A.}~\bibnamefont {{Askar}}}, \bibinfo {author} {\bibfnamefont {Q.}~\bibnamefont {{Baghi}}}, \bibinfo {author} {\bibfnamefont {R.}~\bibnamefont {{Balasov}}}, \bibinfo {author} {\bibfnamefont {I.}~\bibnamefont {{Bartos}}}, \bibinfo {author} {\bibfnamefont {S.~S.}\ \bibnamefont {{Bavera}}}, \bibinfo {author} {\bibfnamefont {J.}~\bibnamefont {{Bellovary}}}, \bibinfo {author} {\bibfnamefont {C.~P.~L.}\ \bibnamefont {{Berry}}}, \bibinfo {author} {\bibfnamefont {E.}~\bibnamefont {{Berti}}}, \bibinfo {author} {\bibfnamefont {S.}~\bibnamefont {{Bianchi}}}, \bibinfo {author} {\bibfnamefont {L.}~\bibnamefont {{Blecha}}}, \bibinfo {author} {\bibfnamefont {S.}~\bibnamefont {{Blondin}}}, \bibinfo {author} {\bibfnamefont {T.}~\bibnamefont {{Bogdanovi{\'c}}}}, \bibinfo {author} {\bibfnamefont {S.}~\bibnamefont {{Boissier}}}, \bibinfo {author} {\bibfnamefont {M.}~\bibnamefont {{Bonetti}}}, \bibinfo {author} {\bibfnamefont {S.}~\bibnamefont {{Bonoli}}}, \bibinfo {author} {\bibfnamefont {E.}~\bibnamefont {{Bortolas}}}, \bibinfo {author} {\bibfnamefont {K.}~\bibnamefont {{Breivik}}}, \bibinfo {author} {\bibfnamefont {P.~R.}\ \bibnamefont {{Capelo}}}, \bibinfo {author} {\bibfnamefont {L.}~\bibnamefont {{Caramete}}}, \bibinfo {author} {\bibfnamefont {F.}~\bibnamefont {{Cattorini}}}, \bibinfo {author} {\bibfnamefont {M.}~\bibnamefont {{Charisi}}}, \bibinfo {author} {\bibfnamefont {S.}~\bibnamefont {{Chaty}}}, \bibinfo {author} {\bibfnamefont {X.}~\bibnamefont {{Chen}}}, \bibinfo {author} {\bibfnamefont {M.}~\bibnamefont {{Chru{\'s}li{\'n}ska}}}, \bibinfo {author} {\bibfnamefont {A.~J.~K.}\ \bibnamefont {{Chua}}}, \bibinfo {author} {\bibfnamefont {R.}~\bibnamefont {{Church}}}, \bibinfo {author} {\bibfnamefont {M.}~\bibnamefont {{Colpi}}}, \bibinfo {author} {\bibfnamefont {D.}~\bibnamefont {{D'Orazio}}}, \bibinfo {author} {\bibfnamefont {C.}~\bibnamefont {{Danielski}}}, \bibinfo {author} {\bibfnamefont {M.~B.}\ \bibnamefont {{Davies}}}, \bibinfo {author} {\bibfnamefont {P.}~\bibnamefont {{Dayal}}}, \bibinfo {author} {\bibfnamefont {A.}~\bibnamefont {{De Rosa}}}, \bibinfo {author} {\bibfnamefont {A.}~\bibnamefont {{Derdzinski}}}, \bibinfo {author} {\bibfnamefont {K.}~\bibnamefont {{Destounis}}}, \bibinfo {author} {\bibfnamefont {M.}~\bibnamefont {{Dotti}}}, \bibinfo {author} {\bibfnamefont {I.}~\bibnamefont {{Du{\c{t}}an}}}, \bibinfo {author} {\bibfnamefont {I.}~\bibnamefont {{Dvorkin}}}, \bibinfo {author} {\bibfnamefont {G.}~\bibnamefont {{Fabj}}}, \bibinfo {author} {\bibfnamefont {T.}~\bibnamefont {{Foglizzo}}}, \bibinfo {author} {\bibfnamefont {S.}~\bibnamefont {{Ford}}}, \bibinfo {author} {\bibfnamefont {J.-B.}\ \bibnamefont {{Fouvry}}}, \bibinfo {author} {\bibfnamefont {A.}~\bibnamefont {{Franchini}}}, \bibinfo {author} {\bibfnamefont {T.}~\bibnamefont {{Fragos}}}, \bibinfo {author} {\bibfnamefont {C.}~\bibnamefont {{Fryer}}}, \bibinfo {author} {\bibfnamefont {M.}~\bibnamefont {{Gaspari}}}, \bibinfo {author} {\bibfnamefont {D.}~\bibnamefont {{Gerosa}}}, \bibinfo {author} {\bibfnamefont {L.}~\bibnamefont {{Graziani}}}, \bibinfo {author} {\bibfnamefont {P.}~\bibnamefont {{Groot}}}, \bibinfo {author} {\bibfnamefont {M.}~\bibnamefont {{Habouzit}}}, \bibinfo {author} {\bibfnamefont {D.}~\bibnamefont {{Haggard}}}, \bibinfo {author} {\bibfnamefont {Z.}~\bibnamefont {{Haiman}}}, \bibinfo {author} {\bibfnamefont {W.-B.}\ \bibnamefont {{Han}}}, \bibinfo {author} {\bibfnamefont {A.}~\bibnamefont {{Istrate}}}, \bibinfo {author} {\bibfnamefont {P.~H.}\ \bibnamefont {{Johansson}}}, \bibinfo {author} {\bibfnamefont {F.~M.}\ \bibnamefont {{Khan}}}, \bibinfo {author} {\bibfnamefont {T.}~\bibnamefont {{Kimpson}}}, \bibinfo {author} {\bibfnamefont {K.}~\bibnamefont {{Kokkotas}}}, \bibinfo {author} {\bibfnamefont {A.}~\bibnamefont {{Kong}}}, \bibinfo {author} {\bibfnamefont {V.}~\bibnamefont {{Korol}}}, \bibinfo {author} {\bibfnamefont {K.}~\bibnamefont {{Kremer}}}, \bibinfo {author} {\bibfnamefont {T.}~\bibnamefont {{Kupfer}}}, \bibinfo {author} {\bibfnamefont {A.}~\bibnamefont {{Lamberts}}}, \bibinfo {author} {\bibfnamefont {S.}~\bibnamefont {{Larson}}}, \bibinfo {author} {\bibfnamefont {M.}~\bibnamefont {{Lau}}}, \bibinfo {author} {\bibfnamefont {D.}~\bibnamefont {{Liu}}}, \bibinfo {author} {\bibfnamefont {N.}~\bibnamefont {{Lloyd-Ronning}}}, \bibinfo {author} {\bibfnamefont {G.}~\bibnamefont {{Lodato}}}, \bibinfo {author} {\bibfnamefont {A.}~\bibnamefont {{Lupi}}}, \bibinfo {author} {\bibfnamefont {C.-P.}\ \bibnamefont {{Ma}}}, \bibinfo {author} {\bibfnamefont {T.}~\bibnamefont {{Maccarone}}}, \bibinfo {author} {\bibfnamefont {I.}~\bibnamefont {{Mandel}}}, \bibinfo {author} {\bibfnamefont {A.}~\bibnamefont {{Mangiagli}}}, \bibinfo {author} {\bibfnamefont {M.}~\bibnamefont {{Mapelli}}}, \bibinfo {author} {\bibfnamefont {S.}~\bibnamefont {{Mathis}}}, \bibinfo {author} {\bibfnamefont {L.}~\bibnamefont {{Mayer}}}, \bibinfo {author} {\bibfnamefont {S.}~\bibnamefont {{McGee}}}, \bibinfo {author} {\bibfnamefont {B.}~\bibnamefont {{McKernan}}}, \bibinfo {author} {\bibfnamefont {M.~C.}\ \bibnamefont {{Miller}}}, \bibinfo {author} {\bibfnamefont {D.~F.}\ \bibnamefont {{Mota}}}, \bibinfo {author} {\bibfnamefont {M.}~\bibnamefont {{Mumpower}}}, \bibinfo {author} {\bibfnamefont {S.~S.}\ \bibnamefont {{Nasim}}}, \bibinfo {author} {\bibfnamefont {G.}~\bibnamefont {{Nelemans}}}, \bibinfo {author} {\bibfnamefont {S.}~\bibnamefont {{Noble}}}, \bibinfo {author} {\bibfnamefont {F.}~\bibnamefont {{Pacucci}}}, \bibinfo {author} {\bibfnamefont {F.}~\bibnamefont {{Panessa}}}, \bibinfo {author} {\bibfnamefont {V.}~\bibnamefont {{Paschalidis}}}, \bibinfo {author} {\bibfnamefont {H.}~\bibnamefont {{Pfister}}}, \bibinfo {author} {\bibfnamefont {D.}~\bibnamefont {{Porquet}}}, \bibinfo {author} {\bibfnamefont {J.}~\bibnamefont {{Quenby}}}, \bibinfo {author} {\bibfnamefont {A.}~\bibnamefont {{Ricarte}}}, \bibinfo {author} {\bibfnamefont {F.~K.}\ \bibnamefont {{R{\"o}pke}}}, \bibinfo {author} {\bibfnamefont {J.}~\bibnamefont {{Regan}}}, \bibinfo {author} {\bibfnamefont {S.}~\bibnamefont {{Rosswog}}}, \bibinfo {author} {\bibfnamefont {A.}~\bibnamefont {{Ruiter}}}, \bibinfo {author} {\bibfnamefont {M.}~\bibnamefont {{Ruiz}}}, \bibinfo {author} {\bibfnamefont {J.}~\bibnamefont {{Runnoe}}}, \bibinfo {author} {\bibfnamefont {R.}~\bibnamefont {{Schneider}}}, \bibinfo {author} {\bibfnamefont {J.}~\bibnamefont {{Schnittman}}}, \bibinfo {author} {\bibfnamefont {A.}~\bibnamefont {{Secunda}}}, \bibinfo {author} {\bibfnamefont {A.}~\bibnamefont {{Sesana}}}, \bibinfo {author} {\bibfnamefont {N.}~\bibnamefont {{Seto}}}, \bibinfo {author} {\bibfnamefont {L.}~\bibnamefont {{Shao}}}, \bibinfo {author} {\bibfnamefont {S.}~\bibnamefont {{Shapiro}}}, \bibinfo {author} {\bibfnamefont {C.}~\bibnamefont {{Sopuerta}}}, \bibinfo {author} {\bibfnamefont {N.~C.}\ \bibnamefont {{Stone}}}, \bibinfo {author} {\bibfnamefont {A.}~\bibnamefont {{Suvorov}}}, \bibinfo {author} {\bibfnamefont {N.}~\bibnamefont {{Tamanini}}}, \bibinfo {author} {\bibfnamefont {T.}~\bibnamefont {{Tamfal}}}, \bibinfo {author} {\bibfnamefont {T.}~\bibnamefont {{Tauris}}}, \bibinfo {author} {\bibfnamefont {K.}~\bibnamefont {{Temmink}}}, \bibinfo {author} {\bibfnamefont {J.}~\bibnamefont {{Tomsick}}}, \bibinfo {author} {\bibfnamefont {S.}~\bibnamefont {{Toonen}}}, \bibinfo {author} {\bibfnamefont {A.}~\bibnamefont {{Torres-Orjuela}}}, \bibinfo {author} {\bibfnamefont {M.}~\bibnamefont {{Toscani}}}, \bibinfo {author} {\bibfnamefont {A.}~\bibnamefont {{Tsokaros}}}, \bibinfo {author} {\bibfnamefont {C.}~\bibnamefont {{Unal}}}, \bibinfo {author} {\bibfnamefont {V.}~\bibnamefont {{V{\'a}zquez-Aceves}}}, \bibinfo {author} {\bibfnamefont {R.}~\bibnamefont {{Valiante}}}, \bibinfo {author} {\bibfnamefont {M.}~\bibnamefont {{van Putten}}}, \bibinfo {author} {\bibfnamefont {J.}~\bibnamefont {{van Roestel}}}, \bibinfo {author} {\bibfnamefont {C.}~\bibnamefont {{Vignali}}}, \bibinfo {author} {\bibfnamefont {M.}~\bibnamefont {{Volonteri}}}, \bibinfo {author} {\bibfnamefont {K.}~\bibnamefont {{Wu}}}, \bibinfo {author} {\bibfnamefont {Z.}~\bibnamefont {{Younsi}}}, \bibinfo {author} {\bibfnamefont {S.}~\bibnamefont {{Yu}}}, \bibinfo {author} {\bibfnamefont {S.}~\bibnamefont {{Zane}}}, \bibinfo {author} {\bibfnamefont {L.}~\bibnamefont {{Zwick}}}, \bibinfo {author} {\bibfnamefont {F.}~\bibnamefont {{Antonini}}}, \bibinfo {author} {\bibfnamefont {V.}~\bibnamefont {{Baibhav}}}, \bibinfo {author} {\bibfnamefont {E.}~\bibnamefont {{Barausse}}}, \bibinfo {author} {\bibfnamefont {A.}~\bibnamefont {{Bonilla Rivera}}}, \bibinfo {author} {\bibfnamefont {M.}~\bibnamefont {{Branchesi}}}, \bibinfo {author} {\bibfnamefont {G.}~\bibnamefont {{Branduardi-Raymont}}}, \bibinfo {author} {\bibfnamefont {K.}~\bibnamefont {{Burdge}}}, \bibinfo {author} {\bibfnamefont {S.}~\bibnamefont {{Chakraborty}}}, \bibinfo {author} {\bibfnamefont {J.}~\bibnamefont {{Cuadra}}}, \bibinfo {author} {\bibfnamefont {K.}~\bibnamefont {{Dage}}}, \bibinfo {author} {\bibfnamefont {B.}~\bibnamefont {{Davis}}}, \bibinfo {author} {\bibfnamefont {S.~E.}\ \bibnamefont {{de Mink}}}, \bibinfo {author} {\bibfnamefont {R.}~\bibnamefont {{Decarli}}}, \bibinfo {author} {\bibfnamefont {D.}~\bibnamefont {{Doneva}}}, \bibinfo {author} {\bibfnamefont {S.}~\bibnamefont {{Escoffier}}}, \bibinfo {author} {\bibfnamefont {P.}~\bibnamefont {{Gandhi}}}, \bibinfo {author} {\bibfnamefont {F.}~\bibnamefont {{Haardt}}}, \bibinfo {author} {\bibfnamefont {C.~O.}\ \bibnamefont {{Lousto}}}, \bibinfo {author} {\bibfnamefont {S.}~\bibnamefont {{Nissanke}}}, \bibinfo {author} {\bibfnamefont {J.}~\bibnamefont {{Nordhaus}}}, \bibinfo {author} {\bibfnamefont {R.}~\bibnamefont {{O'Shaughnessy}}}, \bibinfo {author} {\bibfnamefont {S.}~\bibnamefont {{Portegies Zwart}}}, \bibinfo
  {author} {\bibfnamefont {A.}~\bibnamefont {{Pound}}}, \bibinfo {author} {\bibfnamefont {F.}~\bibnamefont {{Schussler}}}, \bibinfo {author} {\bibfnamefont {O.}~\bibnamefont {{Sergijenko}}}, \bibinfo {author} {\bibfnamefont {A.}~\bibnamefont {{Spallicci}}}, \bibinfo {author} {\bibfnamefont {D.}~\bibnamefont {{Vernieri}}},\ and\ \bibinfo {author} {\bibfnamefont {A.}~\bibnamefont {{Vigna-G{\'o}mez}}},\ }\bibfield  {title} {\bibinfo {title} {{Astrophysics with the Laser Interferometer Space Antenna}},\ }\href {https://doi.org/10.1007/s41114-022-00041-y} {\bibfield  {journal} {\bibinfo  {journal} {Living Reviews in Relativity}\ }\textbf {\bibinfo {volume} {26}},\ \bibinfo {eid} {2} (\bibinfo {year} {2023}{\natexlab{b}})},\ \Eprint {https://arxiv.org/abs/2203.06016} {arXiv:2203.06016 [gr-qc]} \BibitemShut {NoStop}%
\bibitem [{\citenamefont {{De Rosa}}\ \emph {et~al.}(2019)\citenamefont {{De Rosa}}, \citenamefont {{Vignali}}, \citenamefont {{Bogdanovi{\'c}}}, \citenamefont {{Capelo}}, \citenamefont {{Charisi}}, \citenamefont {{Dotti}}, \citenamefont {{Husemann}}, \citenamefont {{Lusso}}, \citenamefont {{Mayer}}, \citenamefont {{Paragi}}, \citenamefont {{Runnoe}}, \citenamefont {{Sesana}}, \citenamefont {{Steinborn}}, \citenamefont {{Bianchi}}, \citenamefont {{Colpi}}, \citenamefont {{del Valle}}, \citenamefont {{Frey}}, \citenamefont {{Gab{\'a}nyi}}, \citenamefont {{Giustini}}, \citenamefont {{Guainazzi}}, \citenamefont {{Haiman}}, \citenamefont {{Herrera Ruiz}}, \citenamefont {{Herrero-Illana}}, \citenamefont {{Iwasawa}}, \citenamefont {{Komossa}}, \citenamefont {{Lena}}, \citenamefont {{Loiseau}}, \citenamefont {{Perez-Torres}}, \citenamefont {{Piconcelli}},\ and\ \citenamefont {{Volonteri}}}]{2019NewAR..8601525D}%
  \BibitemOpen
  \bibfield  {author} {\bibinfo {author} {\bibfnamefont {A.}~\bibnamefont {{De Rosa}}}, \bibinfo {author} {\bibfnamefont {C.}~\bibnamefont {{Vignali}}}, \bibinfo {author} {\bibfnamefont {T.}~\bibnamefont {{Bogdanovi{\'c}}}}, \bibinfo {author} {\bibfnamefont {P.~R.}\ \bibnamefont {{Capelo}}}, \bibinfo {author} {\bibfnamefont {M.}~\bibnamefont {{Charisi}}}, \bibinfo {author} {\bibfnamefont {M.}~\bibnamefont {{Dotti}}}, \bibinfo {author} {\bibfnamefont {B.}~\bibnamefont {{Husemann}}}, \bibinfo {author} {\bibfnamefont {E.}~\bibnamefont {{Lusso}}}, \bibinfo {author} {\bibfnamefont {L.}~\bibnamefont {{Mayer}}}, \bibinfo {author} {\bibfnamefont {Z.}~\bibnamefont {{Paragi}}}, \bibinfo {author} {\bibfnamefont {J.}~\bibnamefont {{Runnoe}}}, \bibinfo {author} {\bibfnamefont {A.}~\bibnamefont {{Sesana}}}, \bibinfo {author} {\bibfnamefont {L.}~\bibnamefont {{Steinborn}}}, \bibinfo {author} {\bibfnamefont {S.}~\bibnamefont {{Bianchi}}}, \bibinfo {author} {\bibfnamefont {M.}~\bibnamefont {{Colpi}}}, \bibinfo {author} {\bibfnamefont {L.}~\bibnamefont {{del Valle}}}, \bibinfo {author} {\bibfnamefont {S.}~\bibnamefont {{Frey}}}, \bibinfo {author} {\bibfnamefont {K.~{\'E}.}\ \bibnamefont {{Gab{\'a}nyi}}}, \bibinfo {author} {\bibfnamefont {M.}~\bibnamefont {{Giustini}}}, \bibinfo {author} {\bibfnamefont {M.}~\bibnamefont {{Guainazzi}}}, \bibinfo {author} {\bibfnamefont {Z.}~\bibnamefont {{Haiman}}}, \bibinfo {author} {\bibfnamefont {N.}~\bibnamefont {{Herrera Ruiz}}}, \bibinfo {author} {\bibfnamefont {R.}~\bibnamefont {{Herrero-Illana}}}, \bibinfo {author} {\bibfnamefont {K.}~\bibnamefont {{Iwasawa}}}, \bibinfo {author} {\bibfnamefont {S.}~\bibnamefont {{Komossa}}}, \bibinfo {author} {\bibfnamefont {D.}~\bibnamefont {{Lena}}}, \bibinfo {author} {\bibfnamefont {N.}~\bibnamefont {{Loiseau}}}, \bibinfo {author} {\bibfnamefont {M.}~\bibnamefont {{Perez-Torres}}}, \bibinfo {author} {\bibfnamefont {E.}~\bibnamefont {{Piconcelli}}},\ and\ \bibinfo {author} {\bibfnamefont {M.}~\bibnamefont {{Volonteri}}},\ }\bibfield  {title} {\bibinfo {title} {{The quest for dual and binary supermassive black holes: A multi-messenger view}},\ }\href {https://doi.org/10.1016/j.newar.2020.101525} {\bibfield  {journal} {\bibinfo  {journal} {New Astron. Rev.}\ }\textbf {\bibinfo {volume} {86}},\ \bibinfo {eid} {101525} (\bibinfo {year} {2019})},\ \Eprint {https://arxiv.org/abs/2001.06293} {arXiv:2001.06293 [astro-ph.GA]} \BibitemShut {NoStop}%
\bibitem [{\citenamefont {{Deane}}\ \emph {et~al.}(2014)\citenamefont {{Deane}}, \citenamefont {{Paragi}}, \citenamefont {{Jarvis}}, \citenamefont {{Coriat}}, \citenamefont {{Bernardi}}, \citenamefont {{Fender}}, \citenamefont {{Frey}}, \citenamefont {{Heywood}}, \citenamefont {{Kl{\"o}ckner}}, \citenamefont {{Grainge}},\ and\ \citenamefont {{Rumsey}}}]{2014Natur.511...57D}%
  \BibitemOpen
  \bibfield  {author} {\bibinfo {author} {\bibfnamefont {R.~P.}\ \bibnamefont {{Deane}}}, \bibinfo {author} {\bibfnamefont {Z.}~\bibnamefont {{Paragi}}}, \bibinfo {author} {\bibfnamefont {M.~J.}\ \bibnamefont {{Jarvis}}}, \bibinfo {author} {\bibfnamefont {M.}~\bibnamefont {{Coriat}}}, \bibinfo {author} {\bibfnamefont {G.}~\bibnamefont {{Bernardi}}}, \bibinfo {author} {\bibfnamefont {R.~P.}\ \bibnamefont {{Fender}}}, \bibinfo {author} {\bibfnamefont {S.}~\bibnamefont {{Frey}}}, \bibinfo {author} {\bibfnamefont {I.}~\bibnamefont {{Heywood}}}, \bibinfo {author} {\bibfnamefont {H.~R.}\ \bibnamefont {{Kl{\"o}ckner}}}, \bibinfo {author} {\bibfnamefont {K.}~\bibnamefont {{Grainge}}},\ and\ \bibinfo {author} {\bibfnamefont {C.}~\bibnamefont {{Rumsey}}},\ }\bibfield  {title} {\bibinfo {title} {{A close-pair binary in a distant triple supermassive black hole system}},\ }\href {https://doi.org/10.1038/nature13454} {\bibfield  {journal} {\bibinfo  {journal} {\nat}\ }\textbf {\bibinfo {volume} {511}},\ \bibinfo {pages} {57} (\bibinfo {year} {2014})},\ \Eprint {https://arxiv.org/abs/1406.6365} {arXiv:1406.6365 [astro-ph.GA]} \BibitemShut {NoStop}%
\bibitem [{\citenamefont {{Comerford}}\ \emph {et~al.}(2015)\citenamefont {{Comerford}}, \citenamefont {{Pooley}}, \citenamefont {{Barrows}}, \citenamefont {{Greene}}, \citenamefont {{Zakamska}}, \citenamefont {{Madejski}},\ and\ \citenamefont {{Cooper}}}]{2015ApJ...806..219C}%
  \BibitemOpen
  \bibfield  {author} {\bibinfo {author} {\bibfnamefont {J.~M.}\ \bibnamefont {{Comerford}}}, \bibinfo {author} {\bibfnamefont {D.}~\bibnamefont {{Pooley}}}, \bibinfo {author} {\bibfnamefont {R.~S.}\ \bibnamefont {{Barrows}}}, \bibinfo {author} {\bibfnamefont {J.~E.}\ \bibnamefont {{Greene}}}, \bibinfo {author} {\bibfnamefont {N.~L.}\ \bibnamefont {{Zakamska}}}, \bibinfo {author} {\bibfnamefont {G.~M.}\ \bibnamefont {{Madejski}}},\ and\ \bibinfo {author} {\bibfnamefont {M.~C.}\ \bibnamefont {{Cooper}}},\ }\bibfield  {title} {\bibinfo {title} {{Merger-driven Fueling of Active Galactic Nuclei: Six Dual and Offset AGNs Discovered with Chandra and Hubble Space Telescope Observations}},\ }\href {https://doi.org/10.1088/0004-637X/806/2/219} {\bibfield  {journal} {\bibinfo  {journal} {\apj}\ }\textbf {\bibinfo {volume} {806}},\ \bibinfo {eid} {219} (\bibinfo {year} {2015})},\ \Eprint {https://arxiv.org/abs/1504.01391} {arXiv:1504.01391 [astro-ph.GA]} \BibitemShut {NoStop}%
\bibitem [{\citenamefont {{M{\"u}ller-S{\'a}nchez}}\ \emph {et~al.}(2015)\citenamefont {{M{\"u}ller-S{\'a}nchez}}, \citenamefont {{Comerford}}, \citenamefont {{Nevin}}, \citenamefont {{Barrows}}, \citenamefont {{Cooper}},\ and\ \citenamefont {{Greene}}}]{2015ApJ...813..103M}%
  \BibitemOpen
  \bibfield  {author} {\bibinfo {author} {\bibfnamefont {F.}~\bibnamefont {{M{\"u}ller-S{\'a}nchez}}}, \bibinfo {author} {\bibfnamefont {J.~M.}\ \bibnamefont {{Comerford}}}, \bibinfo {author} {\bibfnamefont {R.}~\bibnamefont {{Nevin}}}, \bibinfo {author} {\bibfnamefont {R.~S.}\ \bibnamefont {{Barrows}}}, \bibinfo {author} {\bibfnamefont {M.~C.}\ \bibnamefont {{Cooper}}},\ and\ \bibinfo {author} {\bibfnamefont {J.~E.}\ \bibnamefont {{Greene}}},\ }\bibfield  {title} {\bibinfo {title} {{The Origin of Double-peaked Narrow Lines in Active Galactic Nuclei. I. Very Large Array Detections of Dual AGNs and AGN Outflows}},\ }\href {https://doi.org/10.1088/0004-637X/813/2/103} {\bibfield  {journal} {\bibinfo  {journal} {\apj}\ }\textbf {\bibinfo {volume} {813}},\ \bibinfo {eid} {103} (\bibinfo {year} {2015})},\ \Eprint {https://arxiv.org/abs/1509.04291} {arXiv:1509.04291 [astro-ph.GA]} \BibitemShut {NoStop}%
\bibitem [{\citenamefont {{Charisi}}\ \emph {et~al.}(2016)\citenamefont {{Charisi}}, \citenamefont {{Bartos}}, \citenamefont {{Haiman}}, \citenamefont {{Price-Whelan}}, \citenamefont {{Graham}}, \citenamefont {{Bellm}}, \citenamefont {{Laher}},\ and\ \citenamefont {{M{\'a}rka}}}]{2016MNRAS.463.2145C}%
  \BibitemOpen
  \bibfield  {author} {\bibinfo {author} {\bibfnamefont {M.}~\bibnamefont {{Charisi}}}, \bibinfo {author} {\bibfnamefont {I.}~\bibnamefont {{Bartos}}}, \bibinfo {author} {\bibfnamefont {Z.}~\bibnamefont {{Haiman}}}, \bibinfo {author} {\bibfnamefont {A.~M.}\ \bibnamefont {{Price-Whelan}}}, \bibinfo {author} {\bibfnamefont {M.~J.}\ \bibnamefont {{Graham}}}, \bibinfo {author} {\bibfnamefont {E.~C.}\ \bibnamefont {{Bellm}}}, \bibinfo {author} {\bibfnamefont {R.~R.}\ \bibnamefont {{Laher}}},\ and\ \bibinfo {author} {\bibfnamefont {S.}~\bibnamefont {{M{\'a}rka}}},\ }\bibfield  {title} {\bibinfo {title} {{A population of short-period variable quasars from PTF as supermassive black hole binary candidates}},\ }\href {https://doi.org/10.1093/mnras/stw1838} {\bibfield  {journal} {\bibinfo  {journal} {"Mon. Not. Roy. Astron. Soc.",}\ }\textbf {\bibinfo {volume} {463}},\ \bibinfo {pages} {2145} (\bibinfo {year} {2016})},\ \Eprint {https://arxiv.org/abs/1604.01020} {arXiv:1604.01020 [astro-ph.GA]} \BibitemShut {NoStop}%
\end{thebibliography}%

\end{document}